\documentclass[12pt]{iopart}

\usepackage{graphicx}
\usepackage{textcomp}
\usepackage{CJK}
\usepackage[dvipsnames]{xcolor}
\usepackage[colorlinks=true,urlcolor=Blue,linkcolor=Blue]{hyperref}
\usepackage[all]{hypcap}
\newcommand{\be}{\begin{equation}}
\newcommand{\ee}{\end{equation}}
\newcommand{\bea}{\begin{eqnarray}}
\newcommand{\eea}{\end{eqnarray}}

\begin{document}

\title{Storage ring mass spectrometry for nuclear structure and astrophysics research}

\author{
    Y~H ~Zhang${}^{1}$,
    Yu~A~Litvinov${}^{1,2,3}$,
    T~Uesaka${}^{4}$,
    H~S~Xu${}^{1}$
    }
\address{$^{1}$ Key Laboratory of High Precision Nuclear Spectroscopy and Center for Nuclear Matter Science, Institute of Modern Physics, Chinese Academy of Sciences, Lanzhou 730000, China}
\address{$^{2}$ GSI Helmholtzzentrum f\"{u}r Schwerionenforschung, Planckstra{\ss}e 1, 64291 Darmstadt, Germany}
\address{$^{3}$ Max-Planck-Institut f\"{u}r Kernphysik, Saupfercheckweg 1, 69117 Heidelberg, Germany}
\address{$^{4}$ RIKEN Nishina Center, RIKEN, Saitama 351-0198, Japan}

\ead{{yhzhang@impcas.ac.cn},{y.litvinov@gsi.de},{hushan@impcas.ac.cn},{uesaka@riken.jp}}
\vspace{10pt}
\date{today}
\begin{abstract}

In the last two and a half decades ion storage rings have proven to be powerful tools for 
precision experiments with unstable nuclides in realm of nuclear structure and astrophysics. 
There are presently three storage ring facilities in the world at which experiments with stored radioactive ions are possible.
These are the ESR in GSI, Darmstadt/Germany, the CSRe in IMP, Lanzhou/China, and the R3 storage ring in RIKEN, Saitama/Japan. 
In this work, an introduction to the facilities is given. 
Selected characteristic experimental results and their impact in nuclear physics and astrophysics are presented. 
Planned technical developments and the envisioned future experiments are outlined.\\

\noindent{Keywords}: Ion storage ring, storage ring mass spectrometry, nuclear binding energy\\

\noindent{(Some figures may appear in colour only in the online journal)}
\end{abstract}

%
%
%
\maketitle
%
%
\section{Introduction}
Nuclei are many-body quantum systems built out of two types of fermions, protons and neutrons, called the nucleons~\cite{Bohr-NS1969}.
There is a complex interplay of strong, weak and electromagnetic interactions acting between the nucleons.
Likewise for any closed system, a firm understanding of the nuclear force must result in a reliable estimation of the total energy of the nucleus, that is its mass.
In other words, nuclear mass is a fundamental property of each atomic nucleus reflecting its internal structure.
Masses of neighbouring nuclides plotted together form the nuclear mass surface. 
This is in general a smooth surface, on which new structure effects may be seen 
as irregularities~\cite{Novikov-NPA2002}.
Indeed, historically the well-known nucleon-nucleon pairing correlations~\cite{Heisenberg-ZP1932} and shell structure~\cite{Mayer-PR1949, Haxel-PR1949} 
were discovered in stable nuclei by investigating the systematics of nuclear masses.

Nuclear mass measurements began about a century ago with 
pioneering works of sir J.~J.~Thomson~\cite{Thomson-PM1912} and F.~W.~Aston~\cite{Aston-PM1923}.
This is still a broad and fast developing 
field of modern research (see, e.g., \cite{Lunney-RMP2003, Blaum-PR2006, Munzenberg-AIP2010, 100YMS, Wollnik-HI2015}).
The questions addressed today by nuclear physics require the knowledge 
of properties of short-lived nuclei far from the valley of stability.
New measured masses can be used for numerous investigations like, for instance, studying 
the limits of nuclear existence (see, e.g., \cite{Novikov-NPA2002, Rauth-PRL2008}), 
changes in nuclear deformation (see, e.g., \cite{Naimi-PRC2012, Manea-PRC2013}), 
nuclear collectivity (see, e.g., \cite{Cakirli-PRL2009, Casten-PRL2014})
and correlations (see, e.g., \cite{Neidherr-PRL2009, Boehm-PRC2014}),
the structure of halo-nuclei (see, e.g., \cite{Geithner-PRL2008, Nakamura-PRL2009}), 
fundamental symmetries and interactions (see, e.g., \cite{Blaum-PRL2003, Kellerbauer-PRL2004, 
Naimi-PRL2010, Eliseev-PRL2011, Nesterenko-PRC2014, Wienholtz-Nature2013}),
and many others.
One of the present challenges in nuclear structure is to understand the shell structure evolution at extreme neutron-to-proton ratios,
where the well-known magic numbers may disappear and the new shell closures may 
show up (see, e.g., \cite{Dworschak-PRL2008, Kanungo-PRL2009,Kanungo-PS2013,Otsuka-PS2013}).
In nuclear astrophysics, the nuclear masses are an indispensable input 
to the modelling of nucleosynthesis processes in stars
(see, e.g., \cite{Rodriguez-PRL2004, Weber-PRC2008, Baruah-PRL2008,
Haettner-PRL2011, Bertulani-PR2010,  Arcones-JPG2013, Langanke-PS2013, Wolf-PRL2013, Langanke-PS2015}).
Dependent on the specific objective, high mass accuracies from some $10^{-5}$ up to $10^{-8}$ can be required \cite{Blaum-PR2006}.
For a comprehensive review on the applications of nuclear masses in various fields of modern research, the reader is referred to a dedicated issue \cite{100YMS}.

\begin{figure}[h!]
\begin{center}
  \includegraphics[width=16cm]{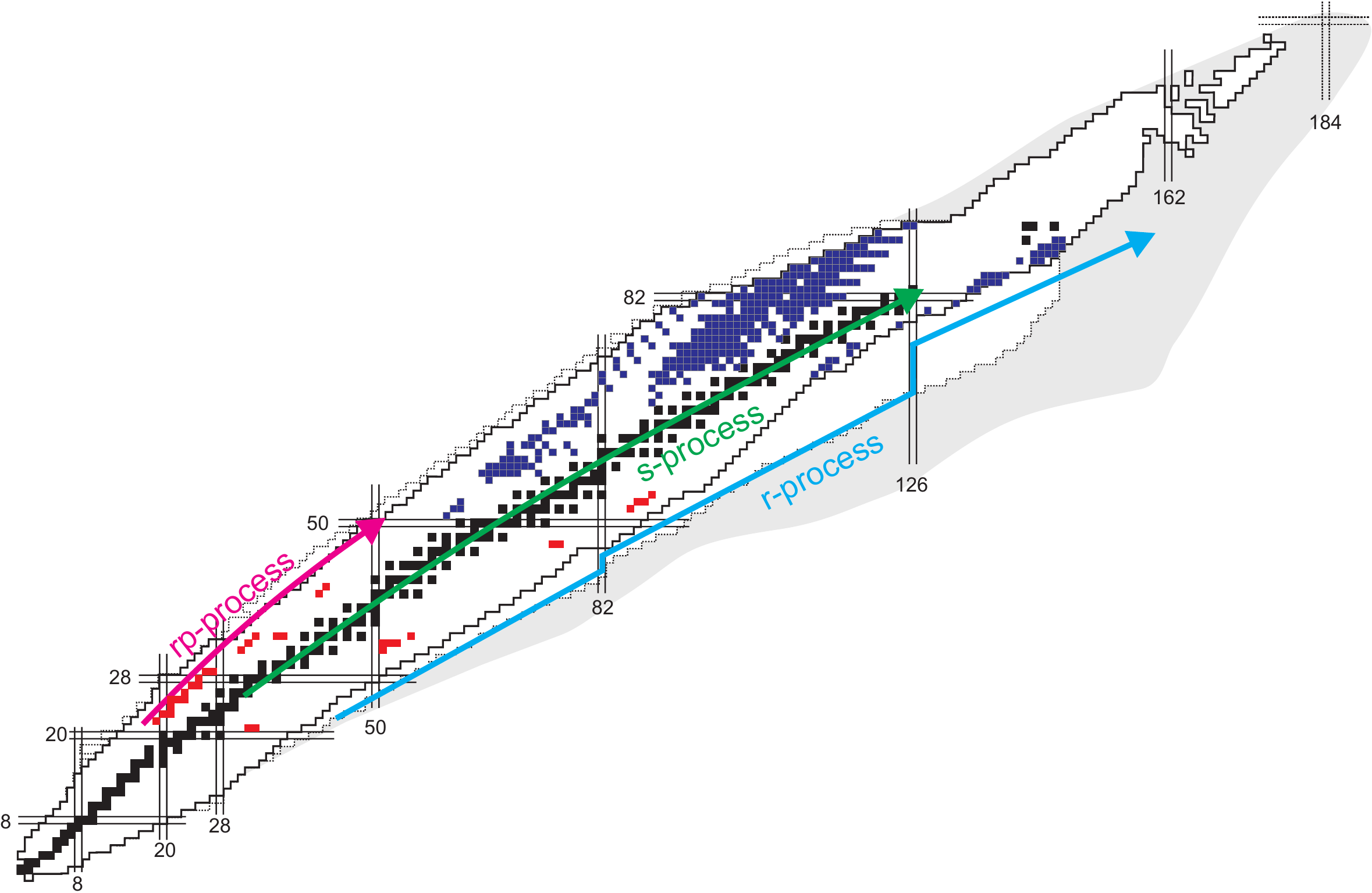}\\
  \caption{(Colour online) The chart of nuclides. Stable nuclides are shown with black colour. White region of nuclei with solid black borders represents the nuclei which 
  were synthesised in laboratory. In grey shown is the expected region of existence of bound nuclei. 
  Nuclei whose masses were determined with help of the storage ring mass spectrometry are indicated with red and blue colours, 
  respectively, for Isochronous and Schottky mass spectrometry, see Section~\ref{section:srms}. 
  Masses obtained indirectly by using the data from $\alpha$ and proton decay spectroscopy on the measured $^{209}$Bi projectiles are also indicated with blue colour. 
  The white region with dotted boundary reflects the nuclei which will be accessible at the next generation 
  radioactive-ion beam facilities like HIAF or FAIR, see Section~\ref{section:future}.
  Locations of slow neutron capture (s-process), rapid proton (rp-process), and rapid neutron (r-process) process of nucleosynthesis are schematically illustrated by arrows.
  Modified from Refs. \cite{Blaum-JPCS2011, Bosch-PPNP2013}.
  }\label{fig:nchart}
\end{center}
\end{figure}

About 7000 nuclides are expected to exist in nature~\cite{Erler-Na2012}, 
see nuclidic chart in Figure~\ref{fig:nchart}, while masses of 2438 nuclides are known experimentally at present~\cite{AME12}.
The nuclides with still unknown masses are difficult to measure due to extremely low production cross sections and short lifetimes.
Most of them are barely within reach at present and will remain inaccessible at the presently in construction next-generation radioactive beam facilities~\cite{Kurcewicz-PLB2012}.
As a consequence, the properties of nuclei, for instance, at the end point of the rapid-neutron capture nucleosynthesis process (r-process) will need to be calculated~\cite{Sun-PRC2008, Sun-FP2015}.
However, accurate predictions of nuclear masses is still a big challenge for theories (see, e.g., \cite{Sobiczewski-PRC2014a, Sobiczewski-PRC2014b}).
Therefore, probably the most important application of newly measured masses is to test and to constrain nuclear theories.

New mass measurements inevitably require very efficient and fast experimental techniques.
Penning traps and storage rings are examples of such techniques (see, e.g., \cite{Blaum-JPCS2011, Blaum-PS2013, Franzke-MSR2008, STORI14}).
For the review on the former see \cite{Blaum-PR2006}, whereas we focus in this paper on the measurements with the latter. 
In this article we present a brief introduction to the facilities, experimental techniques, 
characteristic experiments and their impact in nuclear physics and astrophysics.
Planned technical developments and the envisioned future experiments are outlined.

\section{Storage ring facilities}
To efficiently use the rarely produced exotic nuclides, it is of a straightforward advantage to store them in a trapping facility.
There are presently three laboratories where a combination of a radioactive-ion-beam with a storage-ring facilities is realised.

\begin{figure}[h!]
\begin{center}
  \includegraphics[width=16cm]{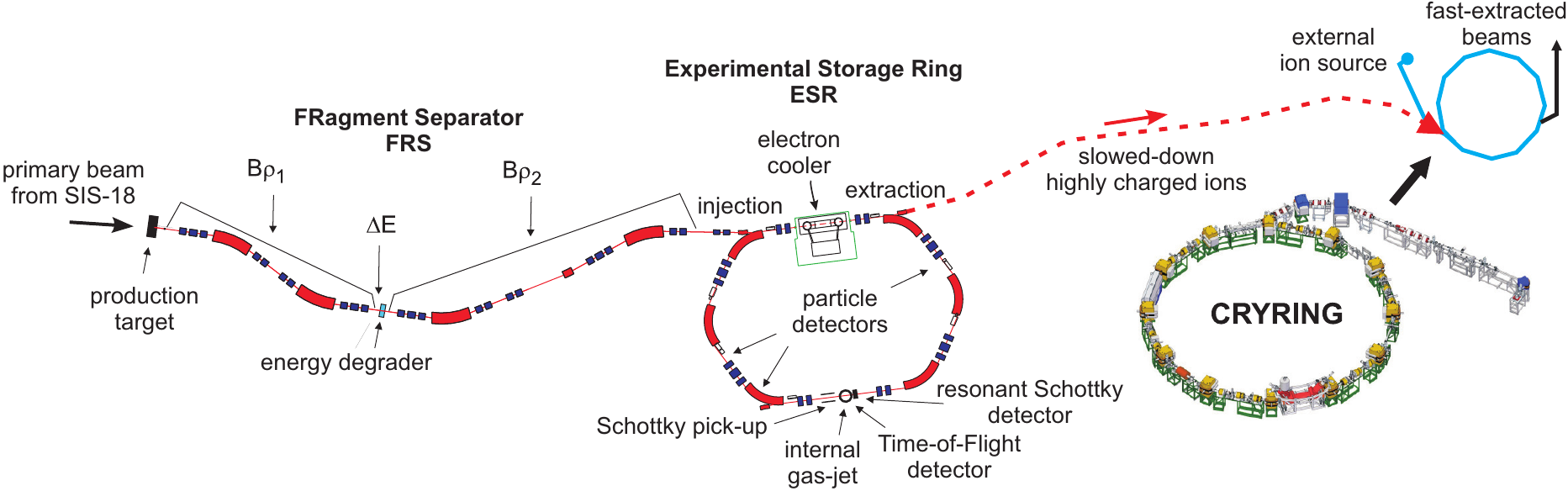}\\
  \caption{(Colour online) The FRS-ESR arrangement at GSI. 
The two stages for in-flight $B\rho-\Delta E-B\rho$ separation and the energy degrader in the FRS 
as well as the electron cooler and the main detection systems in the ESR are shown. 
The foreseen location of the low-energy storage ring CRYRING (see insert) 
is indicated together with the beam-line connecting it to the ESR 
for transporting slowed-down highly-charged stable or radioactive ions. Modified from Ref. \cite{Walker-IJMS2013}.}\label{fig:gsi}
\end{center}
\end{figure}

The facility at GSI Helmholtzzentrum f{\"u}r Schwerionenforschung was chronologically the first one to be taken into operation in 1990.
The facility is schematically illustrated in Figure~\ref{fig:gsi}.
It comprises the heavy-ion synchrotron, SIS, \cite{Blasche-IEEE1985} 
the projectile fragment separator, FRS, \cite{Armbruster-AIPCP1987, Geissel-NIM1992} and the experimental storage ring, ESR, \cite{Franzke-NIM1987}.
A low-energy storage ring, CRYRING, which was until recently in operation at Stockholm university, is being presently installed behind the ESR \cite{Lestinsky-PS2015}. A detailed description of the GSI facilities can be found in Ref.~\cite{Franzke-MSR2008} and references cited therein.

\begin{figure}[h!]
\begin{center}
\includegraphics[width=0.7\textwidth]{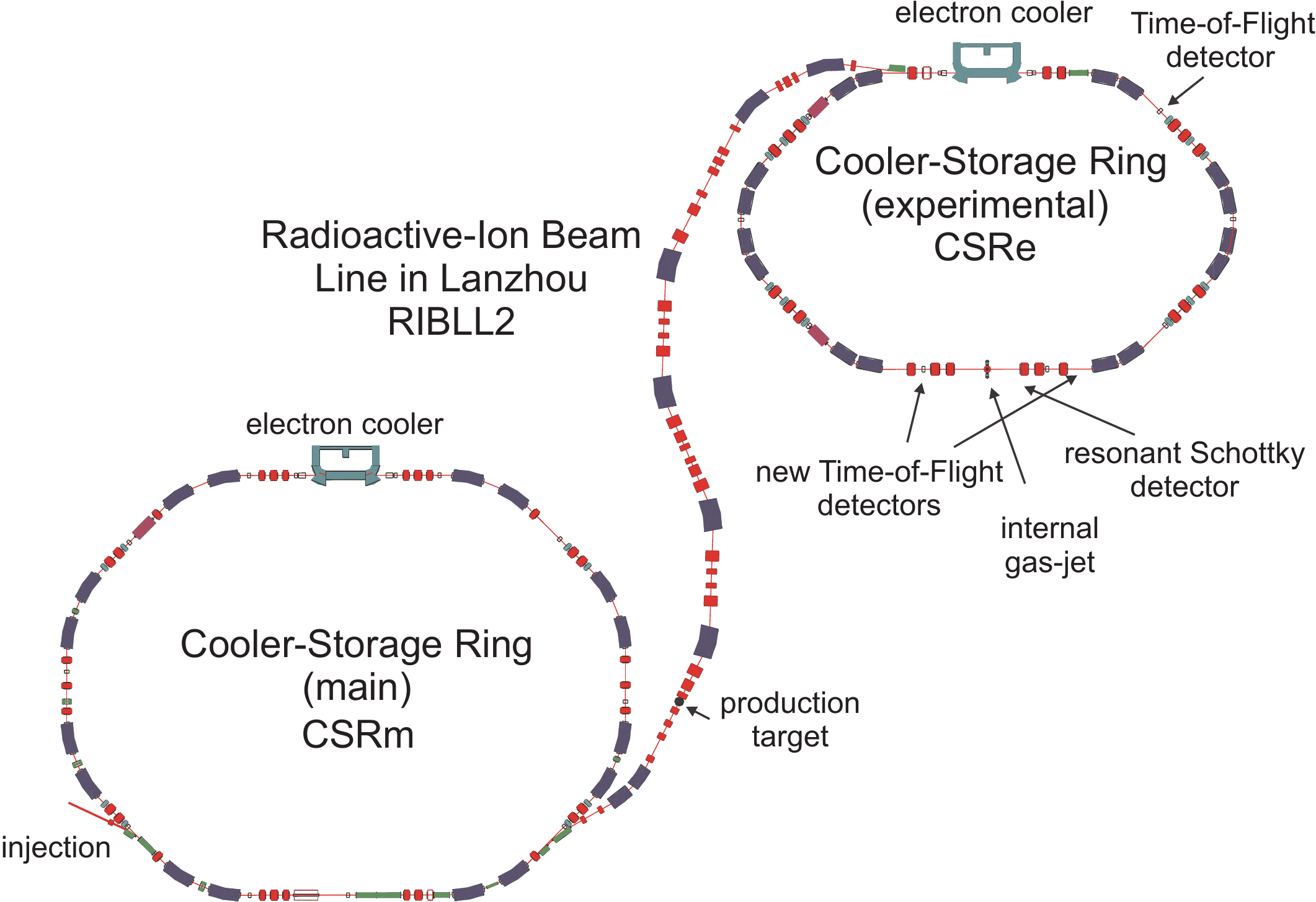}
  \caption{(Colour online) Schematic layout of the high-energy part of the 
Cooler Storage Ring at the Heavy Ion Research Facility in Lanzhou (HIRFL-CSR)
located at the Institute of Modern Physics in Lanzhou (IMP), Chinese Academy of Sciences. 
The heavy ion synchrotron CSRm, the in-flight fragment separator RIBLL2 and the experimental storage ring CSRe are indicated.
Modified from Refs. \cite{Xu-IJMPE2009, Xu-IJMS2013}}\label{fig:imp}
\end{center}
\end{figure}

The second facility, HIRFL-CSR complex, is in operation since 2007 at the Institute of Modern Physics, Chinese Academy of Sciences in Lanzhou.
It is illustrated in Figure~\ref{fig:imp} and comprises the main cooler-storage ring, CSRm, used as a synchrotron, radioactive ion beam line in Lanzhou 2, RIBLL2, used as a fragment separator, and the experimental cooler-storage ring, CSRe.
A detailed description of the HIRFL-CSR acceleration complex can be found in Refs.~\cite{Xia-NIM2002, Xiao-IJMPE2009, Zhan-NPA2010, Yuan-NIM2013}.

\begin{figure}[h!]
\begin{center}
  \includegraphics[width=16cm]{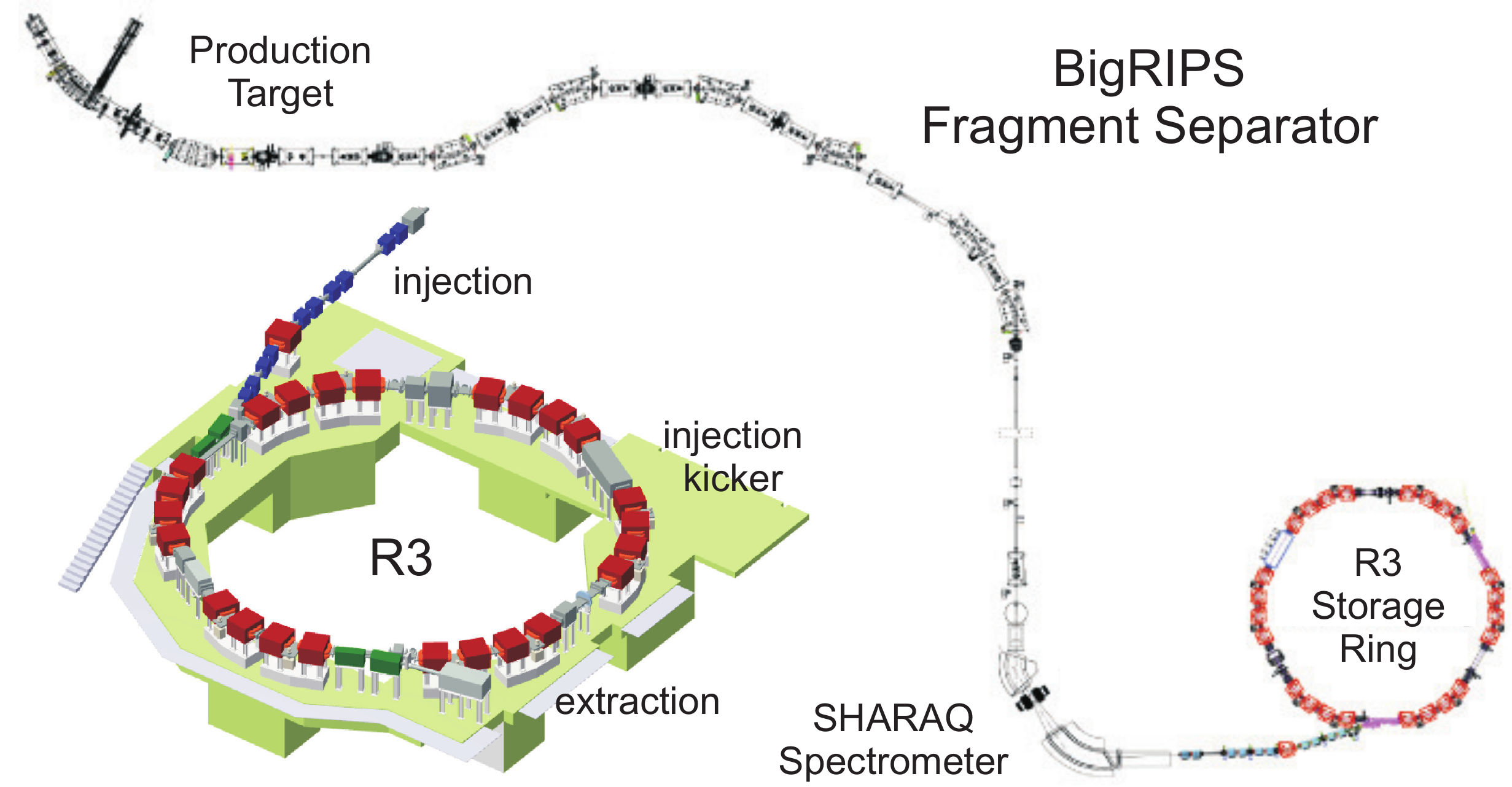}\\
  \caption{(Colour online) Schematic view of the BigRIPS-R3 facility at RIKEN. 
  Exotic nuclei produced in the production target are identified in flight with the BigRIPS separator.
  If an ion of interest is observed, the injection kicker is fired and the particle is stored in the R3 \cite{Ozawa-PTEP2012}.
  The inset shows a 3D model of the R3 ring.}\label{fig:riken}
\end{center}
\end{figure}

The last but not least, the third facility, the Rare RI Ring (R3), 
was commissioned in 2015 at RIKEN Nishina Center in Japan \cite{Ozawa-PTEP2012}. 
Different from GSI and IMP complexes, the driver accelerator in RIKEN is not a synchrotron but a superconducting ring cyclotron, SRC, 
which provides instead of a pulsed beam a quasi-DC beam.
The facility is illustrated in Figure~\ref{fig:riken}.
We note that to date the highest primary-beam intensities worldwide of, e.g., $^{238}$U stable ions, are available at RIKEN.

An indispensable prerequisite for the experimental investigations of short-lived nuclei, 
is their production and cleaning from inevitable--more abundant--contaminations \cite{Geissel-ARNPS1995}. 
All three facilities employ the so-called {\it in-flight separation}, for which the facilities are equipped with the fragment separators.
The intense primary beams are accelerated by the SIS-18/CSRm synchrotrons or by the SRC cyclotron to energies of several hundreds MeV/u.
These beams are focused on the production targets in which the nuclei of interest are produced.
Projectile fragmentation or, in case of uranium beam, also fission reactions are employed for this purpose. 
The secondary particles emerge from the target as highly-charged ions.
Dedicated variable-energy degraders can be employed at focal planes of the separators, 
which, in combination with the magnetic rigidity ($B\rho$) analyses, before and after the degrader, 
enables the so-called $B\rho-\Delta E-B\rho$ separation, where $\Delta E$ reflects the energy loss in the degrader \cite{Geissel-NIM1992}. 
This is a powerful method, which allows for separation of clean mono-isotopic beams. 

After the separation, selected exotic ions can be injected into the corresponding storage ring
for precision experiments in nuclear structure and astrophysics 
(see, e.g, \cite{Zhong-JPCS2010, Bo-PRC2015, Woods-PS2015, Doherty-PS2015, vonSchmid-PS2015, Zamora-PS2015}, 
as well as atomic and fundamental physics 
(see, e.g., \cite{Botermann-PRL2014, Tashenov-PRL2014, Lochmann-PRA2014, Beyer-JPB2015, Nortershauser-PS2015, Sanchez-PS2015, Jungmann-PS2015}). 
In cases of GSI and IMP facilities, the fast extraction of the primary beam from the synchrotron allows all produced and transmitted particles to be stored in the ring.
In the case of the RIKEN setup, the DC-nature of the primary beam results in the storage and investigation of a single particle at a time.
The biggest advantage of such system is that each particle is identified in BigRIPS \cite{Kubo-PTEP2012} 
and no identification is needed in the storage ring itself.

Since the storage times in the ring can reach hours, different manipulations with
the ions like beam cooling, slowing down, or preparation of clean mono-isomeric beams 
can be conducted (see, e.g., \cite{Scheidenberger-HI2006, Meshkov-PS2015, Brandau-HI2010, Doherty-PS2015a}). 

\section{Storage ring mass spectrometry}
\label{section:srms}
Storage ring mass spectrometry was pioneered at GSI in Darmstadt~\cite{Franzke-MSR2008, Geissel-PRL1992}.
Heavy-ion storage rings are complicated facilities which can contain numerous components, 
like, e.g., kicker magnets to inject and extract particles, 
dipole magnets to bend the trajectories, quadrupole magnets to focus the particles, 
sextupole magnets to correct for aberrations, cooling devices, de-/acceleration cavities, various detector and diagnostic setups, etc.
We note that the R3 ring is composed of only dipole magnets \cite{Wakasugi-JPSJ2015}.

Due to the Lorentz force, particles with different momenta are bent differently by the ring magnets and thus travel 
along different paths in the ring. 
Since the revolution time $T$ is
\be
T=\frac{L}{v},
\ee
 where $L$, $v$ are the path length and the velocity of the circulating particle. 
The fractional change of the revolution time or the revolution frequency $f=1/T$ is
\be
\frac{\Delta T}{T} = -\frac{\Delta f}{f} = \frac{\Delta L}{L} - \frac{\Delta v}{v}.
\label{equ1}
\ee
For particles that vary only in momentum, $p$, the velocity difference is:
\be
\frac{\Delta v}{v}=\frac{1}{\gamma^2}\frac{\Delta p}{p}.
\ee
Here $\gamma=1/\sqrt{1-\beta^2}$ is the relativistic Lorentz factor, $\beta=v/c$, and $c$ speed of light in vacuum.
Therefore, we can rewrite Eq.~(\ref{equ1}) as
\be
\frac{\Delta T}{T} = \bigg( \alpha_{p} - \frac{1}{\gamma^2}\bigg)\frac{\Delta p}{p} = -\eta \frac{\Delta p}{p},
\ee
where $\eta$ and $\alpha_p$ are the so-called {\it phase-slip factor} and {\it momentum compaction factor}, respectively, connected as \cite{Franzke-MSR2008}
\be
\eta=\frac{1}{\gamma^2}-\alpha_p=\frac{1}{\gamma^2}-\frac{1}{\gamma_t^2},
\ee
where the {\it transition energy}, $\gamma_t$ is defined as
\be
\gamma_t\equiv\frac{1}{\sqrt{\alpha_p}}.
\ee
The physical meaning of $\alpha_p$ \cite{Franzke-MSR2008} is that it reflects the ratio between the relative change in the orbital length
and the relative change in the magnetic rigidity $B\rho$ of the stored ions. 
The $\alpha_p$ can also be deduced from the dispersion function $D(s)$ of the ring. 
\be
\alpha_p\equiv\frac{dL/L}{d(B\rho)/(B\rho)}=\frac{1}{L_0}\oint\frac{D(s)}{\rho}ds.
\ee
Here $s$ denotes the coordinate along the reference orbit $L_0$ in the ring, 
and $\rho$ is the radius of the curvature of this reference orbit in the bending sections~\cite{Franzke-MSR2008}. 
Detailed investigations of the $\alpha_p$ as a function of the ESR magnetic rigidity can be found in \cite{Yan-PS2015}.

If a motion of particles with different mass-to-charge ratios, $m/q$, and momenta is considered, 
then the fractional momentum change can be expressed as
\be
\frac{\Delta p}{p} = \frac{\Delta(m/q)}{m/q}+\gamma^2 \frac{\Delta v}{v}.
\ee
and Eq.~(\ref{equ1}) can be written as
\be
\frac{\Delta f}{f} = -\frac{\Delta T}{T} = -\frac{1}{\gamma_t^2}\cdot\frac{\Delta (m/q)}{m/q}+
\bigg(1-\frac{\gamma^2}{\gamma_t^2}\bigg)\frac{\Delta v}{v}.
\label{main}
\ee
Equation~(\ref{main}) is the basic equation for the storage ring mass spectrometry \cite{Radon-PRL1997, Radon-NPA2000}.
By inspecting this equation one can see that the measurement of the revolution times / revolution frequencies
of the stored ions allows the determination of their $m/q$ values only if the second term on the right hand side can be neglected. 
However, secondary beams inevitably--mainly due to the production process--have 
a velocity spread $\Delta v/v$ of the order of a few percent \cite{Geissel-ARNPS1995}.
The magnitude of the second term on the right hand side of Eq.~(\ref{main}) 
directly affects the achievable mass resolving power and thus has to be made as small as possible.
There are two complementary experimental methods, namely {\it Schottky} (SMS) and {\it Isochronous} (IMS) {\it Mass Spectrometry}, 
which have been developed for accurate mass measurements \cite{Franzke-MSR2008, Munzenberg-IJMS2013}.

\subsection{Schottky Mass Spectrometry}
In the SMS, the velocity spread can be reduced by stochastic~\cite{Nolden-NIM2004} and electron~\cite{Steck-NIM2004} cooling, 
which force all stored ions towards the same mean velocity. 
Intra-beam scattering acts against the cooling process and the velocity spread of the stored ions is proportional to their number.
For beam intensities of below about a thousand ions, the average distance between 
the particles increases to a few centimetres and a significant reduction of velocity 
spread is observed \cite{Steck-PRL1996}, which is understood 
as if the intra-beam scattering between electron-cooled ions is just disabled \cite{Hasse-JPB2003}.
Since we deal with exotic nuclides with small production rates, the above situation represents the typical conditions after the electron cooling.
The velocity spread of the cooled ions of roughly $5\cdot10^{-7}$ can be assumed.
Thus the second term on the right hand side of Eq.~(\ref{main}) can be neglected.
The revolution frequencies are measured non-destructively by applying the {\it Schottky-noise spectroscopy} \cite{Borer-1974}. 
Two detector types are employed for frequency measurements.
Their basic working principle can be compared to the one of an antenna.

\begin{figure}[htb]
\begin{center}
\includegraphics*[width=0.57\textwidth]{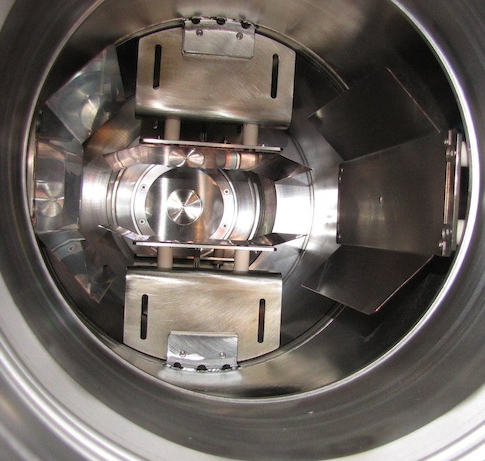}
\includegraphics*[width=0.414\textwidth]{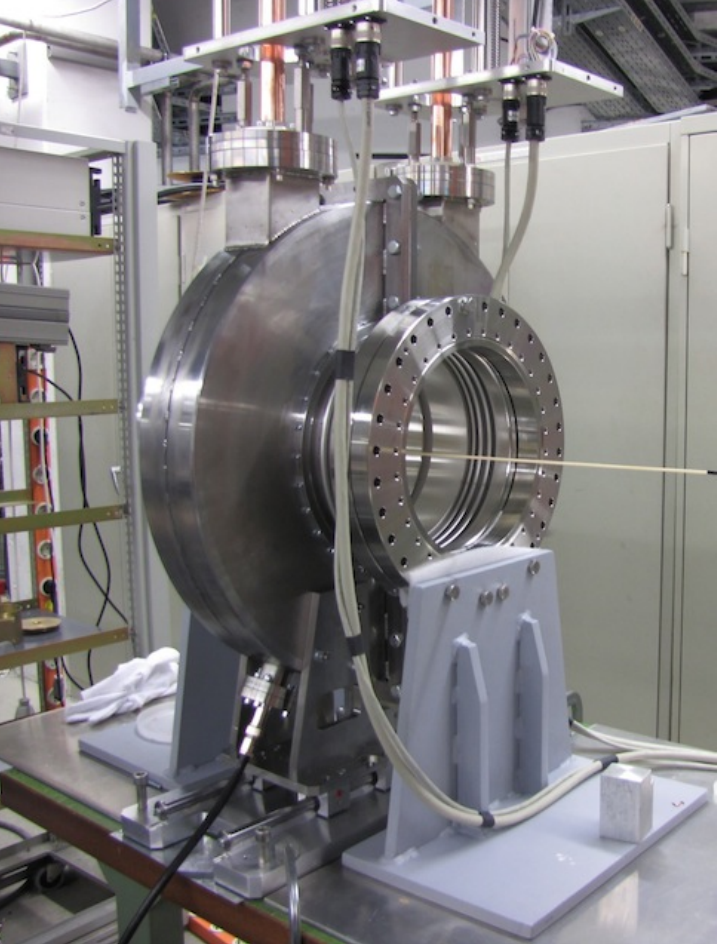}
\end{center}
{\caption{(Colour online) Left: a photo of a capacitive Schottky detector inside the ESR aperture. Two pairs of metal plates are used for vertical and horizontal analyses. 
A zoomed version from Ref. \cite{Bosch-PPNP2013}. Right: a photo of a resonant-cavity Schottky detector installed on a test stand outside of ring \cite{Sanjari-PS2013}. 
Taken from Ref. \cite{Bosch-PPNP2013}. Photos are courtesy to Shahab Sanjari and Peter Petri, GSI, Darmstadt.} 
\label{fig:schottkies}}
\end{figure}

The first detector type represents a pair of electrostatic pick-up electrodes \cite{Schaaf-PHD}, 
which are copper plates installed in a parallel geometry inside the ring aperture, see left panel of Figure~\ref{fig:schottkies}.
To comply with the ultra-high vacuum (UHV) conditions of a storage ring with the rest gas pressure of $10^{-10}-10^{-12}$~mbar,
the detector components have to be made out of dedicated materials and have to stand baking up to $150-200~^\circ$C.

\begin{figure}[htb]
\begin{center}
\includegraphics*[width=\textwidth]{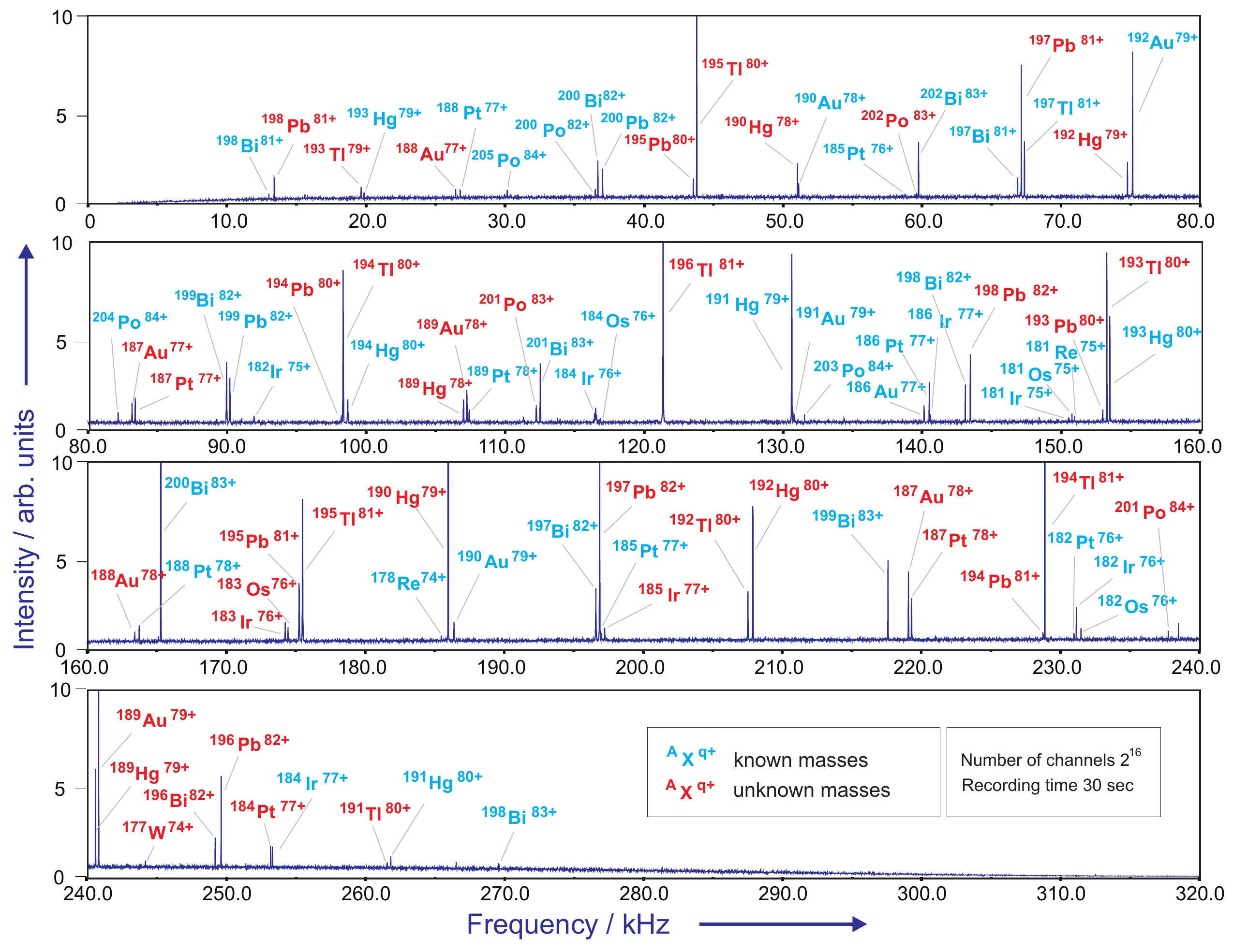}
\end{center}
{\caption{(Colour online) Schottky frequency spectrum of $^{209}$Bi projectile fragments stored in the ESR and recorded by the capacitive Schottky detector, see left panel of Figure~\ref{fig:schottkies}.
The spectrum is subdivided in 4 parts and covers the entire storage acceptance of the ESR. 
Please note that the acceptance of the ESR ($\Delta(m/q)/(m/q)=\pm1.2$\%) allows for storing same nuclides 
in up to three charge states, see $^{198}$Bi$^{81+,82+,83}$ at about 12 kHz, 143 kHz and 270 kHz respectively.
The spectrum was accumulated for 30 seconds. 
It was taken at the 30$^{th}$ harmonics of the revolution frequency and the 60~MHz offset frequency was subtracted prior to the Fourier analysis.
Taken from Ref. \cite{Litvinov-NPA2005}.
} 
\label{fig:sms_spectrum}}
\end{figure}

The stored ions circulate in the ring about a million times per second. 
Each ion induces at each revolution a mirror charge on the electrodes. 
The noise from the detector is amplified and analysed \cite{Nolden-AIP2006}.
Since the signals from stored ions are periodic, the Fourier transform of the noise yields a revolution frequency spectrum or {\it Schottky frequency spectrum}.
An example of such Schottky frequency spectrum is depicted in Figure~\ref{fig:sms_spectrum} \cite{Litvinov-NPA2005}.
Typically, the noise at high harmonic numbers of the revolution frequency is analysed.
For instance the detector at the ESR is operated at $28^{th}-34^{th}$ harmonics, 
which corresponds, taken the revolution frequency of the ions of about 2~MHz, to about 60~MHz.
This is done to increase the achievable frequency resolution for a fixed recording time, see reference~\cite{Franzke-MSR2008, Nolden-AIP2006} for more details.
The quality factor of such a detector is small (in the case of the ESR it is only about 8) but allows to 
simultaneously cover the overall frequency acceptance of the storage ring.

The second detector type is based on a resonance cavity with a much higher quality factor than for the previous detector type \cite{Nolden-NIM2011}.
Presently the detector is not bakeable and is separated from the UHV by using a ceramic gap, which allows for electromagnetic waves to pass through.
The particles passing the detector excite resonance modes in the resonator volume.
The example of such detector, placed on the test bench outside of the ring, is shown in the right panel of Figure~\ref{fig:schottkies}.
The noise from the cavity is also Fourier analysed resulting in the frequency spectrum.
The higher quality factor allows for analysing data at higher harmonic numbers.
For instance at the ESR the detector is operated at about 250~MHz, corresponding to the 125$^{th}$ harmonics of the revolution frequency.
If comparing two detector types: 
For the same data taken with both detectors in the ESR, 
the signal-to-noise characteristics of the cavity-based one is by about a factor of 100 better than of the capacitive detector.
However, the higher sensitivity is achieved in a smaller frequency range that can simultaneously be monitored.

The only disadvantage of SMS is that the cooling process lasts a few seconds. 
Thus, only the nuclides with half-lives longer than about 1~s can be investigated \cite{Litvinov-HI2001}.
The mass resolving power $m/\Delta m\sim750~000$ is routinely achieved in SMS measurements.

The intensity of the frequency peaks in the Schottky spectra is proportional to the corresponding number of stored ions \cite{Litvinov-RPP2011}.
Thus, by tracing in time the intensities of the stored ions ({\it time-resolved SMS} \cite{Litvinov-NPA2004, Geissel-NPA2004}), their half-lives can be measured. 
If an ion decays in the storage ring, then its $m/q$ changes, so the radius of its orbit changes 
(for instance, it increases for $\beta ^+$ decay and decreases for $\beta ^-$ decay) \cite{Ohtsubo-PRL2005, Atanasov-JPB2015}. 
Depending on the difference in $m/q$ ratios of the parent and daughter ions, 
this may cause the ion to further circulate in the ring and be monitored by the SMS or to leave the ring acceptance.
In the latter case, judicious placement of heavy-ion detectors behind bending magnets 
in the ring allows for the interception of daughter ions, and thus they can be counted with good detection efficiency.
As a result, decay branchings can accurately be determined. 

\subsection{Isochronous Mass Spectrometry}
The IMS is ideally suited for mass measurements of the shortest-lived nuclides~\cite{Hausmann-HI2001}.
In the cases of the ESR and CSRe, it is based on the isochronous ion-optical mode of the ring, 
in which $\Delta{v}/v$ of the ions -- injected with energies corresponding to $\gamma\approx\gamma_t$ -- is nearly exactly compensated 
by the lengths of their closed orbits in the ring~\cite{Hausmann-NIM2000}. 
In a simplified description this means, that a faster particle moves on a relatively longer orbit as compared 
to a slower particle of the same type, which moves on the correspondingly shorter orbit.
Thus the second term on the right hand side of Eq.~(\ref{main}) is (in first order) zero.

\begin{figure}[htb]
\begin{center}
\includegraphics*[width=0.46\textwidth]{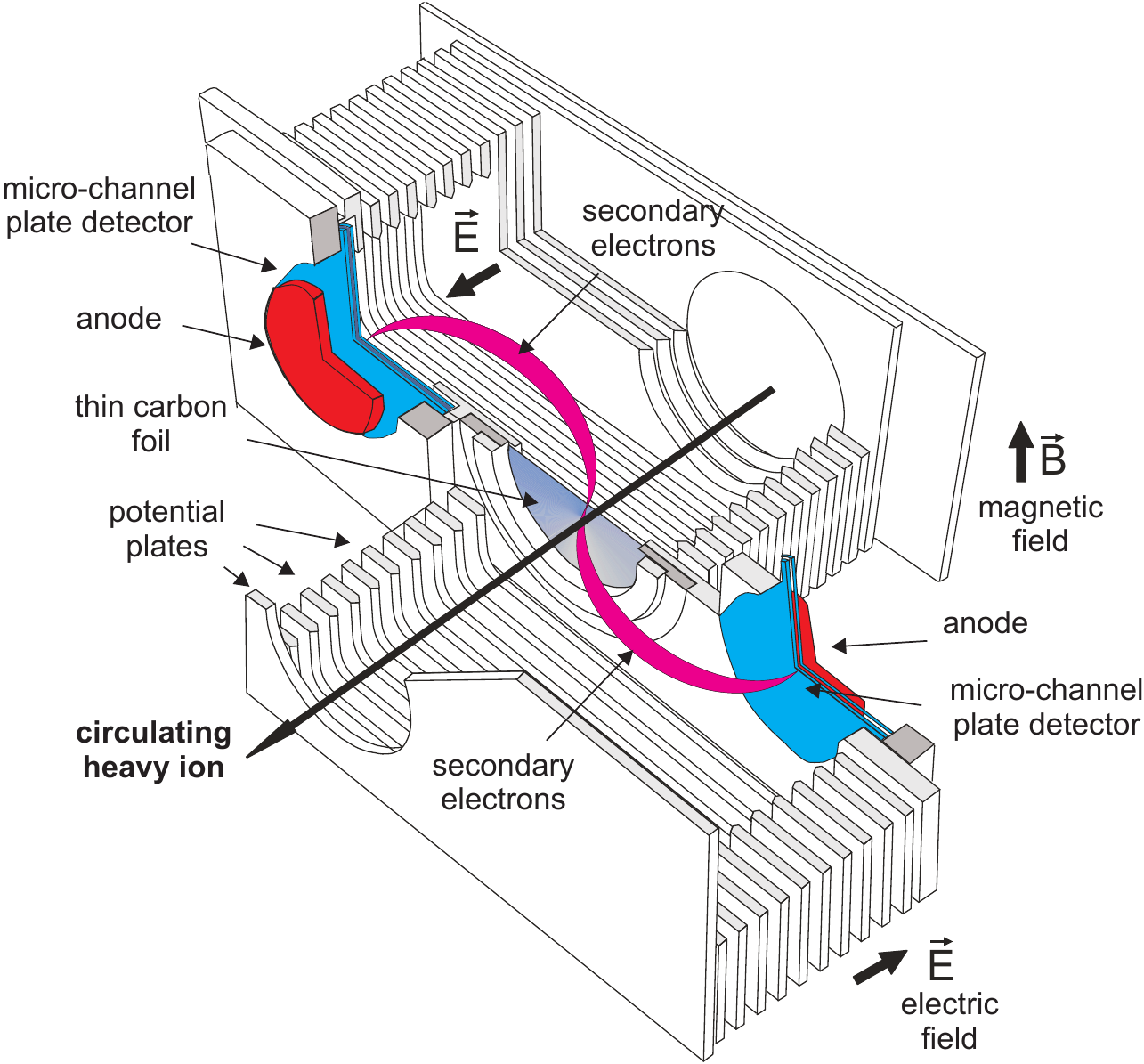}
\includegraphics*[width=0.53\textwidth]{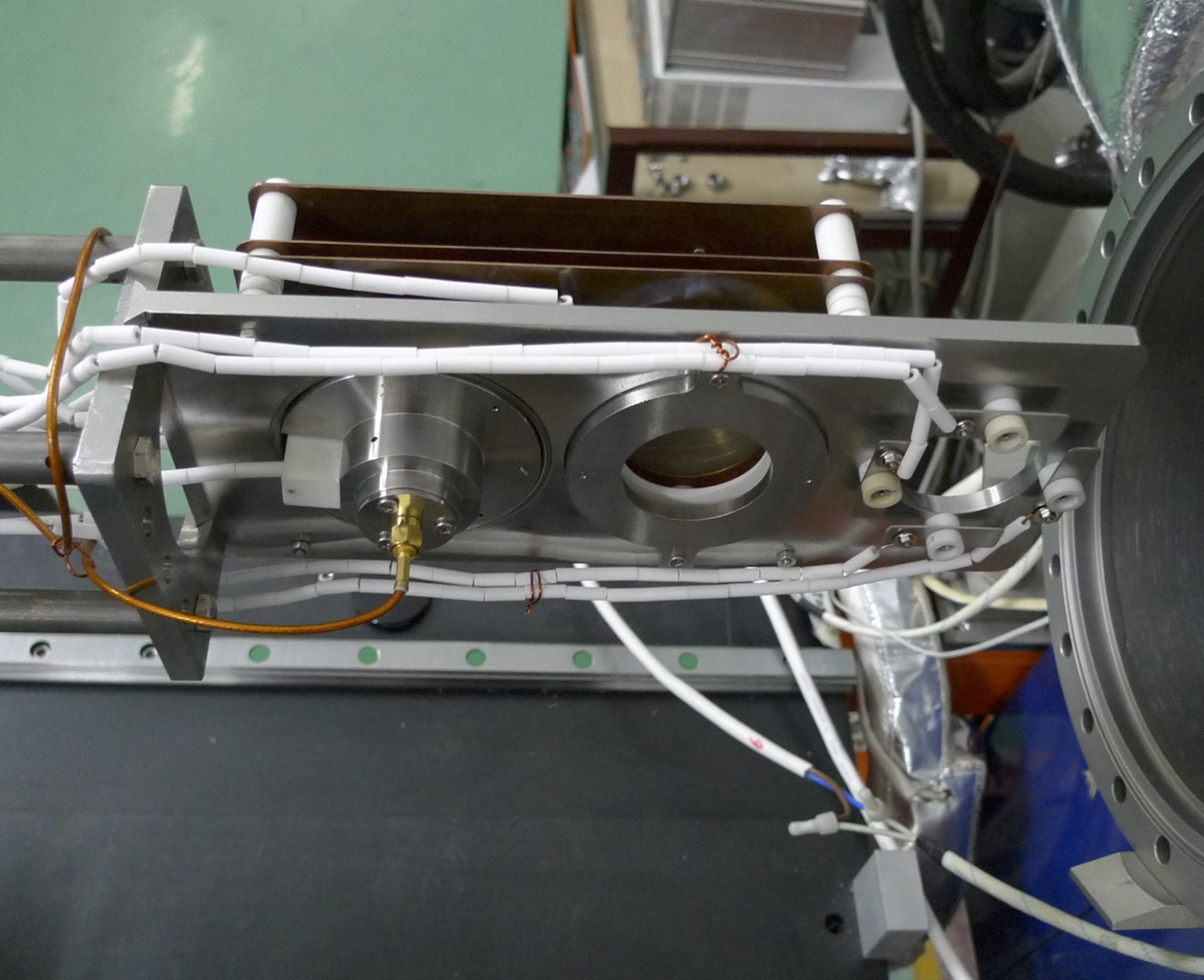}
\end{center}
{\caption{(Colour online) Left: a schematic view of the UHV-capable time-of-flight detector installed in the ESR. 
The ions penetrate the foil at each revolution in the ring and release secondary electrons from both sides of the foil.
The electrons are guided by electric ($\vec{\bf E}$) and magnetic ($\vec{\bf B}$) fields to micro-channel plate detectors, 
which provide fast timing signals. The latter are then used to determine the revolution frequencies (revolution times) of the ions.
Modified from Ref. \cite{Bosch-PPNP2013}.
Right: a photograph of the detector installed in the CSRe. Different from the detector at the ESR, 
only the secondary electrons emitted in the forward direction are measured \cite{Bo-NIM2010}.
Photo is courtesy to Xiaolin Tu and Wang Meng, IMP, Lanzhou. Taken from Ref. \cite{Xu-IJMS2013}.
} 
\label{fig:tofdet}}
\end{figure}

The technique employed at the R3 is directly based on the well-known isochronicity property of a cyclotron device. 
Made out of dipoles only, the R3 is a cyclotron-type storage ring that maintains the isochronous condition in a wide momentum range. 
A radial dependence of the magnetic field is adjusted by 10 trim coils \cite{Abe-PS2015}
so that the second term in the right hand side of Eq. (9) is made as small as possible. 

In the ESR and CSRe, the revolution times of the ions are measured by a dedicated time-of-flight (ToF) detector
inserted directly into the UHV of the rings~\cite{Bo-NIM2010, Trotscher-NIM1992, Nagae-NIM2013, Zhang-NIM2014b}, see Figure~\ref{fig:tofdet}.
In the detector, ions pass at each revolution through a thin carbon foil inserted into the ring aperture and release secondary electrons from the foil surface. 
These electrons are guided by weak electric and magnetic fields to multi-channel plate detectors providing a timing stamp.
This enables the passage of each ion to be recorded, and hence its revolution frequency to be determined. 
The number of released secondary electrons depends on the charge of the ion 
and in some cases can be used to resolve ions with very close revolution frequencies \cite{Shuai-PLB2014}.
After several hundreds of passages through the carbon foil, the ions lose too much energy and are no longer stored. 
Typically, the ions which accomplish a few hundreds of revolutions are considered in the analysis \cite{Sun-NPA2008, Tu-NIM2011}.
This allows for the determination of the revolution times/revolution frequencies 
with statistical uncertainties that can be neglected in comparison to other uncertainties.
In the CSRe, the minimal number of revolutions is set to 400, which corresponds to about 300~$\mu$s.
For a reliable ion identification at least 20 passes through the foil are needed, which takes about 10 $\mu$s. 

At the R3 ring, the foil detector at the injection and a plastic scintillation 
detector at the extraction from the ring provide the start and stop signals. 
The ions conduct about a thousand revolutions, which is sufficient to deduce the revolution times with a high accuracy \cite{Yamaguchi-IJMS2013}.
The fast kicker system \cite{Yamaguchi-NIM2013, Yamaguchi-PS2015} used to inject the particles is also used to extract the particle. 

\begin{figure}[htb]
\begin{center}
\includegraphics*[width=\textwidth]{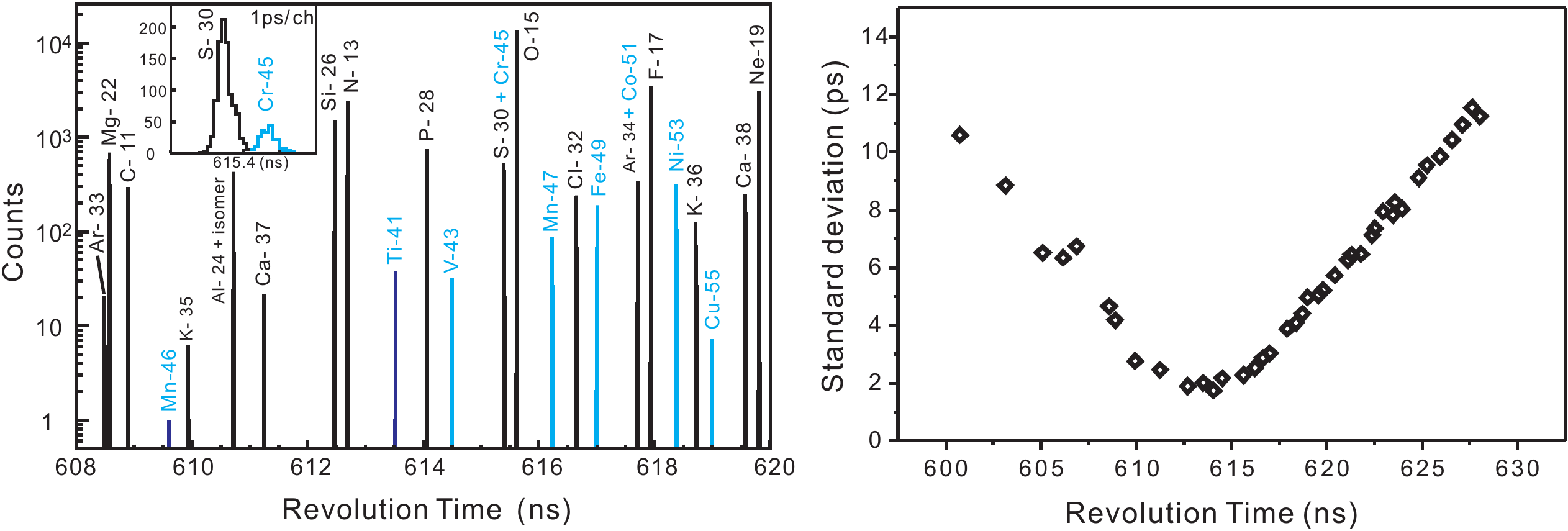}
\end{center}
{\caption{(Colour online) Left: acquired at the CSRe the revolution time spectrum of $^{58}$Ni projectile fragments
zoomed in the time window of 608~ns $\le t \le 620$~ns.
The insert shows the well-resolved peaks of $^{30}$S$^{16+}$ and $^{45}$Cr$^{24+}$ nuclei with very similar $m/q$ values.
Nuclei with masses obtained in that experiment and those used as references are indicated with light blue and black colour, respectively.
Taken from Ref. \cite{Zhang-PRL2012}.
Right: standard deviations of the revolution time peaks for the full spectrum. 
Please note different horizontal scales in both panels. 
Taken from Ref. \cite{Xu-IJMS2013}.} 
\label{fig:ims_spectrum}}
\end{figure}

The deduced revolution times are put into a histogram thus forming the revolution time spectrum.
An example of such spectrum acquired at the CSRe is illustrated in the left panel of Figure~\ref{fig:ims_spectrum} \cite{Zhang-PRL2012}.
The acceptance of all rings allows for simultaneous storage of $\Delta(m/q)/(m/q)>10$\% \cite{Litvinov-APP2010}
and only a small zoom of the entire spectrum is shown in Figure~\ref{fig:ims_spectrum}.
The isochronous conditions $\gamma\approx\gamma_t$ is fulfilled in a small region of the spectrum \cite{Geissel-JPG2005}.
A mass resolving power of $m/\delta{m}\sim100~000-200~000$ is typically achievable in this isochronous region.
However, the resolving power is reduced quickly if moving away from this region.
For the spectrum shown in the left panel of Figure~\ref{fig:ims_spectrum}, the latter is illustrated in the right panel of the same figure,
where the dependence of the standard deviations versus the revolution time is clearly seen.

\begin{figure}[htb]
\includegraphics[angle=-0,width=0.45\textwidth]{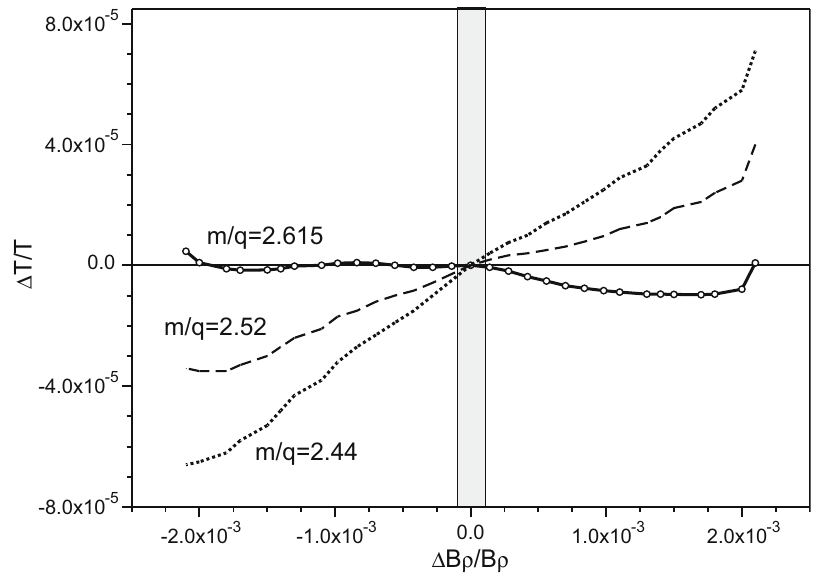}
\includegraphics[angle=-0,width=0.45\textwidth]{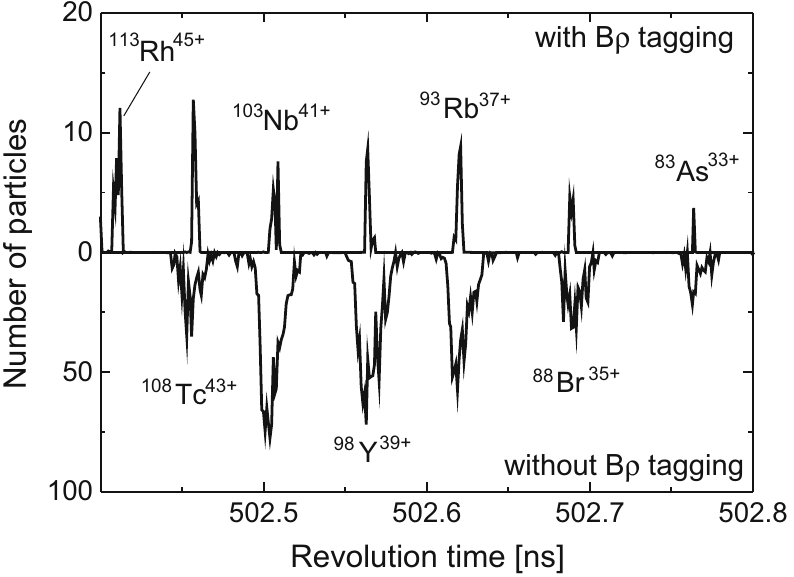}
\caption{(Colour online) Left: revolution time resolving power $\Delta{T}/T$ for nuclides with different mass-over-charge ratios $m/q$. 
The curve for $m/q=2.615$ corresponding to $^{238}$U$^{90+}$ ions has been measured in the ESR. 
The curves for other $m/q$ were scaled assuming that the particles with the same $B\rho$ have the same orbit length in the ESR.
The grey region illustrates the range of magnetic rigidities determined in the FRS in the so-called $B\rho$-tagging method~\cite{Geissel-AIP2006}.
Right: examples of measured revolution-time spectra in the ESR with and without applying the $B\rho$-tagging method. 
A dramatic effect on the resolving power is evident. Taken from Ref.~\cite{Geissel-HI2006}.}
\label{hg1}
\end{figure}

It is possible to correct for the above {\it non-isochronicity} effect if a velocity or a magnetic rigidity of each particle is known in addition to its revolution time \cite{Geissel-JPG2005}.
This has been proven in a test experiment at GSI \cite{Geissel-HI2006, Geissel-EPJST2007}, where only the particles 
with a well-defined magnetic rigidity ($\Delta(B\rho)/(B\rho)=\pm1.5\cdot10^{-5}$) were injected into the ESR.
This is the so-called {\it $B\rho$-tagging method}.
Nearly a constant mass resolving power was observed over the entire time of flight spectrum (see Figure~\ref{hg1}), however, at a dramatically reduced transmission to the ESR.
Since the particles of interest have tiny production rates, the latter cannot be tolerated and other solutions are needed.

At the CSRe, to measure the velocity of each stored ion a pair of time-of-flight detectors 
is installed in one of the straight sections \cite{Zhang-NIM2014a, Xing-PS2015}, see Figure~\ref{fig:imp}.
In the case of the R3, the particle is identified with BigRIPS prior to its injection into the 
ring which requires the measurements of both the $B\rho$ and $v$ \cite{Ozawa-PTEP2012}.
First data from CSRe and R3 with $v$ and/or $B\rho$ information for each ion are expected soon.

\subsection{Schottky spectrometry in the isochronously tuned storage ring}

\begin{figure}[htb]
\centering\includegraphics[angle=-0,width=\textwidth]{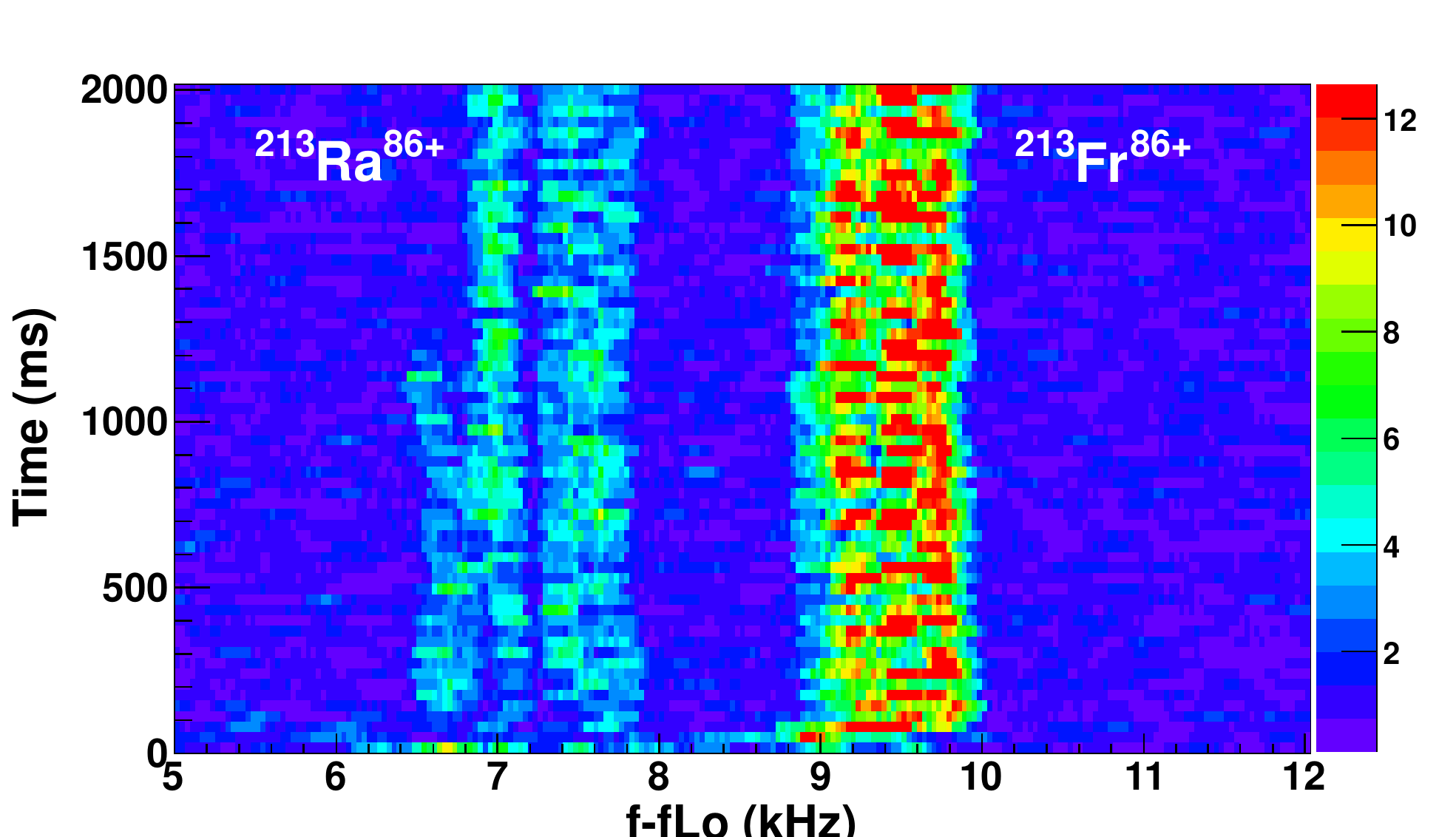}
\caption{(Colour online) Time-resolved Schottky frequency spectra acquired with the new resonant Schottky detector~\cite{Nolden-NIM2011} 
in the ESR tuned into the isochronous mode. 
The ions are highly charged (86+) while the atomic numbers are $Z=88$ and 87 for Ra and Fr, respectively. 
Traces from individual ions of $^{213}$Ra can be distinguished. Taken from Refs.~\cite{Walker-IJMS2013, Sun-GSI2011}. 
\label{fig:bh1}}
\end{figure}

The recent development of a resonant Schottky detector~\cite{Nolden-NIM2011} allows for precision frequency  
determination of a single stored ion within a few tens of milliseconds, which raises the question whether the SMS method can be applied in the isochronous mode of the ring.
The advantage is clearly the non-destructive detection technique which enables simultaneous lifetime measurement for each short-lived nuclide in addition to its mass.
An illustration of the first test measurement in the ESR is shown in Figure~\ref{fig:bh1}.
In this example, revolution frequencies of helium-like $^{213}$Ra and hydrogen-like $^{213}$Fr isobars are measured.
Resonant Schottky detectors are now installed in all three storage rings \cite{Sanjari-PS2013, Zang-CPC2011, Suzaki-NIM2013, Suzaki-JPSJ2015, Suzaki-PS2015} and will be used in the forthcoming experiments.

\section{Selected Results}
Overall more than 1000 mass values were measured with storage ring mass spectrometry.
The mass surface obtained for the first time is illustrated in Figure~\ref{fig:nchart}.
More than 200 masses were measured for the first time by employing SMS at the ESR and IMS at the ESR and CSRe.
More than 100 nuclear masses were determined in addition indirectly by using data from $\alpha$- and proton-decay spectroscopy.
R3 ring was taken into operation in 2015 with stable ions and the first measurements on nuclei with presently unknown masses are expected in 2016.
Although SMS is available at CSRe, all SMS measurements to date were conducted at the ESR.
To date $^{238}$U \cite{Chen-NPA2012}, $^{209}$Bi \cite{Radon-NPA2000, Litvinov-NPA2005}, $^{208}$Hg \cite{Kurcewicz-APP2010}, 
$^{197}$Au \cite{Reed-PRC2012, Shubina-PRC2013}, $^{152}$Sm \cite{Litvinov-HI2006}, and $^{84}$Kr \cite{Stadlmann-PLB2004}
projectile fragments as well as $^{238}$U \cite{Sun-NPA2008} fission fragments were addressed at the ESR, and
$^{78}$Kr \cite{Tu-PRL2011}, $^{86}$Kr \cite{Zhang-JPSJ2015}, 
$^{58}$Ni \cite{Zhang-PRL2012, Yan-APJL2013}, and $^{112}$Sn \cite{Zhang-JPSJ2015} projectile fragments were addressed at the CSRe.
Typical relative mass uncertainties of $\delta m/m = 10^{-6}-10^{-7}$ are achieved.
The new and improved mass values were included into the latest Atomic Mass Evaluation AME'12~\cite{AME12}.
The analyses of the data on neutron-rich $^{86}$Kr and $^{208}$Hg fragments and neutron-deficient $^{112}$Sn and $^{152}$Sm fragments is presently in progress.
In the following, we present some selected results and their implications in nuclear structure and astrophysics applications.

\subsection{Testing of the predictive powers of mass models}
\label{models}
\begin{figure}[t]
\begin{center}
\includegraphics[width=1.0\textwidth]{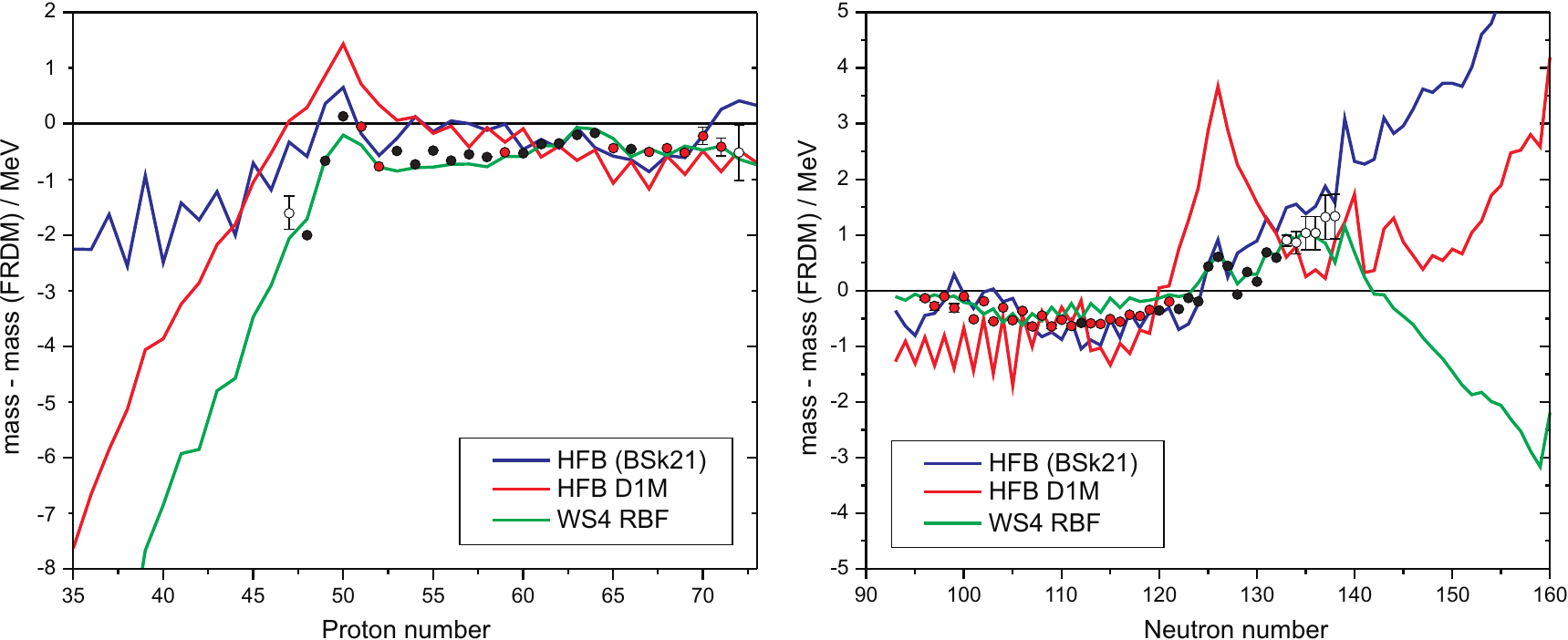}
\caption{(Colour online) Comparison of experimental mass values with predictions of several mass models for $N=82$ (Left) and $Z=82$ (Right) shell closures. 
The description of the models is given in the text.
Experimental data (filled symbols) are taken from AME'12 \cite{AME12} except for the value of $^{130}$Cd, which is taken from \cite{Atanasov-PRL2015}.
Values which contain contributions from storage ring experiments are indicated in red colour.
Open symbols indicate extrapolated values from AME'12.
Error bars if not shown are within the symbol size.
} \label{fig:shells}
\end{center}
\vspace*{-0.5cm}
\end{figure}

\begin{figure}[t]
\begin{center}
\includegraphics[width=0.8\textwidth]{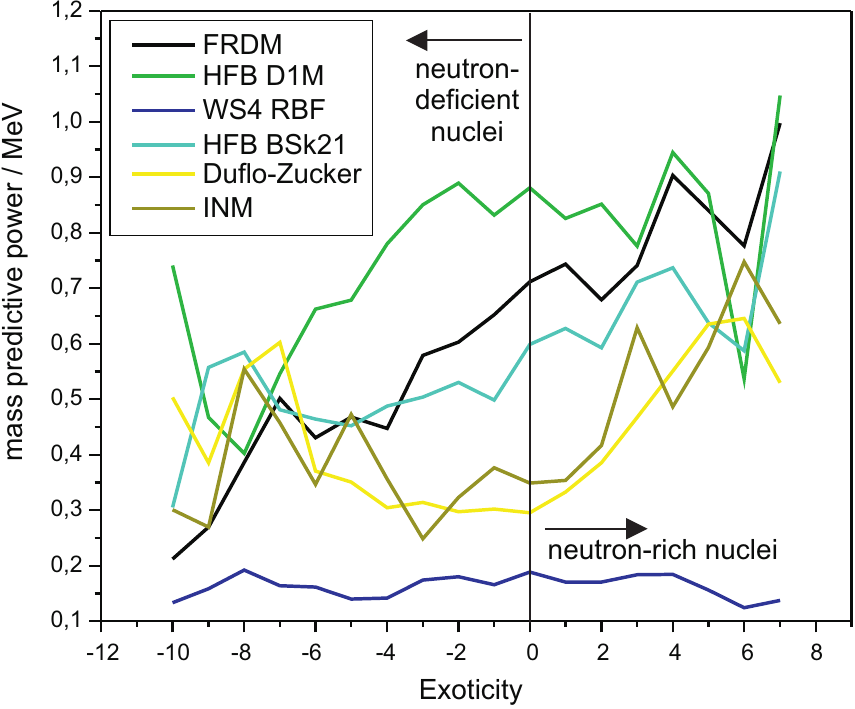}
\caption{(Colour online) Mass resolving powers of several mass models versus the remoteness from stability, exoticity.
All experimentally known mass values from AME'12 \cite{AME12} are taken for the analysis. Zero exoticity reflects the stable nuclei.
For the description of the models and the definition of exoticity see text.
} \label{fig:exoticity}
\end{center}
\vspace*{-0.5cm}
\end{figure}

The developments in nuclear theory were enormous in the recent years resulting in a range of new mass models.
The models aim at predicting masses and other nuclear properties for nuclei yet inaccessible in experiment.
For instance, the peak in Solar element abundances at $A\sim195$ 
depends critically on the strength of the $N=126$ neutron shell closure in very neutron-rich nuclei \cite{Dillmann-PPNP2011}.
It is interesting to note that for $\beta$-decay half-lives strong deviations to model predictions in neutron-rich nuclei below $Z=82$ were observed \cite{Caballero-Folch-arXiv2015}.
It is essential to understand that many of nuclear properties will not be feasible to measure and they will have to be calculated.

Considerable regions of the nuclear mass surface is simultaneously covered in a typical storage ring experiment, 
which allows for testing the predictions of various mass models.
For instance, the left and right panels of Figure~\ref{fig:shells} illustrate the comparison of experimental 
and theoretical masses for $N=82$ and $Z=82$ shell closures, respectively.
Four frequently used mass models are employed, 
namely the Finite-Range Droplet Model (FRDM) \cite{Moller-ADNDT1995}, the Hartree-Fock-Bogoliubov calculations 
with Skyrme force BSk21\cite{Goriely-PRC2010} and Gogny force D1M \cite{Goriely-PRL2009}, and the WS4 version of the model by Wang {\it et al.} 
with radial basis function correction (RBF) \cite{Wang-PLB2014}.
Experimental values are taken from AME'12 \cite{AME12} except for a new mass value of $^{130}$Cd taken from \cite{Atanasov-PRL2015}.
Experimental mass values with contributions by storage ring measurements are indicated with red colour.
Apart from FRDM, which was fixed in 1995, there is an overall qualitative agreement between theory and experiment.
However, the predictions in presently unknown regions deviate dramatically for different theories.

For a mass model, the quality of its calculations can be characterised by the predictive power, $\sigma_{\rm rms}$. 
The latter is defined as \cite{Maripuu-ADNDT1976}
\be
\sigma_{\rm rms}^2 = \frac{1}{N}\sum_{i=1}^{\rm N}(m_{\rm th}-m_{\rm exp})^2,
\label{srms}
\ee
where $m_{\rm th}$ and $m_{\rm exp}$ are the theoretical and experimental mass values, respectively, and $N$ is the number of compared values.
Let is define ``exoticity'' as a relative distance of a nucleus to stability.
The proton number $Z_0$ corresponding to the most stable nucleus in a chain of isobars with mass $A$ can be deduced \cite{Heyde-1999}
\be
Z_0=\frac{A/2}{1+0.0077\cdot A^{2/3}}.
\ee
The value $\epsilon = Z_0 - Z$ reflects the remoteness, exoticity, of a nucleus $(Z,A)$ from the valley of stability for which $\epsilon = 0$.
Figure \ref{fig:exoticity} shows the predictive powers, $\sigma_{\rm rms}$ of several mass models versus exoticity.
In addition to the above four models, the 28-parameter version of the model by Duflo and Zuker \cite{Duflo-PRC1995} and the Infinite Nuclear Matter model (INM) 
by Nayak and Satpathy \cite{Nayak-ADNDT2012} are added.
Striking is the excellent description of the known masses by the WS4+RBF model.
Such powers are otherwise achieved by models based on Garvey-Kelson relation only for stable nuclei.
Compared to the previous study in \cite{Litvinov-NPA2007}, with the maximum $\epsilon=6$, 
values for $\epsilon=7$ are available in this new analysis.
Although the absolute values of predictive powers improved, the 
trend of degrading predictive power towards more neutron-rich nuclei remains.

\subsection{Isospin dependence of odd-even staggering of nuclear binding energies}
Masses of hundreds of nuclides can be addressed in a single experiment. 
The advantage of this is that a big region of the mass surface is measured with the same method and thus with the same systematic uncertainty.
In two runs, described in detail in Refs.~\cite{Radon-NPA2000, Litvinov-NPA2005}, 
a large number of previously unknown masses have been determined in the ESR with a high relative mass accuracy  $\Delta m/m$ of a few $10^{-7}$.
This enabled a systematic investigation of nuclear odd-even staggering (OES) values 
in proton-rich nuclei between $Z=50$ and $Z=82$ closed proton shells \cite{Litvinov-PRL2005}.

The OES of nuclear binding energies was detected in the early days of nuclear physics~\cite{Heisenberg-ZP1932}.
It was explained by the presence of pairing correlations between nucleons in the nucleus. 
Although pairing contribution to the total nuclear binding energy is tiny, its effect on nuclear structure is enormous~\cite{Bohr-NS1969}.
Different formulas exist to extract the OES values from nuclear masses of several neighbouring nuclei~\cite{Madland-NPA1988}. 
One of them is the so-called 5-mass formula:
\begin{eqnarray}
\nonumber
{\rm OES}(n)=-\frac{1}{8}\biggr[m(Z,N+2)-4m(Z,N+1)+\\
~~~~~~~~~~~~~~6m(Z,N)-4m(Z,N-1)+m(Z,N-2)\biggl]\\
\nonumber
{\rm OES}(p)=-\frac{1}{8}\biggr[m(Z+2,N)-4m(Z+1,N)+\\
~~~~~~~~~~~~~~6m(Z,N)-4m(Z-1,N)+m(Z-2,N)\biggl],
\end{eqnarray}
where OES($n$) and OES($p$) stand for neutron and proton OES values, respectively.

\begin{figure*}[h!]
\begin{center}
\includegraphics*[width=\textwidth]{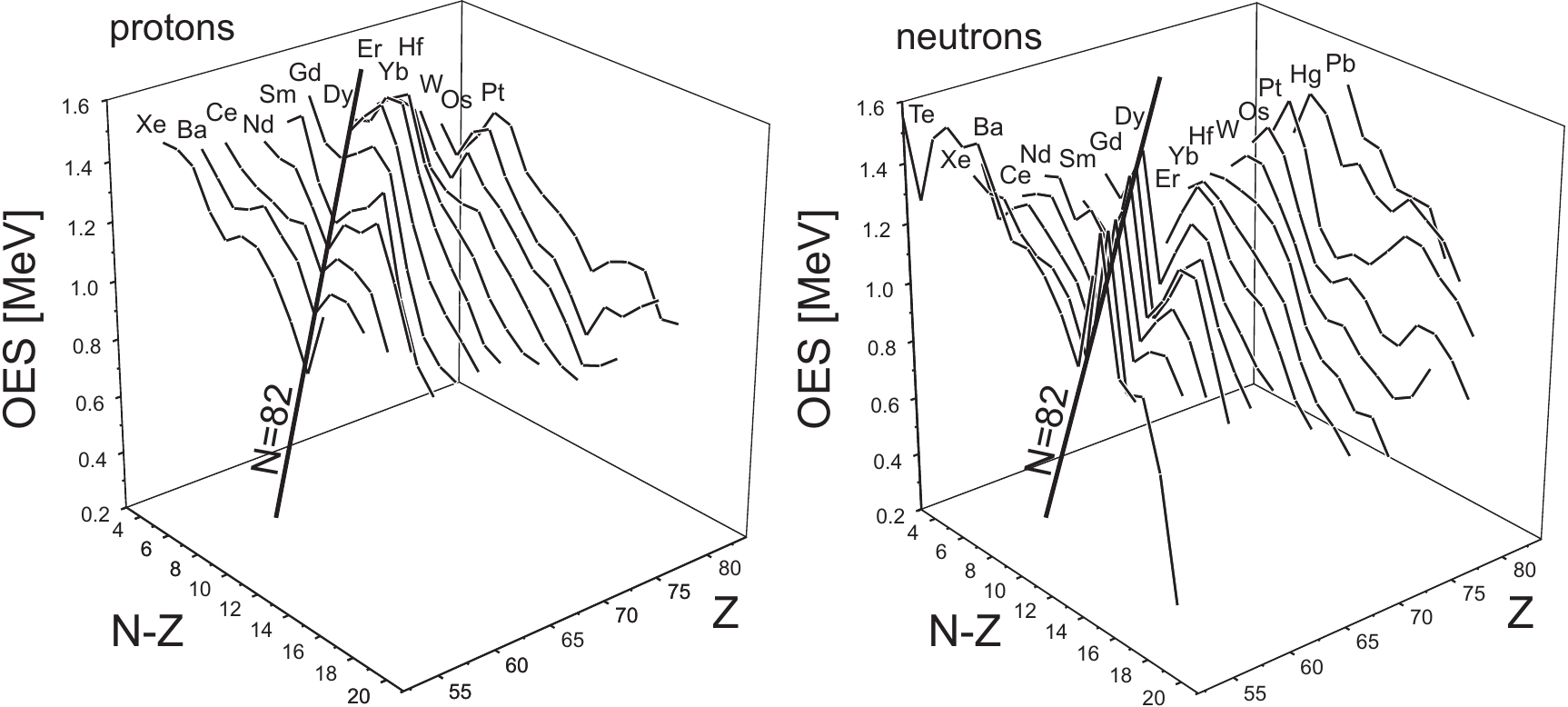}
\end{center}
{\caption{Odd-even staggering of nuclear masses (OES) for even-even isotopes between $Z=50$ and $Z=82$ closed proton shells.
It is clearly seen, that OES values for both, protons and neutrons, increase towards the proton drip-line for all presented isotopic chains. 
The neutron shell-closure at $N=82$ is indicated. Taken from Ref.~\cite{Litvinov-NPA2008}.} \label{ors}}
\end{figure*}

\begin{figure*}[h!]
\begin{center}
\includegraphics*[width=\textwidth]{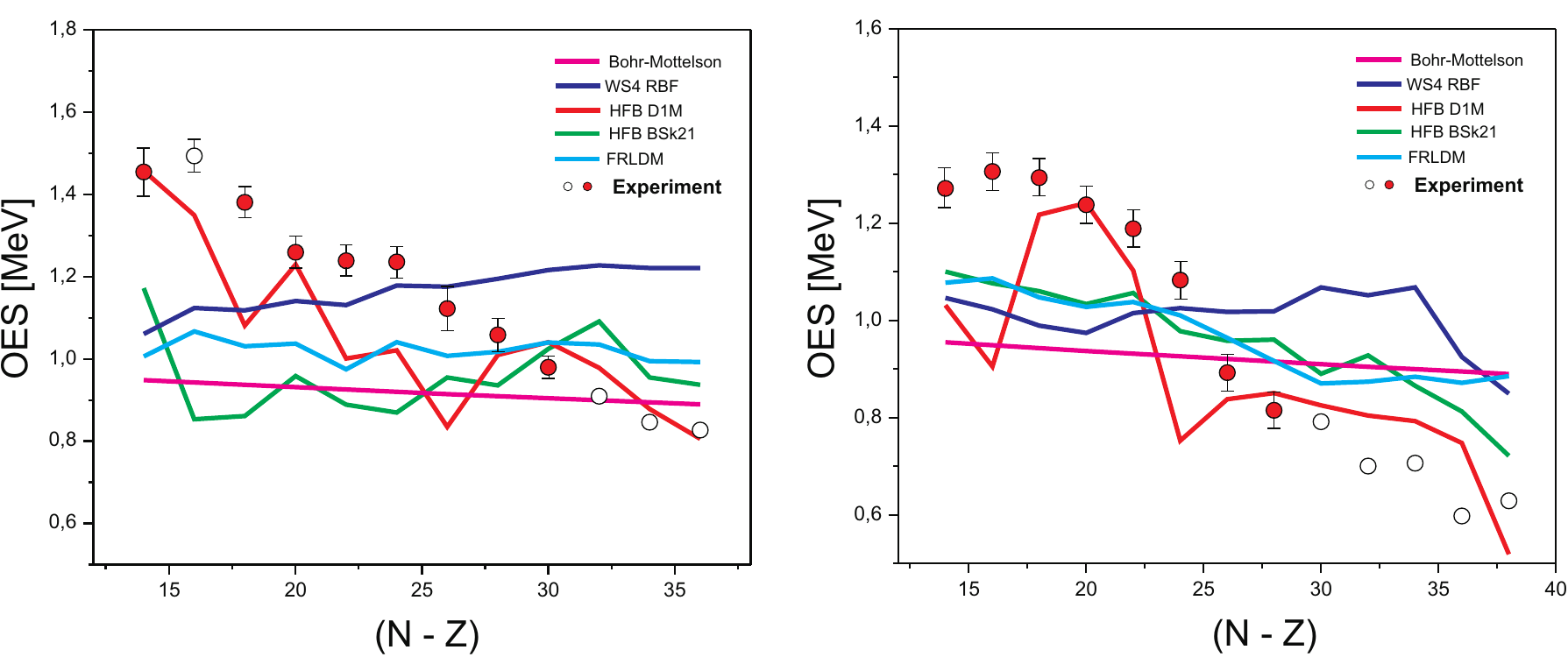}
\end{center}
{\caption{Odd-even staggering of nuclear masses (OES) for even-even hafnium isotopes compared to several mass models.
OES values for protons (left) and neutrons (right) increase towards the proton drip-line. 
Mass models are defined in Section~\ref{models} and
Borh-Mottelson parametrisation, OES$=12/\sqrt{A}$~MeV, is from \cite{Bohr-NS1969}.
Experimental data are from the latest atomic mass evaluation AME'12 \cite{AME12}.
The data points coloured in red represent the values which could be obtained for the first time with storage ring measurements.
} \label{hfpp}}
\end{figure*}

Experimental OES values for even-even nuclei are illustrated in Figure~\ref{ors}~\cite{Litvinov-NPA2008}.
Apart from the strong shell effect at $N=82$,  
the OES values clearly increase towards the 
proton drip-line for both protons and neutrons~\cite{Litvinov-PRL2005, Alkhazov-ZPA1983}. 
The origin of this isospin-dependent trend is still not clear and cannot be reproduced by the modern mass models.
Th latter is illustrated in Figure~\ref{hfpp} on the example of OES-values for even-even Hf isotopes compared to several
mass models defined in Section~\ref{models} and the famous parametrisation of Bohr and Mottelson \cite{Bohr-NS1969} OES$=12/\sqrt{A}$~MeV. 
One of the reasons could be the onset of deformation in these neutron-deficient nuclides~\cite{Bohm-PRC2014}.
Such strong isospin dependence of the OES values is not observed for nuclei below $Z=50$ and above $Z=82$ shells, see, e.g., \cite{Kreim-PRC2014}.


\subsection{The mass of $^{208}$Hg nuclide and the proton-neutron interaction strength around doubly-magic $^{208}$Pb}
\begin{figure}[htb]
\begin{center}
\includegraphics*[width=0.49\textwidth]{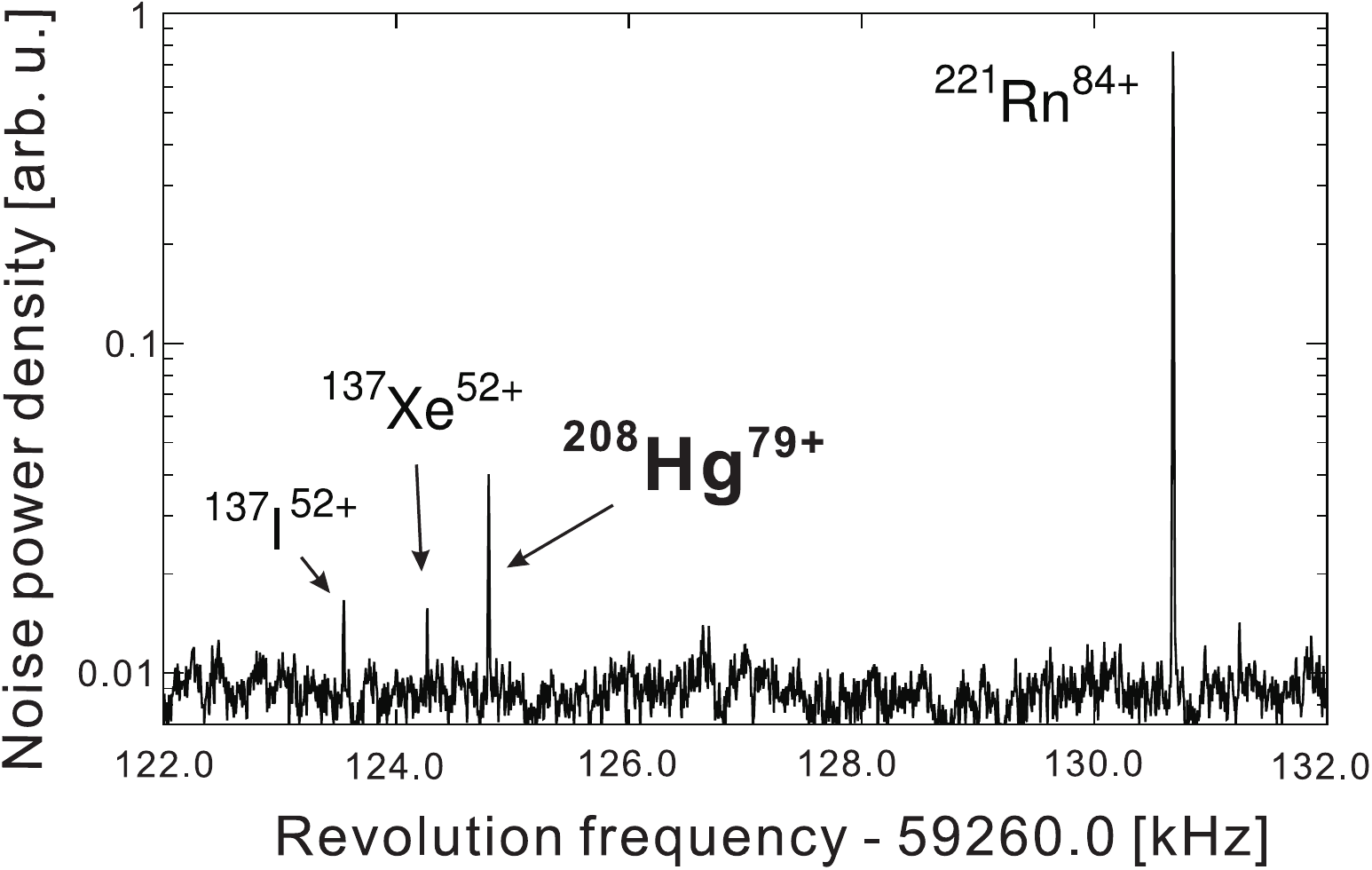}
\includegraphics*[width=0.49\textwidth]{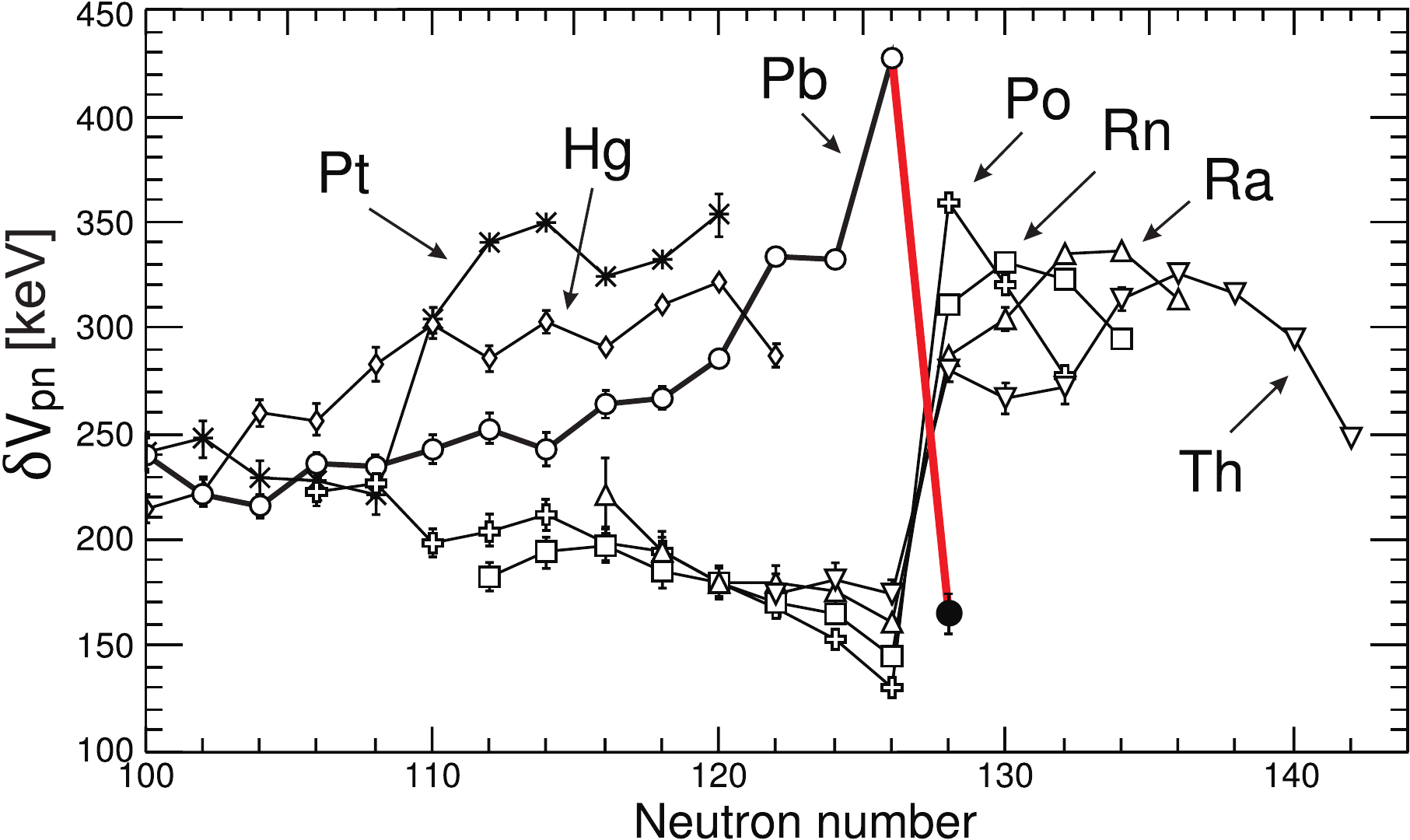}
\end{center}
{\caption{(Colour online) Left: a 10~kHz part of the Schottky frequency spectrum of neutron-rich $^{238}$U projectile fragments. 
A peak of a single hydrogen-like $^{208}$Hg$^{79+}$ ion is at about 125~kHz. 
The known masses of $^{137}$I$^{52+}$, $^{137}$Xe$^{52+}$, and $^{221}$Rn$^{84+}$ ions were used as calibrants. Taken from Ref.~\cite{Chen-PRL2009}.
Right: experimentally known $\delta V_{pn}$ values for isotopic chains below and above the $Z=82$ proton closed shell. 
Only even-even nuclei are considered. The $\delta V_{pn} (^{210}{\rm Pb})$ value is indicated by a filled symbol.
The sharp drop from $^{208}$Pb to $^{210}$Pb is obvious. Taken from Ref.~\cite{Chen-PRL2009}.} 
\label{fig:hg208}}
\end{figure}

Owing to the ultimate sensitivity of the SMS to single stored ions, the masses of very rarely produced nuclei can be addressed.
Furthermore, a single stored ion is sufficient to determine its mass with a high accuracy.
One example of such a measurement is illustrated in Figure~\ref{fig:hg208}, 
where a zoom of a Schottky frequency spectrum of stored $^{238}$U projectile fragments is shown~\cite{Chen-PRL2009}.
The frequency peak at about 125~kHz corresponds to a single $^{208}$Hg$^{79+}$ ion which was stored only once within a two-weeks long experiment.
The obtained mass excess value ($ME=m-A$) is $ME(^{208}{\rm Hg})=-13265(31)$~keV~\cite{Chen-PRL2009}.
The mass of this even-even nucleus enabled addressing a long-standing question on proton-neutron interaction ($\delta V_{pn}$) around the doubly-magic $^{208}$Pb nucleus.

For even-even nuclei, the average $p-n$ interaction of the last two protons with the last two neutrons can be defined as~\cite{Zhang-PLB1989}:
\begin{eqnarray}
\nonumber
\delta V_{pn} = \frac{1}{4} \biggl[ m(Z,N) + m(Z-2,N-2) - 
m(Z,N-2) - m(Z-2,N) \biggr].
\end{eqnarray}

Thus the $\delta V_{pn}$ for $^{210}$Pb nucleus could be extracted.
The experimental $\delta V_{pn}$ values for Hg, Pt, Pb, Po, Rn, Ra and Th isotopes are illustrated in the right panel of Figure~\ref{fig:hg208}.
The doubly-magic $^{208}_{~82}$Pb nucleus is in the symmetric hole-hole region, 
where both protons and neutrons fill low-$j$ high-$n$ orbits below the closed shell, 
and therefore it should (and does) have a very large $\delta V_{pn}$ value. 
In contrast, $^{210}$Pb has two extra valence neutrons which occupy orbits just above the $N=126$ closed shell, 
giving an asymmetric particle-hole case where one would expect a low $\delta V_{pn}$ value. 
These expectations--based on the spatial overlap of the valence orbits in this region~\cite{Cakirli-PRL2006}--are confirmed experimentally.

\subsection{Discovery of new isotopes and investigations of nuclear isomers}
\begin{figure*}[t]
\begin{center}
\includegraphics*[width=0.8\textwidth]{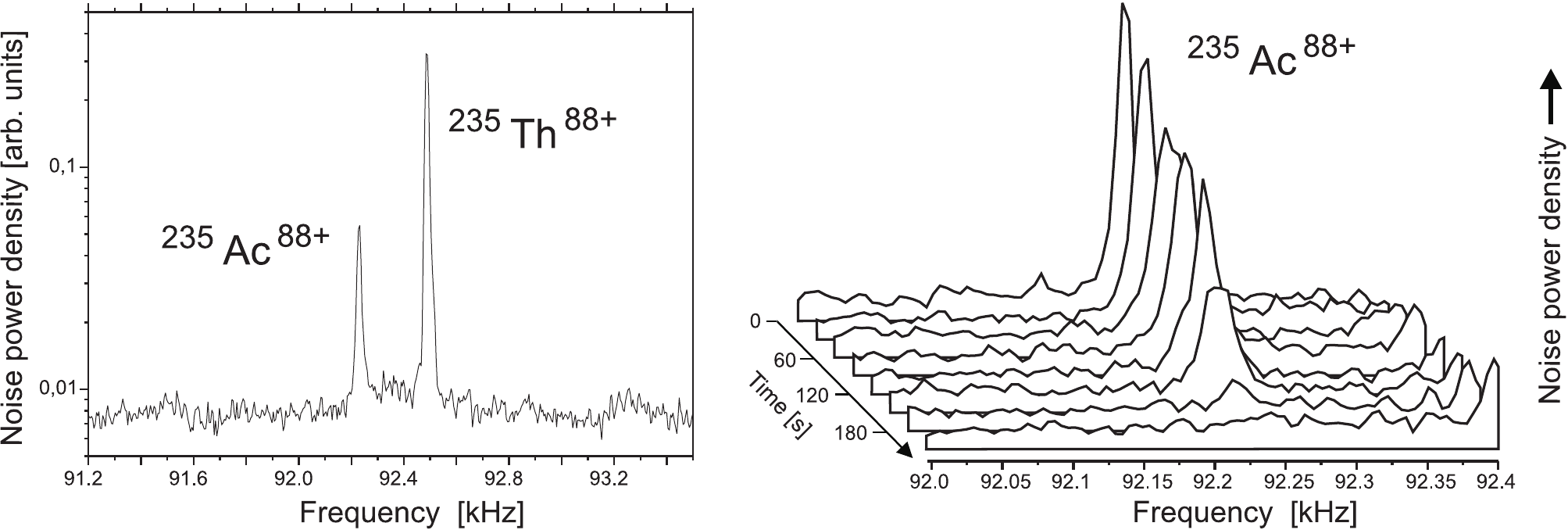}
\end{center}
{\caption{Left: Schottky frequency spectrum of the new isotope $^{235}$Ac present as hydrogen-like ions. 
The frequency peak of $^{235}$Th$^{88+}$ can be used to calibrate the frequency scale. Taken from Ref.~\cite{Geissel-AIP2006}
Right: The half-life of $^{235}$Ac$^{88+}$ ions can be extracted from the time evolution of the peak-area. Taken from Ref.~\cite{Bosch-IJMS2006}} 
\label{235ac}}
\end{figure*}
The single-particle sensitivity of the SMS can be used as a tool to search for new isotopes.
In addition to the discovery of the isotope itself, the measurement of its mass and half-life is performed.
One example is illustrated in Figure~\ref{235ac}.
We note, that the measured value for hydrogen-like $^{235}$Ac$^{88+}$ ions (see Figure~\ref{235ac})  
has to be corrected for missing bound electrons to obtain the half-life for the neutral $^{235}$Ac atoms. 
To date, six isotopes, $^{235,236}$Ac, $^{224}$At, $^{221,222}$Po, and $^{213}$Tl, have been discovered in the ESR~\cite{Chen-PLB2010}.

\begin{figure}[htb]
\begin{center}
\includegraphics*[width=0.45\textwidth]{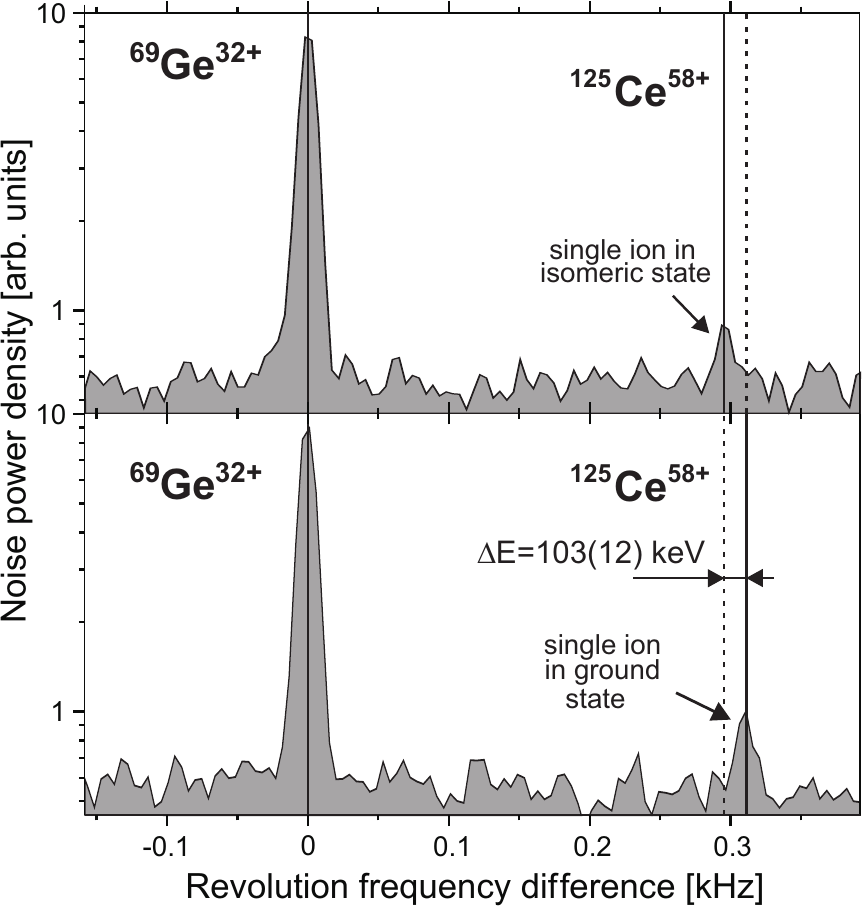}
\end{center}
{\caption{Schottky frequency spectra of single stored fully-ionised $^{125}$Ce$^{58+}$ ions in the isomeric (upper panel) and ground (lower panel) states.
The frequency peak of $^{69}$Ge$^{32+}$ ions can be used as a reference. The frequency difference between the two single $^{125}$Ce ions corresponds
to the isomeric excitation energy of $E^*=103(12)$~keV. Taken from Ref.~\cite{Sun-EPJ2007}. } 
\label{125ce}}
\end{figure}

Nuclear metastable states not decaying promptly are called {\it isomers} \cite{Walker-Nature1999}.
The existence of isomers points to specific nuclear structure which leads to their formation.
Typically isomers are studied with $\gamma$-spectroscopy methods.
However, if the isomeric state lives longer than about a second and is rarely produced, these methods can no longer be applied.
Here, the storage ring mass spectrometry turns out to be a unique tool to study such states.

Figure~\ref{125ce} illustrates a discovery of a long-lived isomeric state in neutron-deficient $^{125}$Ce nuclide by means of the SMS~\cite{Sun-EPJ2007}.
In two different injections into the ESR, two fully-stripped $^{125}$Ce atoms were stored.
A single $^{125}$Ce ion is present in the isomeric state and in the ground state in the upper and lower panels, respectively.
The difference of the revolution frequencies can be measured relative to the reference frequency of $^{69}$Ge$^{32+}$ ions, which 
yields the excitation energy of the isomer of $E^*=103(12)$~keV.
If ions in both states would be present in the ring at the same time it would not be possible to resolve them.

Combined with the capability of storage rings to cover a wide range of different nuclides in one frequency spectrum, 
the single ion sensitivity can be used for a broadband search of nuclear isomers on the chart of the nuclides.

\begin{figure}[htb]
\begin{center}
\includegraphics*[width=0.45\textwidth]{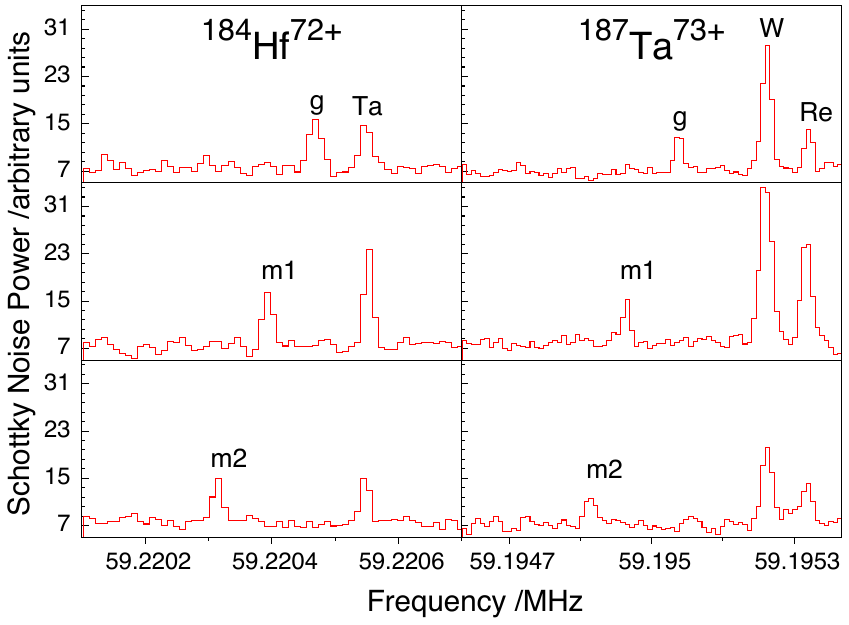}
\includegraphics*[width=0.49\textwidth]{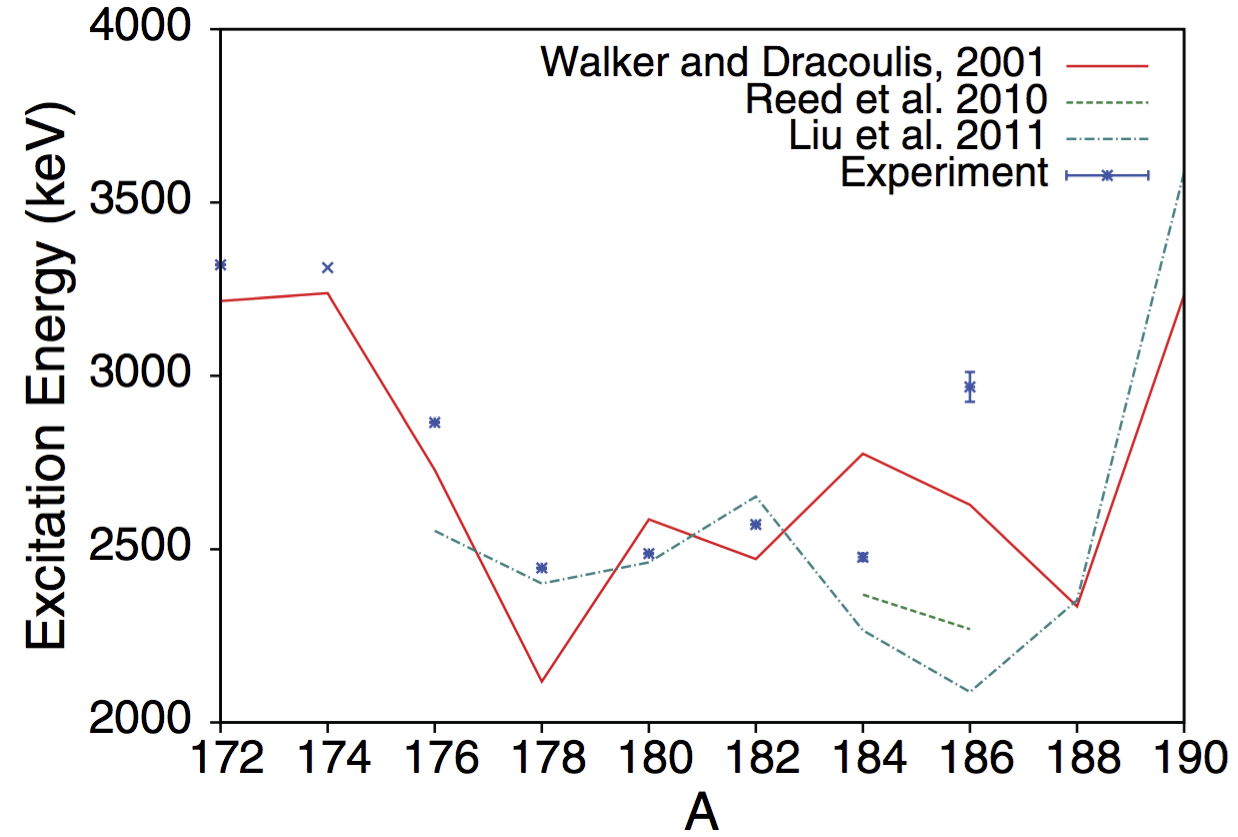}
\end{center}
{\caption{(Colour online) Left: Schottky frequency spectra of $A = 184$ (left) and $A=187$ (right) isobars and isomers. 
Each spectrum corresponds to a different injection of the ions in the ESR. 
The $^{184}$Hf and $^{187}$Ta isomers (m1 and m2) and ground states (g) are identified as individual bare ions. 
The ground states of $^{184}$Ta$^{72+}$, $^{187}$W$^{73+}$ and $^{187}$Re$^{73+}$ are used to calibrate the frequency scale. 
Taken from Ref.~\cite{Reed-PRL2010}.
Right: systematics of experimental and calculated excitation energies of four-quasiparticle isomers for even-even hafnium isotopes. Taken from Ref. \cite{Reed-PRC2012}.
}

\label{213bi}}
\end{figure}

A region of neutron-rich nuclei around $^{188}$Hf is predicted to have exceptional isomer properties~\cite{Walker-HI2001, Dracoulis-PS2013}.
It was investigated with the SMS~\cite{Reed-PRC2012, Reed-PRL2010, Reed-JPCS2012, Akber-PRC2015}.
Examples of detected new isomers are illustrated in the left panel of Figure~\ref{213bi}.
In the right panel the systematics of excitation energies of four-quasiparticle isomers for even-even hafnium isotopes is plotted together with several dedicated calculations.
$^{186}$Hf is the most neutron-rich isotope where a long-lived isomeric state ($T_{1/2}>20$~s, $E^*=2.968(43)$~MeV)~\cite{Reed-PRL2010} is experimentally known. 
A new experiment is presently being prepared aiming at the predicted $18^+$ four-quasiparticle isomer in $^{188}$Hf, which is expected to 
have a very long half-life with respect to $\gamma$-de-excitation.

\begin{figure}[htb]
\includegraphics[width=0.49\textwidth]{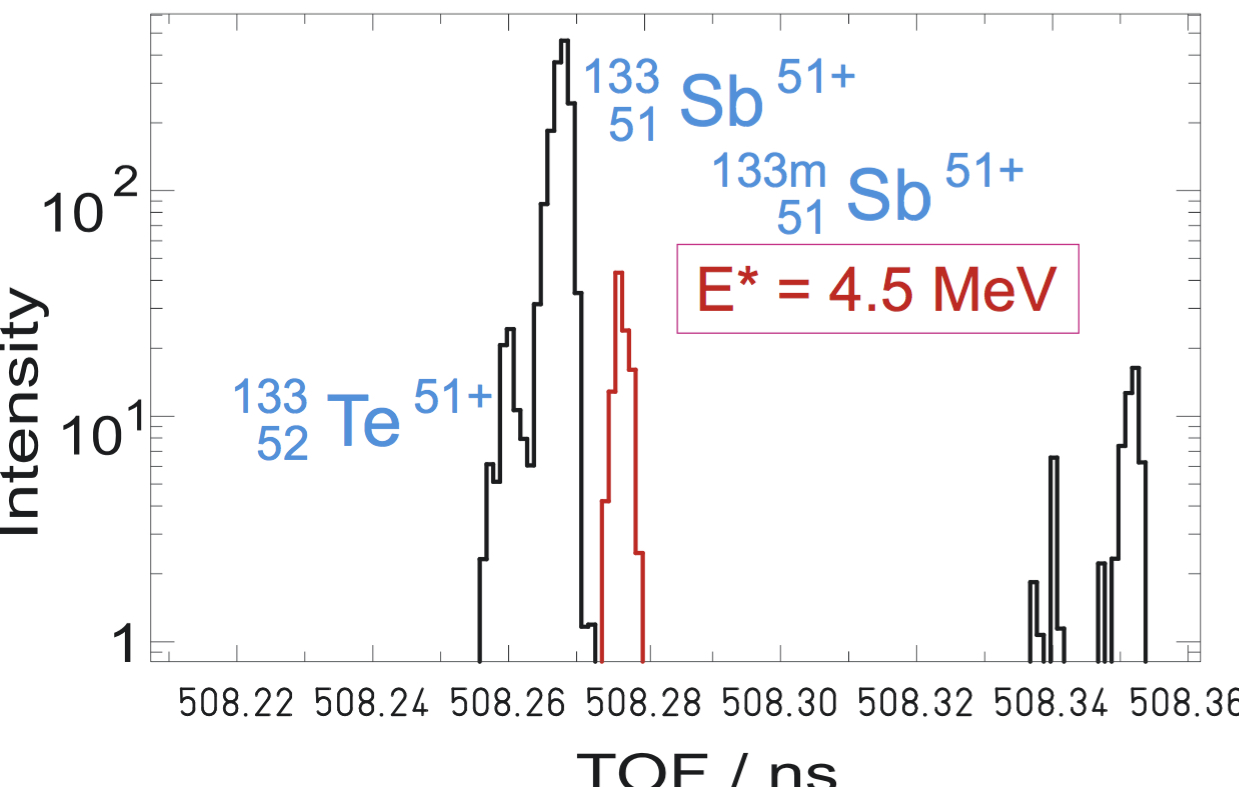}
\includegraphics[width=0.45\textwidth]{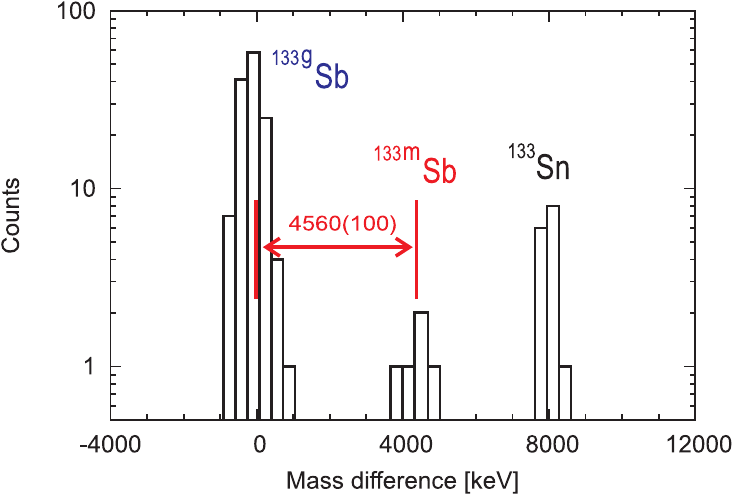}
\caption{(Colour online) Left: uncorrected revolution frequency spectrum of $^{238}$U fission fragments zoomed on $A=133$ isobars. 
The figure is modified from Ref. \cite{Scheidenberger-APH2004}.
Right: distribution of determined mass values for $A =133$ isobars relative to the $^{133}$Sb ground state. 
The isomeric state in $^{133}$Sb with an excitation energy $E^*= 4560(100)$~keV is clearly resolved from the corresponding ground state. 
Here, all $^{133}$Sn events were observed with the charge state $q=50+$. The figure is from Ref.~\cite{Sun-PLB2010}.
The left (right) spectrum was acquired without (with) the $B\rho$-tagging technique.
This explains the different numbers of observed isomers and different resolving powers.
}
\label{ims-exa}
\end{figure}

While the SMS is well suited for the search of long-lived isomers \cite{Chen-PRL2013}, 
the investigations of the shorter-lived isomers can be pursued by applying IMS~\cite{Sun-NPA2010}.
For instance, an isomeric state with excitation energy $E^*=4.56(10)$~MeV has been observed in $^{133}$Sb 
nucleus on the basis of a few stored particles~\cite{Sun-PLB2010}, see left panel of Figure~\ref{ims-exa}. 
The half-life of this isomer in neutral atoms is $T_{1/2}^{\rm atom}=16.54(19)$~$\mu$s~\cite{AME12}.
However, due to the fact that bare $^{133}$Sb nuclei were stored in the ESR, 
all decay modes involving bound electrons were disabled~\cite{Litvinov-PLB2003} and, 
assuming the conversion coefficient of $\alpha\approx991$, a much longer half-life $(T_{1/2}\approx17)$~ms is expected. 
The determined $E^*$ and $T_{1/2}$ values solved the uncertainty related 
to the interpretation of the $21/2^+$ isomeric state in this closed-shell $N=82$ nucleus.


The $B\rho$-tagging method in the IMS improves the resolving power but reduces dramatically the efficiency. 
Application of the corresponding IMS analysis is illustrated in the right panel of Figure~\ref{ims-exa}. 
With just five ions of $^{133m}$Sb, the isomer excitation energy is determined to an accuracy of 100~keV~\cite{Sun-PLB2010}.

\subsection{Test of the Isobaric Multiplet Mass Equation in $pf$-shell nuclei}
The isobaric multiplet mass equation (IMME) is based on the fundamental concept 
of the isospin symmetry in nuclear physics~\cite{Wigner-1957,Weinberg-PR1959}.
The IMME connects the members of an isobaric multiplet via a parabolic equation
\begin{equation}
\ {\rm ME} (A,T,T_z)=a(A,T)+b(A,T)\cdot T_z+c(A,T)\cdot T^2_z, \label{eq3}
\end{equation}
where $a$, $b$, and $c$ are parameters depending on the atomic mass number $A$
and the total isospin $T$, and $T_z=(N-Z)/2$.
Thus, to test the validity of the quadratic form of the IMME, 
the energies of four members of the isobaric multiplet are required.
Extensive tests in the $sd$-shell nuclei were conducted previously (see Ref.~\cite{Lam-ADNDT2013} and references therein)
and no significant deviations were found except for slight disagreements at $A=$8, 9, 32, and 33.

Measurements of neutron-deficient $^{58}$Ni projectile fragments at the CSRe provided masses for $^{41}$Ti, $^{43}$V,
$^{45}$Cr, $^{47}$Mn, $^{49}$Fe, $^{51}$Co, $^{53}$Ni, and $^{55}$Cu $T_z=-3/2$ nuclei~\cite{Shuai-PLB2014, Zhang-PRL2012,Yan-APJL2013}.
By using these new masses and the known energies of the isobaric analog states the validity
of the IMME could be tested for the first time in the $fp$-shell nuclei~\cite{Zhang-PRL2012,Zhang-JPCS2013,Tu-JPG2014}.
The data for four $T=3/2$ isospin quartets with $A=41,45,49$, and 53 are completed.
To characterise the deviation from the quadratic form of the IMME, an extra term $d\cdot T_z^3$ is added to Eq~(\ref{eq3}).

\begin{figure}
  \centering
  \includegraphics[width=0.8\textwidth]{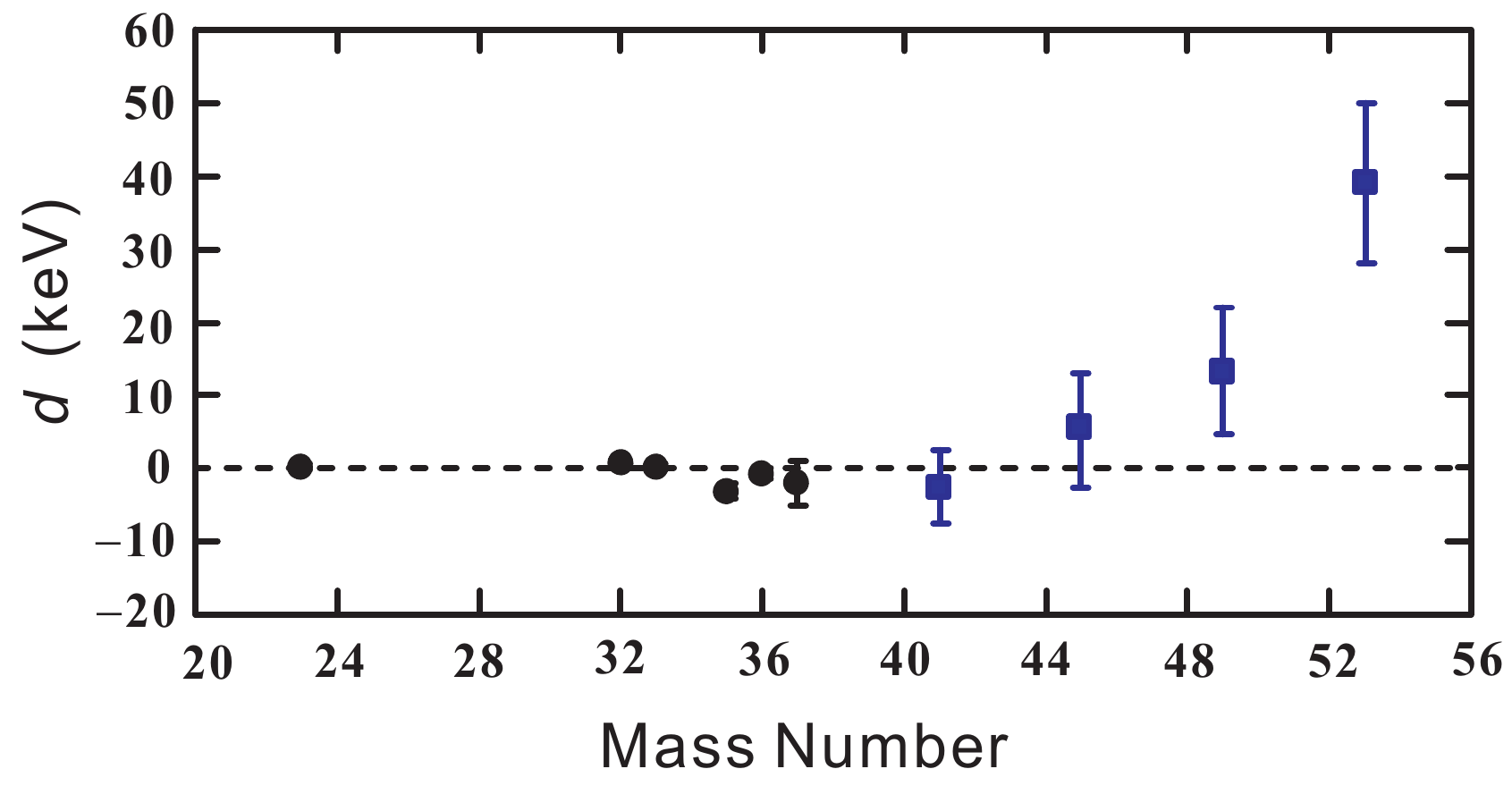}\\
  \caption{(Colour online) $d$ coefficients for the four $T = 3/2$ isobaric multiplets in $pf$-shell (squares).
Experimental data since 2001 (circles)~\cite{Pyle-PRL2002,Ringle-PRC2007,Yazidjian-PRC2007,Saastamoinen-PRC2009,Kankainen-PRC2010}
are shown for comparison.
Note, that albeit the large uncertainties,
there seems to be a trend of gradual increase of $d$ with $A$ in $fp$-shell.
Taken from Ref. \cite{Zhang-PRL2012}.}
  \label{fig06}
\end{figure}

The obtained $d$-coefficients are plotted in Figure~\ref{fig06} 
together with recent results for the $sd$-shell nuclei~\cite{Pyle-PRL2002,Ringle-PRC2007,Yazidjian-PRC2007,Saastamoinen-PRC2009,Kankainen-PRC2010}.
It can be seen that for all mass numbers except for $A=53$ the corresponding $d$-coefficients are comparable with zero, 
thus confirming the quadratic form of the IMME.
However, there is a $3.5\sigma$ deviation for the $A=53$ isobaric multiplet.
The corresponding $d=39\pm 11$~keV indicates a dramatic breakdown of the quadratic form of the IMME.
Previous as well as new dedicated theoretical calculations of isospin mixing cannot explain this rather large $d$ coefficient~\cite{Zhang-PRL2012}.
If this breakdown is confirmed by improved experimental data,
it may point to new effects like enhanced isospin mixing or charge-dependent nuclear forces.

\subsection{The mass of $^{65}$As and the $rp$-process waiting point $^{64}$Ge}
Difference of nuclear masses provides the $Q$-values for the corresponding nuclear reactions.
Typically the reaction rates exponentially depend on the $Q$-values \cite{Langanke-PS2015}.
Therefore, accurate mass values are necessary for reliable rate calculations especially 
in the cases where the reaction rate measurements are complicated or are presently impossible.
The latter situation occurs, for instance, 
in nuclear astrophysics where the explosive $rp$-process runs along the proton drip-line, see Figure~\ref{fig:nchart}.
The $rp$-process is a sequence of proton capture and $\beta^+$ decays which can produce neutron-deficient nuclides up to Sn-Te region~\cite{Schatz-PRL2001}.
Key nuclei on the $rp$-process path are the so-called waiting points (WP), 
which are relatively long-lived nuclei and where the process stalls until the $\beta^+$ occurs.
Here the $^{64}$Ge, $^{68}$Se, and $^{72}$Kr are the major waiting-point nuclei,
with half-lives of about 64~s, 36~s, and 17~s, respectively, 
which are significant if compared to the duration of observed Type I X-ray bursts of about 100-200~s~\cite{Wallace-AJSS1981,Schatz-PR1998,Parikh-PRC2009}.
The $S_p$($^{69}$Br) and $S_p$($^{73}$Rb) are known to be negative from experiments indicating
that both nuclides are fast proton emitters~\cite{Blank-PRL1995,Jokinen-ZP1996}.
This makes it likely that $^{68}$Se and $^{72}$Kr are strong waiting points,
although two-proton (2p) captures can somewhat reduce their effective lifetimes~\cite{Schatz-PR1998}.

To investigate the WP-nature of $^{64}$Ge, the mass of $^{65}$As was measured with high precision with the IMS at the CSRe~\cite{Tu-NIM2011,Tu-PRL2011}.
Combined with a Penning trap measurement of the mass of $^{64}$Ge~\cite{Schury-PRC2007}, 
the proton separation energy of $^{65}$As has been determined to be $S_p$($^{65}$As)=$-90(85)$ keV.

First of all, the new result showed that $^{65}$As is unbound against the proton emission at the 68.3\% confidence level,
which fixes experimentally the location of the proton drip-line for arsenic isotopes.
Furthermore, the new $S_p$($^{65}$As) value turned out to affect the modelling of the $rp$-process.
Before this measurement, only a model dependent lower limit of $S_p$($^{65}$As)$ >-250$~keV existed
from the observation of the $\beta^+$-decay of $^{65}$As~\cite{Winger-PRC1993}.
The new  $S_p$($^{65}$As)-value was used in one-zone X-ray burst model~\cite{Schatz-PRL2001} calculations,
which allowed for the definition of the temperatures and densities needed to bypass the $^{64}$Ge waiting point.

\begin{figure*}
\begin{center}
\includegraphics[width=0.47\textwidth]{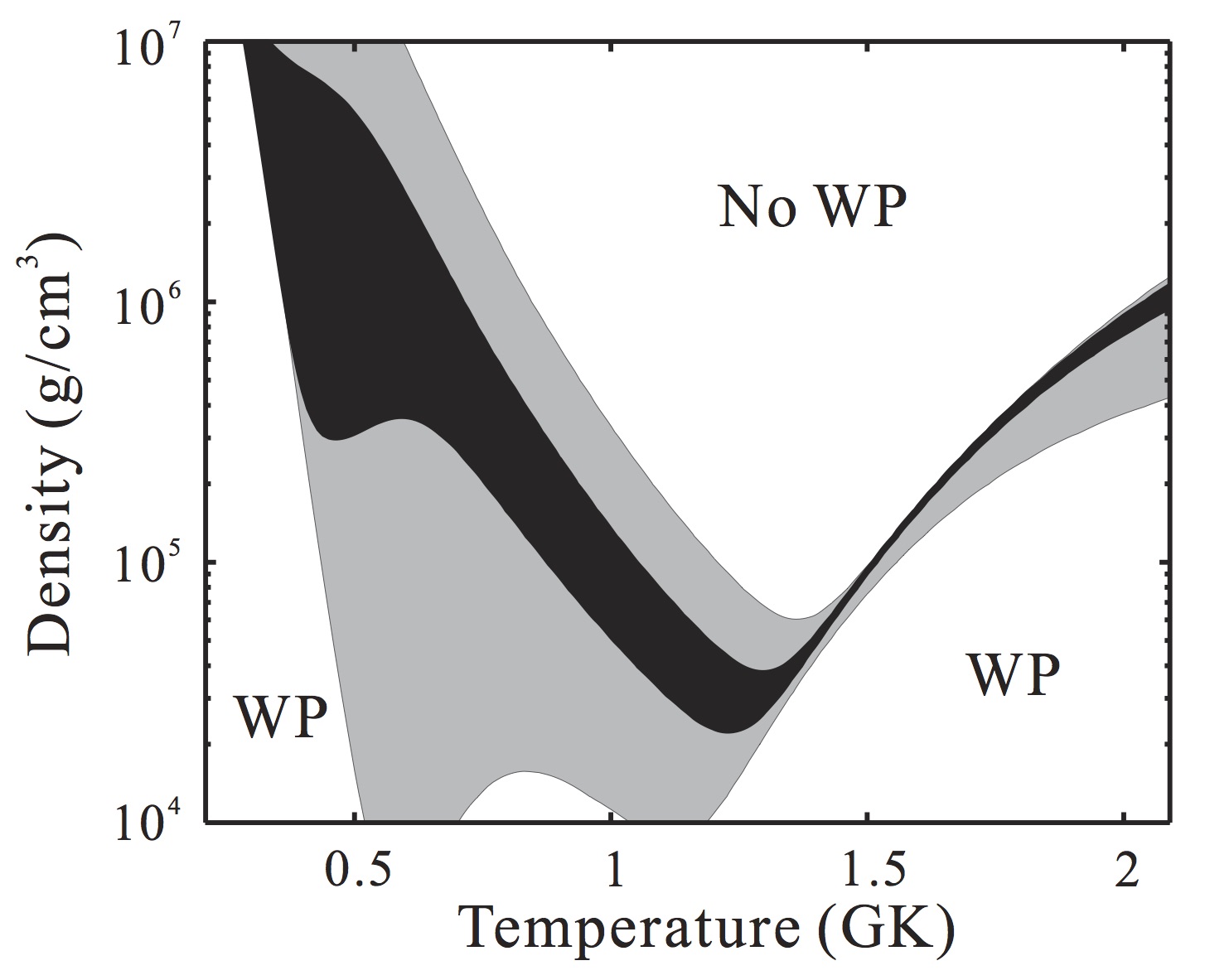}
\includegraphics[width=0.52\textwidth]{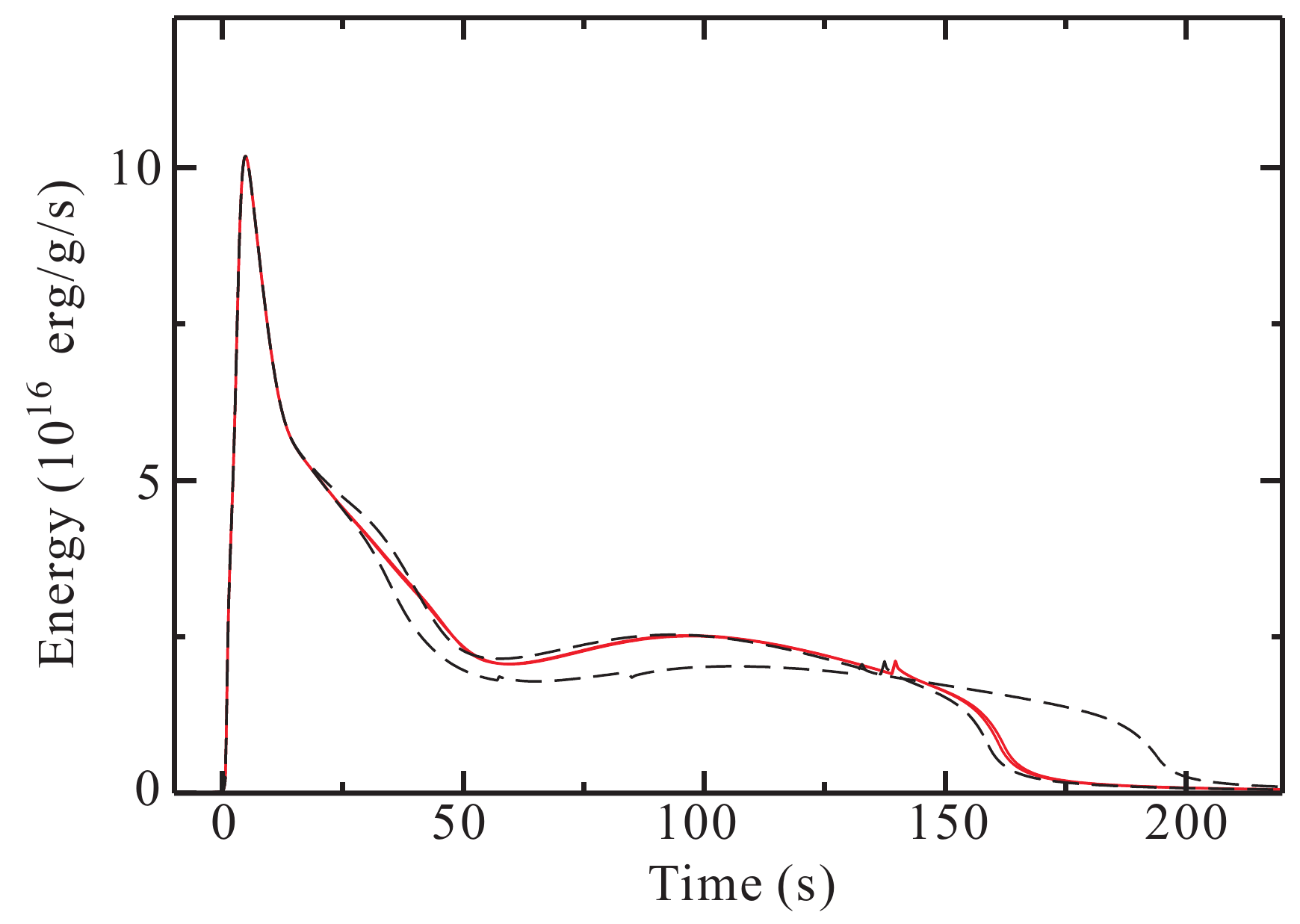}
\caption{
(Colour online) Left: regions in the temperature and density plane where $^{64}$Ge is a strong waiting point (WP)
with an effective lifetime longer than 50\% of the $\beta$-decay lifetime.
Shaded areas are deduced assuming 1$\sigma$ variation of the AME'03~\cite{Audi-NPA2003}
mass for $^{65}$As (grey) and the $^{65}$As mass from our work (black).
Right: calculated X-ray luminosity in a one zone X-ray burst model varying $S_p(^{65}$As)
within 2$\sigma$  for masses from AME'03 (dashed black lines) and from this work (solid red lines).
More details can be found in Refs.~\cite{Tu-NIM2011,Tu-PRL2011}. Taken from Ref. \cite{Tu-PRL2011}.
}
\label{fig04}
\vspace*{-0.5cm}
\end{center}
\end{figure*}

Left panel of Figure~\ref{fig04} shows regions in the temperature-density plane where proton capture reduces
the effective lifetime of $^{64}$Ge to less than 50\% of the $\beta^+$-decay lifetime,
thus resulting in a less effective waiting point.
It can be seen that for densities below $2\times 10^5$~g/cm$^3$
the required temperature range is rather narrow around 1.3~GK.
For lower temperatures, $^{64}$Ge can only be bypassed at higher densities.
Varying the new $S_p$($^{65}$As) value within $2\sigma$ provides essentially identical light curves (see right panel of Figure~\ref{fig04})
demonstrating that the achieved mass accuracy
is sufficient to eliminate the effect of the $^{65}$As mass uncertainty in X-ray burst calculations~\cite{Tu-PRL2011}.
It was found that $89-90$\% of the reaction flow passes through $^{64}$Ge via proton capture
thus indicating that $^{64}$Ge is not a significant $rp$-process waiting point.
In contrast, using the estimated upper $2\sigma$ limit for $^{65}$As from the previously considered values taken from the 2003 Atomic Mass Evaluation (AME'03) \cite{Audi-NPA2003}
($S_p$($^{65}$As)$\approx -650$ keV) leads to a reduction of the proton capture
flow through $^{64}$Ge to 54\% with a significant effect on the calculated light curve as shown in Figure~\ref{fig04} (right).

\subsection{The mass of $^{45}$Cr and the Ca-Sc cycle in the $rp$-process}
CSRe experiments yielded a more precise mass excess ME($^{45}$Cr)$=-19515(35)$ keV~\cite{Yan-APJL2013}
as compared to the previous value ME($^{45}$Cr)$=-18970(500)$ keV~\cite{Audi-NPA2003}.
This resulted in an enhanced proton separation energy $S_p(^{45}$Cr$)=2.69\pm0.13$~MeV
instead of known before $S_p(^{45}$Cr)$=2.1\pm0.5$~MeV.

\begin{figure}[t]
\begin{center}
\includegraphics[width=0.45\textwidth]{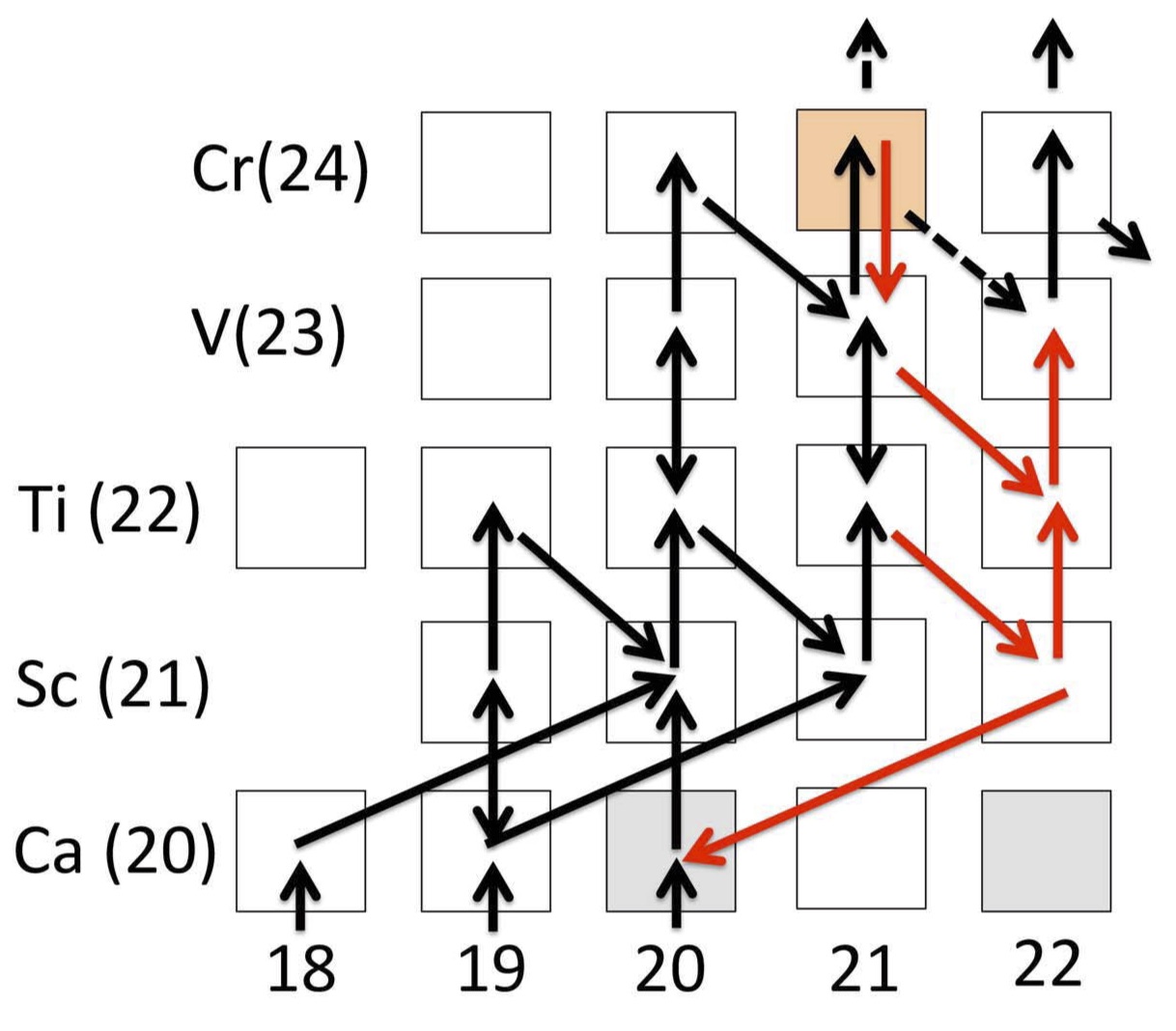}
\includegraphics[width=0.54\textwidth]{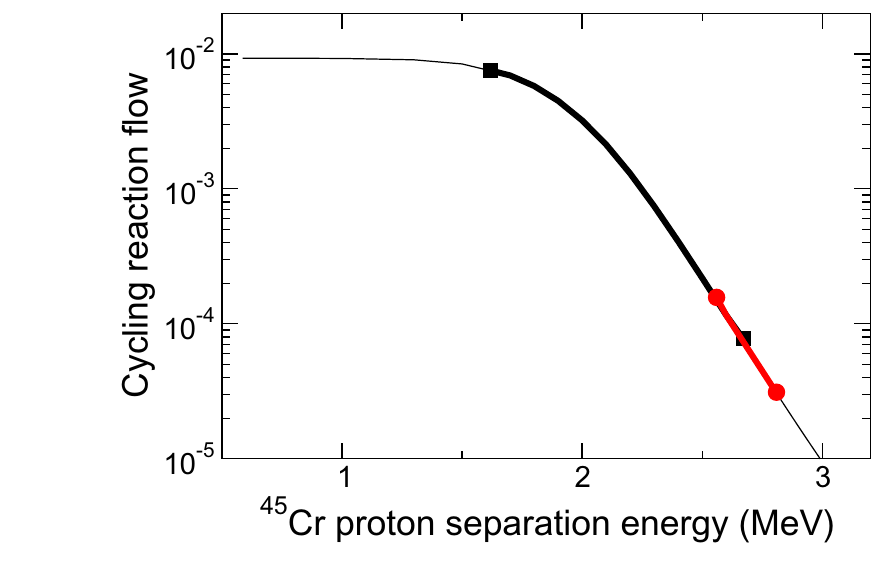}
\caption{(Colour online) Left: Integrated reaction flow in the Ca-Cr region during $rp$-process in X-ray bursts.
Black arrows show the reaction flow for our new $S_p(^{45}$Cr),
while the red arrows indicate the reaction flow for a low $S_p(^{45}$Cr) value from AME'03~\cite{Audi-NPA2003}.
Flows that disappear for the latter case are indicated as black dashed arrows.
Right: Integrated reaction flow through the Ca-Sc cycle during an X-ray burst as a function of $S_p(^{45}$Cr).
The graph spans the $3\sigma$ uncertainty of $S_p(^{45}$Cr) value from AME'03.
The thick black line limited by filled squares indicates the 1$\sigma$ uncertainty of $S_p(^{45}$Cr) in AME'03,
while the thick red line limited by filled circles indicates the $1\sigma$ uncertainty when using our new experimental result.
Taken from Ref.~\cite{Yan-APJL2013}.
} \label{FigFlow}
\end{center}
\vspace*{-0.5cm}
\end{figure}

The effect of the proton separation energy of $^{45}$Cr on the $rp$-process path is illustrated in the left panel of Figure \ref{FigFlow}.
$^{44}$V and $^{43}$Ti nuclei are in $(p,\gamma)-(\gamma,p)$ equilibrium.
The net proton capture flow at $^{43}$Ti is determined by the leakage
out of this equilibrium via the $^{44}$V$(p,\gamma)^{45}$Cr reaction,
the so-called 2p-capture process on $^{43}$Ti~\cite{Schatz-PR1998}.
For a low $S_p(^{45}$Cr) value allowed by the $3\sigma$ uncertainties of the AME'03 data~\cite{Audi-NPA2003},
the $^{45}$Cr($\gamma$, p)$^{44}$V reaction becomes effective.
This reduces the proton capture flow at $^{43}$Ti leading to a significant $\beta^+$-decay branch in $^{43}$Ti,
which drives the reaction flow into $^{43}$Sc, and then via a large (p,$\alpha$) branch into $^{40}$Ca.
This is the so-called Ca-Sc cycle \cite{VanWormer-APJ1994}, which limits strongly the reaction flow towards heavier elements
and affects significantly the X-ray burst observables~\cite{VanWormer-APJ1994}.
The right panel of Figure~\ref{FigFlow} shows the integrated reaction flow in the Ca-Sc cycle as a function of $S_p(^{45}$Cr).
Up to a proton separation energy of about 2~MeV, a strong cyclic reaction flow occurs.
The extrapolated $^{45}$Cr mass from AME'03 compilation (black solid line) allows
for a strong cycle which introduces a significant uncertainty in X-ray burst models.

To address the formation of the Ca-Sc cycle in the $rp$-process using the new mass values, 
one-zone X-ray burst model calculations were performed.
The new $S_p(^{45}$Cr) value basically excludes the formation of a significant cycle (see red line in Figure~\ref{FigFlow} (Right))
and therefore removes this uncertainty~\cite{Yan-APJL2013}.
Masses of $^{44}$V, $^{52}$Co and $^{56}$Cu were recently investigated indirectly
by using the mirror symmetry and known data from beta-delayed proton spectroscopy \cite{Tu-NPA2016}.
These new results result in even less significant Ca-Sc cycle.
 
\section{Summary and outlook}
\label{section:future}
In this contribution we showed that storage ring mass spectrometry, which was pioneered 25 years ago, 
has become an excellent and unique tool to address masses of short-lived rarely produced nuclides.
Mass surface containing more than 1000 nuclei was covered in several experiments.
Masses of more than 200 nuclei were obtained for the first time which allowed for investigating a range of applications in nuclear structure and astrophysics.

The successful mass measurement program will be continued in the future.
At the ESR, in addition to the classical SMS and IMS, the Schottky spectrometry in isochronous mode will be extended.
In order to correct for ``non-isochronicity'', a special Schottky detector is 
being designed which aims at in-ring measurement of a position of each particle at a dispersive place of the ring \cite{Sanjari-PS2015}.
The latter will allow for determining the magnetic rigidity of the particles.
Furthermore, to increase the sensitivity of the resonant detectors, it is planned to remove the ceramic gap,
which will require the detectors to be bakeable. 
At the CSRe, the correction of the ``non-isochronicity'' will be approached in a different way, namely
timing signals from a pair of time-of-flight detectors installed in a straight section of the ring
will provide the velocities of each stored ion \cite{Zhang-NIM2014a, Xing-PS2015}.
The necessary correction at the R3 setup was considered from the beginning.
The BigRips will be used to provide the identification of each injected ion together with its velocity and magnetic rigidity \cite{Yamaguchi-PS2015a}.
Resonant Schottky detectors are installed at all three facilities which will allow for simultaneous lifetime studies of the stored ions.

Storage ring mass spectrometry is in the focus of the new generation radioactive-ion beam facilities \cite{Litvinov-NIM2013,Yan-JPCS2016}.

\begin{figure}[t]
\begin{center}
\includegraphics[width=0.5\textwidth]{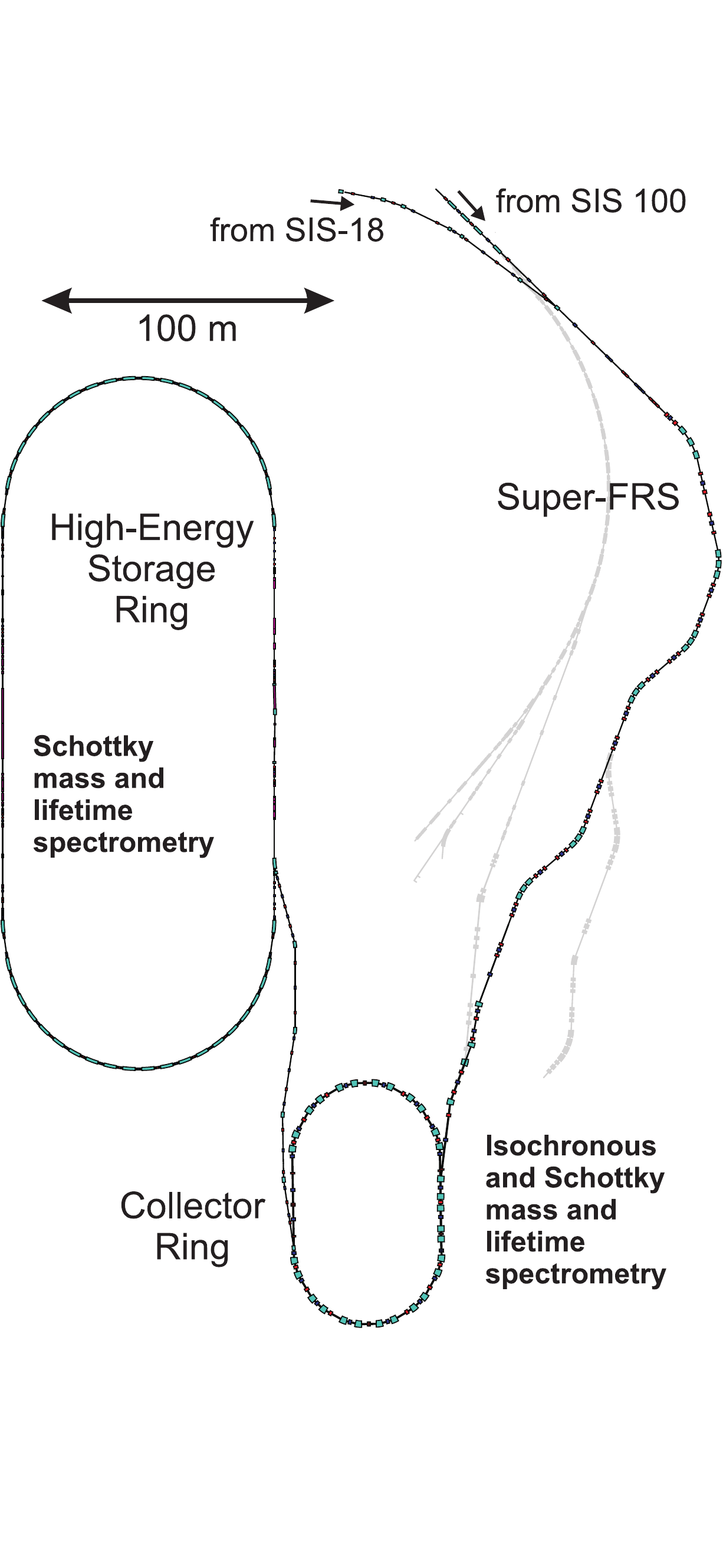}
\caption{The radioactive ion beam facility for storage ring mass measurements at FAIR.
The intense primary beams from the new accelerator SIS-100 
or the existing SIS-18 will be used to produce and separate exotic nuclei with the Super-FRS.
Mass and lifetime measurements of short-lived nuclei will be performed in the dedicated isochronous CR. 
Experiments on longer-lived nuclides will be conducted in the HESR.
Taken from Ref. \cite{Walker-IJMS2013}.
} \label{fig:fair}
\end{center}
\vspace*{-0.5cm}
\end{figure}

At the Facility for Antiproton and Ion Research, FAIR, which is in construction in Darmstadt in Germany \cite{Nilsson-PS2015}, 
several complementary research programs are planned at several storage rings 
(see, e.g., \cite{Krucken-AIP2006, Moeini-NIM2011, Antonov-NIM2011, Walker-IJMS2013, Stohlker-NIM2015}).
The new driver accelerator, 100-Tm heavy ion synchrotron, SIS-100, 
will provide bunched high intensity primary beams of up to $5\cdot10^{11}$ $^{238}$U per bunch each 1/5~s.
The secondary beams produced from these beams will be separated by the new high-acceptance fragment separator, Super-FRS, \cite{Geissel-NIM2003} which 
is designed to achieve a high transmission of fragments to the Collector Ring, CR, \cite{Dolinskii-PS2015} .
The CR is a dedicated isochronous ring \cite{Dolinskii-NIM2007, SLitvinov-NIM2013} 
where lifetime and mass measurements of short-lived nuclei will be performed \cite{Walker-IJMS2013}, see Figure \ref{fig:fair}.
At the present FRS-ESR, the transmission of fragments from the FRS to the ESR in isochronous mode is less than a percent.
The matching of the Super-FRS and the CR is taken into consideration in the design 
and about the unity transmission is expected from the exit of the Super-FRS to the CR.
Furthermore, a high-energy storage ring, HESR, will be coupled to the CR \cite{Kovalenko-PS2015}.
Schottky spectroscopy on electron-cooled ions will be possible,
which will allow for precision mass and lifetime measurements on longer-lived nuclei.
Here of interest, for instance, are the long-lived exotic isomeric states.

\begin{figure}[t]
\begin{center}
\includegraphics[width=0.8\textwidth]{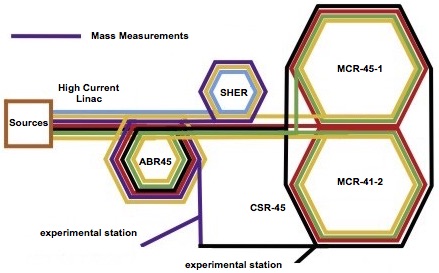}
\caption{(Colour online) A concept of the proposed HIAF facility in China.
Primary beams accelerated by synchrotron ABR-45 will be used to produce radioactive ions which will then be stored for precision experiments in the 
dedicated experimental storage ring, SHER.
Several experimental programs in medical, material, atomic, 
nuclear, plasma and high-energy physics are planned at HIAF which is indicated by different colours.
The line responsible for the mass measurements of exotic nuclei is indicated with blue colour.
Modified from Ref. \cite{Yang-NIM2013}.
} \label{fig:hiaf}
\end{center}
\vspace*{-0.5cm}
\end{figure}

In China, a High-Intensity Accelerator Facility, HIAF, is planned at Huizhou \cite{Yang-NIM2013}.
The design of the facility is being optimised. 
Figure \ref{fig:hiaf} illustrates a proposed concept.
The source terminal will host an Electron Cyclotron Resonance and a Laser Ion sources.
The ions will be accelerated by a high current linac, HISCL, to energies of about 25 MeV/u, 
and then further accelerated by a heavy synchrotron ABR-45, 
which will have the maximum magnetic rigidity $B\rho=45$~Tm.
The experiments with stored exotic nuclei will be performed in the experimental SHER ring \cite{Gao-CPC2014},
which is connected to ABR-45 by an in-flight fragment separator.
The MCR-45-1/2 and CSR-45 rings are a dedicated complex for high-energy physics experiments.

Although mass measurements are not planned at the proposed low-energy storage ring at ISOLDE in CERN, the TSR@ISOLDE project \cite{Grieser-EPJST2012},
the developed techniques will be employed for sensitive diagnosis and detection purposes. 
For instance, mass-spectrometry assisted lifetime measurements or identification of high-K isomers are proposed.
Also a suggested storage ring facility coupled to a source of thermal neutrons for studying neutron-induced nuclear reactions \cite{Reifarth-PRST2014, Glorius-PS2015}
relies on experimental methods used in storage ring mass spectrometry.

As a final conclusion we like to stress that the ability to determine the mass (and in many cases also the lifetime) of a short-lived nucleus on the basis of just a single particle 
is the unique feature of the storage-ring mass spectrometry.
This technique will provide information on the properties of the most exotic nuclei inaccessible by other methods.

\ack
This review is entirely based on the work and on previous publications of our colleagues from the international collaborations performing storage ring experiments
at the ESR, CSRe and R3 facilities. To all of them we are deeply obliged.
This work is supported in part by the 973 Program of China (No.2013CB834401), 
the NSFC grants 11035007, U1232208, 11135005, 11075103, 
the BMBF grant in the framework of the Internationale Zusammenarbeit in Bildung und Forschung Projekt-Nr. 01DO12012, 
the Helmholtz-CAS JointResearch Group HCJRG-108, 
by the External Cooperation Program of the Chinese Academy Sciences Grant No. GJHZ1305 and by
the European Research Council (ERC) under the European Union's Horizon 2020 research and innovation programme (grant agreement No 682841 ``ASTRUm'').
Y.A.L. thanks Chinese Academy of Sciences for support within the visiting professorship for senior international scientists program (Grant No.2009J2-23), 
and ESF for support within the Euro GENESIS program.

\section*{References}
\bibliographystyle{iopart-num}
\bibliography{Atomic,Mine,Rings,Structure,Cryring,Theses,Traps}

\providecommand{\newblock}{}
\begin{thebibliography}{100}
\expandafter\ifx\csname url\endcsname\relax
  \def\url#1{{\tt #1}}\fi
\expandafter\ifx\csname urlprefix\endcsname\relax\def\urlprefix{URL }\fi
\providecommand{\eprint}[2][]{\url{#2}}

\bibitem{Bohr-NS1969}
Bohr A and Mottelson B 1969 {\em Nuclear Structure\/} (World Scientific) ISBN
  9789810239794

\bibitem{Novikov-NPA2002}
Novikov Y~N, Attallah F, Bosch F, Falch M, Geissel H, Hausmann M, Kerscher T,
  Klepper O, Kluge H~J, Kozhuharov C, Litvinov Y~A, L\"obner K~E~G,
  M\"unzenberg G, Patyk Z, Radon T, Scheidenberger C, Wapstra A~H and Wollnik H
  2002 {\em Nuclear Physics A\/} {\bf 697} 92--106
  \urlprefix\url{http://www.sciencedirect.com/science/article/pii/S0375947401012337}

\bibitem{Heisenberg-ZP1932}
Heisenberg W 1932 {\em Zeitschrift f{\"u}r Physik\/} {\bf 78} 156--164
  \urlprefix\url{http://dx.doi.org/10.1007/BF01337585}

\bibitem{Mayer-PR1949}
Mayer M~G 1949 {\em Phys. Rev.\/} {\bf 75}(12) 1969--1970
  \urlprefix\url{http://link.aps.org/doi/10.1103/PhysRev.75.1969}

\bibitem{Haxel-PR1949}
Haxel O, Jensen J~H~D and Suess H~E 1949 {\em Phys. Rev.\/} {\bf 75}(11)
  1766--1766 \urlprefix\url{http://link.aps.org/doi/10.1103/PhysRev.75.1766.2}

\bibitem{Thomson-PM1912}
Thomson J 1912 {\em Philosophical Magazine Series 6\/} {\bf 24} 209--253
  \urlprefix\url{http://www.tandfonline.com/doi/abs/10.1080/14786440808637325}

\bibitem{Aston-PM1923}
Aston F 1923 {\em Philosophical Magazine Series 6\/} {\bf 45} 934--945
  \urlprefix\url{http://www.tandfonline.com/doi/abs/10.1080/14786442308636307}

\bibitem{Lunney-RMP2003}
Lunney D, Pearson J~M and Thibault C 2003 {\em Reviews of Modern Physics\/}
  {\bf 75}(3) 1021--1082
  \urlprefix\url{http://link.aps.org/doi/10.1103/RevModPhys.75.1021}

\bibitem{Blaum-PR2006}
Blaum K 2006 {\em Physics Reports\/} {\bf 425} 1--78
  \urlprefix\url{http://www.sciencedirect.com/science/article/pii/S0370157305004643}

\bibitem{Munzenberg-AIP2010}
M{\"u}nzenberg G, Geissel H and Litvinov Y~A 2010 {\em AIP Conference
  Proceedings\/} {\bf 1224} 28--46
  \urlprefix\url{http://scitation.aip.org/content/aip/proceeding/aipcp/10.1063/1.3431427}

\bibitem{100YMS}
Blaum K and Litvinov Y~A (eds) 2013 {\em International Journal of Mass
  Spectrometry\/} vol 349-350
  \urlprefix\url{http://www.sciencedirect.com/science/journal/13873806/349}

\bibitem{Wollnik-HI2015}
Wollnik H 2015 {\em Hyperfine Interactions\/}  1--15
  \urlprefix\url{http://dx.doi.org/10.1007/s10751-015-1191-3}

\bibitem{Rauth-PRL2008}
Rauth C, Ackermann D, Blaum K, Block M, Chaudhuri A, Di Z, Eliseev S, Ferrer R,
  Habs D, Herfurth F, He\ss{}berger F~P, Hofmann S, Kluge H~J, Maero G,
  Mart\'{\i}n A, Marx G, Mukherjee M, Neumayr J~B, Pla\ss{} W~R, Rahaman S,
  Rodr\'{\i}guez D, Scheidenberger C, Schweikhard L, Thirolf P~G, Vorobjev G
  and Weber C 2008 {\em Phys. Rev. Lett.\/} {\bf 100}(1) 012501
  \urlprefix\url{http://link.aps.org/doi/10.1103/PhysRevLett.100.012501}

\bibitem{Naimi-PRC2012}
Naimi S, Audi G, Beck D, Blaum K, B\"ohm C, Borgmann C, Breitenfeldt M, George
  S, Herfurth F, Herlert A, Kellerbauer A, Kowalska M, Lunney D, Minaya~Ramirez
  E, Neidherr D, Rosenbusch M, Schweikhard L, Wolf R~N and Zuber K 2012 {\em
  Phys. Rev. C\/} {\bf 86}(1) 014325
  \urlprefix\url{http://link.aps.org/doi/10.1103/PhysRevC.86.014325}

\bibitem{Manea-PRC2013}
Manea V, Atanasov D, Beck D, Blaum K, Borgmann C, Cakirli R~B, Eronen T, George
  S, Herfurth F, Herlert A, Kowalska M, Kreim S, Litvinov Y~A, Lunney D,
  Neidherr D, Rosenbusch M, Schweikhard L, Wienholtz F, Wolf R~N and Zuber K
  2013 {\em Phys. Rev. C\/} {\bf 88}(5) 054322
  \urlprefix\url{http://link.aps.org/doi/10.1103/PhysRevC.88.054322}

\bibitem{Cakirli-PRL2009}
Cakirli R~B, Casten R~F, Winkler R, Blaum K and Kowalska M 2009 {\em Phys. Rev.
  Lett.\/} {\bf 102}(8) 082501
  \urlprefix\url{http://link.aps.org/doi/10.1103/PhysRevLett.102.082501}

\bibitem{Casten-PRL2014}
Casten R~F, Cakirli R~B, Blaum K and Couture A 2014 {\em Phys. Rev. Lett.\/}
  {\bf 113}(11) 112501
  \urlprefix\url{http://link.aps.org/doi/10.1103/PhysRevLett.113.112501}

\bibitem{Neidherr-PRL2009}
Neidherr D, Audi G, Beck D, Blaum K, B\"ohm C, Breitenfeldt M, Cakirli R~B,
  Casten R~F, George S, Herfurth F, Herlert A, Kellerbauer A, Kowalska M,
  Lunney D, Minaya-Ramirez E, Naimi S, Noah E, Penescu L, Rosenbusch M, Schwarz
  S, Schweikhard L and Stora T 2009 {\em Phys. Rev. Lett.\/} {\bf 102}(11)
  112501 \urlprefix\url{http://link.aps.org/doi/10.1103/PhysRevLett.102.112501}

\bibitem{Boehm-PRC2014}
B\"ohm C, Borgmann C, Audi G, Beck D, Blaum K, Breitenfeldt M, Cakirli R~B,
  Cocolios T~E, Eliseev S, George S, Herfurth F, Herlert A, Kowalska M, Kreim
  S, Lunney D, Manea V, Minaya~Ramirez E, Naimi S, Neidherr D, Rosenbusch M,
  Schweikhard L, Stanja J, Wang M, Wolf R~N and Zuber K 2014 {\em Phys. Rev.
  C\/} {\bf 90}(4) 044307
  \urlprefix\url{http://link.aps.org/doi/10.1103/PhysRevC.90.044307}

\bibitem{Geithner-PRL2008}
Geithner W, Neff T, Audi G, Blaum K, Delahaye P, Feldmeier H, George S,
  Gu\'enaut C, Herfurth F, Herlert A, Kappertz S, Keim M, Kellerbauer A, Kluge
  H~J, Kowalska M, Lievens P, Lunney D, Marinova K, Neugart R, Schweikhard L,
  Wilbert S and Yazidjian C 2008 {\em Phys. Rev. Lett.\/} {\bf 101}(25) 252502
  \urlprefix\url{http://link.aps.org/doi/10.1103/PhysRevLett.101.252502}

\bibitem{Nakamura-PRL2009}
Nakamura T, Kobayashi N, Kondo Y, Satou Y, Aoi N, Baba H, Deguchi S, Fukuda N,
  Gibelin J, Inabe N, Ishihara M, Kameda D, Kawada Y, Kubo T, Kusaka K, Mengoni
  A, Motobayashi T, Ohnishi T, Ohtake M, Orr N~A, Otsu H, Otsuka T, Saito A,
  Sakurai H, Shimoura S, Sumikama T, Takeda H, Takeshita E, Takechi M, Takeuchi
  S, Tanaka K, Tanaka K~N, Tanaka N, Togano Y, Utsuno Y, Yoneda K, Yoshida A
  and Yoshida K 2009 {\em Phys. Rev. Lett.\/} {\bf 103}(26) 262501
  \urlprefix\url{http://link.aps.org/doi/10.1103/PhysRevLett.103.262501}

\bibitem{Blaum-PRL2003}
Blaum K, Audi G, Beck D, Bollen G, Herfurth F, Kellerbauer A, Kluge H~J, Sauvan
  E and Schwarz S 2003 {\em Phys. Rev. Lett.\/} {\bf 91}(26) 260801
  \urlprefix\url{http://link.aps.org/doi/10.1103/PhysRevLett.91.260801}

\bibitem{Kellerbauer-PRL2004}
Kellerbauer A, Audi G, Beck D, Blaum K, Bollen G, Brown B~A, Delahaye P,
  Gu\'enaut C, Herfurth F, Kluge H~J, Lunney D, Schwarz S, Schweikhard L and
  Yazidjian C 2004 {\em Phys. Rev. Lett.\/} {\bf 93}(7) 072502
  \urlprefix\url{http://link.aps.org/doi/10.1103/PhysRevLett.93.072502}

\bibitem{Naimi-PRL2010}
Naimi S, Audi G, Beck D, Blaum K, B\"ohm C, Borgmann C, Breitenfeldt M, George
  S, Herfurth F, Herlert A, Kowalska M, Kreim S, Lunney D, Neidherr D,
  Rosenbusch M, Schwarz S, Schweikhard L and Zuber K 2010 {\em Phys. Rev.
  Lett.\/} {\bf 105}(3) 032502
  \urlprefix\url{http://link.aps.org/doi/10.1103/PhysRevLett.105.032502}

\bibitem{Eliseev-PRL2011}
Eliseev S, Roux C, Blaum K, Block M, Droese C, Herfurth F, Kretzschmar M,
  Krivoruchenko M~I, Minaya~Ramirez E, Novikov Y~N, Schweikhard L, Shabaev V~M,
  \ifmmode~\check{S}\else \v{S}\fi{}imkovic F, Tupitsyn I~I, Zuber K and Zubova
  N~A 2011 {\em Phys. Rev. Lett.\/} {\bf 107}(15) 152501
  \urlprefix\url{http://link.aps.org/doi/10.1103/PhysRevLett.107.152501}

\bibitem{Nesterenko-PRC2014}
Nesterenko D~A, Eliseev S, Blaum K, Block M, Chenmarev S, D\"orr A, Droese C,
  Filianin P~E, Goncharov M, Minaya~Ramirez E, Novikov Y~N, Schweikhard L and
  Simon V~V 2014 {\em Phys. Rev. C\/} {\bf 90}(4) 042501
  \urlprefix\url{http://link.aps.org/doi/10.1103/PhysRevC.90.042501}

\bibitem{Wienholtz-Nature2013}
Wienholtz F, Beck D, Blaum K, Borgmann C, Breitenfeldt M, Cakirli R~B, George
  S, Herfurth F, Holt J~D, Kowalska M, Kreim S, Lunney D, Manea V, Men{\'e}ndez
  J, Neidherr D, Rosenbusch M, Schweikhard L, Schwenk A, Simonis J, Stanja J,
  Wolf R~N and Zuber K 2013 {\em Nature\/} {\bf 498} 346--349
  \urlprefix\url{http://dx.doi.org/10.1038/nature12226}

\bibitem{Dworschak-PRL2008}
Dworschak M, Audi G, Blaum K, Delahaye P, George S, Hager U, Herfurth F,
  Herlert A, Kellerbauer A, Kluge H~J, Lunney D, Schweikhard L and Yazidjian C
  2008 {\em Phys. Rev. Lett.\/} {\bf 100}(7) 072501
  \urlprefix\url{http://link.aps.org/doi/10.1103/PhysRevLett.100.072501}

\bibitem{Kanungo-PRL2009}
Kanungo R, Nociforo C, Prochazka A, Aumann T, Boutin D, Cortina-Gil D, Davids
  B, Diakaki M, Farinon F, Geissel H, Gernh\"auser R, Gerl J, Janik R, Jonson
  B, Kindler B, Kn\"obel R, Kr\"ucken R, Lantz M, Lenske H, Litvinov Y~A,
  Lommel B, Mahata K, Maierbeck P, Musumarra A, Nilsson T, Otsuka T, Perro C,
  Scheidenberger C, Sitar B, Strmen P, Sun B, Szarka I, Tanihata I, Utsuno Y,
  Weick H and Winkler M 2009 {\em Phys. Rev. Lett.\/} {\bf 102}(15) 152501
  \urlprefix\url{http://link.aps.org/doi/10.1103/PhysRevLett.102.152501}

\bibitem{Kanungo-PS2013}
Kanungo R 2013 {\em Physica Scripta\/} {\bf T152} 014002
  \urlprefix\url{http://stacks.iop.org/1402-4896/2013/i=T152/a=014002}

\bibitem{Otsuka-PS2013}
Otsuka T 2013 {\em Physica Scripta\/} {\bf T152} 014007
  \urlprefix\url{http://stacks.iop.org/1402-4896/2013/i=T152/a=014007}

\bibitem{Rodriguez-PRL2004}
Rodr\'{\i}guez D, Kolhinen V~S, Audi G, \"Ayst\"o J, Beck D, Blaum K, Bollen G,
  Herfurth F, Jokinen A, Kellerbauer A, Kluge H~J, Oinonen M, Schatz H, Sauvan
  E and Schwarz S 2004 {\em Phys. Rev. Lett.\/} {\bf 93}(16) 161104
  \urlprefix\url{http://link.aps.org/doi/10.1103/PhysRevLett.93.161104}

\bibitem{Weber-PRC2008}
Weber C, Elomaa V~V, Ferrer R, Fr\"ohlich C, Ackermann D, \"Ayst\"o J, Audi G,
  Batist L, Blaum K, Block M, Chaudhuri A, Dworschak M, Eliseev S, Eronen T,
  Hager U, Hakala J, Herfurth F, He\ss{}berger F~P, Hofmann S, Jokinen A,
  Kankainen A, Kluge H~J, Langanke K, Mart\'{\i}n A, Mart\'{\i}nez-Pinedo G,
  Mazzocco M, Moore I~D, Neumayr J~B, Novikov Y~N, Penttil\"a H, Pla\ss{} W~R,
  Popov A~V, Rahaman S, Rauscher T, Rauth C, Rissanen J, Rodr\'{\i}guez D,
  Saastamoinen A, Scheidenberger C, Schweikhard L, Seliverstov D~M, Sonoda T,
  Thielemann F~K, Thirolf P~G and Vorobjev G~K 2008 {\em Phys. Rev. C\/} {\bf
  78}(5) 054310
  \urlprefix\url{http://link.aps.org/doi/10.1103/PhysRevC.78.054310}

\bibitem{Baruah-PRL2008}
Baruah S, Audi G, Blaum K, Dworschak M, George S, Gu\'enaut C, Hager U,
  Herfurth F, Herlert A, Kellerbauer A, Kluge H~J, Lunney D, Schatz H,
  Schweikhard L and Yazidjian C 2008 {\em Phys. Rev. Lett.\/} {\bf 101}(26)
  262501 \urlprefix\url{http://link.aps.org/doi/10.1103/PhysRevLett.101.262501}

\bibitem{Haettner-PRL2011}
Haettner E, Ackermann D, Audi G, Blaum K, Block M, Eliseev S, Fleckenstein T,
  Herfurth F, He\ss{}berger F~P, Hofmann S, Ketelaer J, Ketter J, Kluge H~J,
  Marx G, Mazzocco M, Novikov Y~N, Pla\ss{} W~R, Rahaman S, Rauscher T,
  Rodr\'{\i}guez D, Schatz H, Scheidenberger C, Schweikhard L, Sun B, Thirolf
  P~G, Vorobjev G, Wang M and Weber C 2011 {\em Phys. Rev. Lett.\/} {\bf
  106}(12) 122501
  \urlprefix\url{http://link.aps.org/doi/10.1103/PhysRevLett.106.122501}

\bibitem{Bertulani-PR2010}
Bertulani C~A and Gade A 2010 {\em Physics Reports\/} {\bf 485} 195
  \urlprefix\url{http://www.sciencedirect.com/science/article/pii/S037015730900252X}

\bibitem{Arcones-JPG2013}
Arcones A and Thielemann F~K 2013 {\em Journal of Physics G: Nuclear and
  Particle Physics\/} {\bf 40} 013201
  \urlprefix\url{http://stacks.iop.org/0954-3899/40/i=1/a=013201}

\bibitem{Langanke-PS2013}
Langanke K and Schatz H 2013 {\em Physica Scripta\/} {\bf T152} 014011
  \urlprefix\url{http://stacks.iop.org/1402-4896/2013/i=T152/a=014011}

\bibitem{Wolf-PRL2013}
Wolf R~N, Beck D, Blaum K, B\"ohm C, Borgmann C, Breitenfeldt M, Chamel N,
  Goriely S, Herfurth F, Kowalska M, Kreim S, Lunney D, Manea V, Minaya~Ramirez
  E, Naimi S, Neidherr D, Rosenbusch M, Schweikhard L, Stanja J, Wienholtz F
  and Zuber K 2013 {\em Phys. Rev. Lett.\/} {\bf 110}(4) 041101
  \urlprefix\url{http://link.aps.org/doi/10.1103/PhysRevLett.110.041101}

\bibitem{Langanke-PS2015}
Langanke K and Marti­nez-Pinedo G 2015 {\em Physica Scripta\/} {\bf T166}
  014001 \urlprefix\url{http://stacks.iop.org/1402-4896/2015/i=T166/a=014001}

\bibitem{Erler-Na2012}
Erler J, Birge N, Kortelainen M, Nazarewicz W, Olsen E, Perhac A~M and Stoitsov
  M 2012 {\em Nature\/} {\bf 486} 509--512
  \urlprefix\url{http://dx.doi.org/10.1038/nature11188}

\bibitem{AME12}
Audi G, Wang M, Wapstra A~H, Kondev F~G, MacCormick M, Xu X and Pfeiffer B 2012
  {\em Chinese Physics C\/} {\bf 36} 1287--1602
  \urlprefix\url{http://stacks.iop.org/1674-1137/36/i=12/a=002}

\bibitem{Blaum-JPCS2011}
Blaum K, Block M, Cakirli R~B, Eliseev S, Kowalska M, Kreim S, Litvinov Y~A,
  Nagy S, N{\"o}rtersh{\"a}user W and Yordanov D~T 2011 {\em Journal of
  Physics: Conference Series\/} {\bf 312} 092001
  \urlprefix\url{http://stacks.iop.org/1742-6596/312/i=9/a=092001}

\bibitem{Bosch-PPNP2013}
Bosch F, Litvinov Y~A and St{\"o}hlker T 2013 {\em Progress in Particle and
  Nuclear Physics\/} {\bf 73} 84--140
  \urlprefix\url{http://www.sciencedirect.com/science/article/pii/S0146641013000744}

\bibitem{Kurcewicz-PLB2012}
Kurcewicz J, Farinon F, Geissel H, Pietri S, Nociforo C, Prochazka A, Weick H,
  Winfield J, Estrad{\'e} A, Allegro P, Bail A, B{\'e}lier G, Benlliure J,
  Benzoni G, Bunce M, Bowry M, Caballero-Folch R, Dillmann I, Evdokimov A, Gerl
  J, Gottardo A, Gregor E, Janik R, Keli{\'c}-Heil A, Kn{\"o}bel R, Kubo T,
  Litvinov Y, Merchan E, Mukha I, Naqvi F, Pf{\"u}tzner M, Pomorski M,
  Podoly{\'a}k Z, Regan P, Riese B, Ricciardi M, Scheidenberger C, Sitar B,
  Spiller P, Stadlmann J, Strmen P, Sun B, Szarka I, Taieb J, Terashima S,
  Valiente-Dob{\'o}n J, Winkler M and Woods P 2012 {\em Physics Letters B\/}
  {\bf 717} 371--375
  \urlprefix\url{http://www.sciencedirect.com/science/article/pii/S0370269312009689}

\bibitem{Sun-PRC2008}
Sun B, Montes F, Geng L~S, Geissel H, Litvinov Y~A and Meng J 2008 {\em Phys.
  Rev. C\/} {\bf 78}(2) 025806
  \urlprefix\url{http://link.aps.org/doi/10.1103/PhysRevC.78.025806}

\bibitem{Sun-FP2015}
Sun B, Litvinov Y, Tanihata I and Zhang Y 2015 {\em Frontiers of Physics\/}
  {\bf 10} 1--25 \urlprefix\url{http://dx.doi.org/10.1007/s11467-015-0503-z}

\bibitem{Sobiczewski-PRC2014a}
Sobiczewski A and Litvinov Y~A 2014 {\em Phys. Rev. C\/} {\bf 89}(2) 024311
  \urlprefix\url{http://link.aps.org/doi/10.1103/PhysRevC.89.024311}

\bibitem{Sobiczewski-PRC2014b}
Sobiczewski A and Litvinov Y~A 2014 {\em Phys. Rev. C\/} {\bf 90}(1) 017302
  \urlprefix\url{http://link.aps.org/doi/10.1103/PhysRevC.90.017302}

\bibitem{Blaum-PS2013}
Blaum K, Dilling J and N{\"o}rtersh{\"a}user W 2013 {\em Physica Scripta\/}
  {\bf T152} 014017
  \urlprefix\url{http://stacks.iop.org/1402-4896/2013/i=T152/a=014017}

\bibitem{Franzke-MSR2008}
Franzke B, Geissel H and M{\"u}nzenberg G 2008 {\em Mass Spectrometry
  Reviews\/} {\bf 27} 428--469
  \urlprefix\url{http://dx.doi.org/10.1002/mas.20173}

\bibitem{STORI14}
Egelhof P, Litvinov Y~A and Steck M (eds) 2015 {\em Physica Scripta\/} vol T166
  \urlprefix\url{http://iopscience.iop.org/1402-4896/2015/T166}

\bibitem{Blasche-IEEE1985}
Blasche K, B{\"o}hne D, Franzke B and Prange H 1985 {\em IEEE Transactions on
  Nuclear Science\/} {\bf NS-32} 2657--2661
  \urlprefix\url{http://accelconf.web.cern.ch/AccelConf/p85/PDF/PAC1985_2657.PDF}

\bibitem{Armbruster-AIPCP1987}
Armbruster P, Clerc H~G, Dufour J~P, Franczak B, Geissel H, Hanelt E, Klepper
  O, Langenbeck B, M{\"u}nzenberg G, Nickel F, Pravikoff M~S, Roeckl E, Schardt
  D, Schmidt K~H, Sch{\"u}ll D, Schwab T, Sherrill B, S{\"u}mmerer K and
  Wollnik H 1987 {\em AIP Conference Proceedings\/} {\bf 164} 839--844
  \urlprefix\url{http://scitation.aip.org/content/aip/proceeding/aipcp/10.1063/1.37048}

\bibitem{Geissel-NIM1992}
Geissel H, Armbruster P, Behr K, Br{\"u}nle A, Burkard K, Chen M, Folger H,
  Franczak B, Keller H, Klepper O, Langenbeck B, Nickel F, Pfeng E,
  Pf{\"u}tzner M, Roeckl E, Rykaczewski K, Schall I, Schardt D, Scheidenberger
  C, Schmidt K~H, Schr{\"o}ter A, Schwab T, S{\"u}mmerer K, Weber M,
  M{\"u}nzenberg G, Brohm T, Clerc H~G, Fauerbach M, Gaimard J~J, Grewe A,
  Hanelt E, Kn{\"o}dler B, Steiner M, Voss B, Weckenmann J, Ziegler C, Magel A,
  Wollnik H, Dufour J, Fujita Y, Vieira D and Sherrill B 1992 {\em Nuclear
  Instruments and Methods in Physics Research Section B: Beam Interactions with
  Materials and Atoms\/} {\bf 70} 286--297
  \urlprefix\url{http://www.sciencedirect.com/science/article/pii/0168583X9295944M}

\bibitem{Franzke-NIM1987}
Franzke B 1987 {\em Nuclear Instruments and Methods in Physics Research Section
  B: Beam Interactions with Materials and Atoms\/} {\bf 24/25} 18--25
  \urlprefix\url{http://www.sciencedirect.com/science/article/pii/0168583X87905830}

\bibitem{Lestinsky-PS2015}
Lestinsky M, Br{\"a}uning-Demian A, Danared H, Engstr{\"o}m M, Enders W,
  Fedotova S, Franzke B, Heinz A, Herfurth F, K{\"a}llberg A, Kester O,
  Litvinov Y, Steck M, Reistad D, Simonsson A, Skeppstedt {\"O}, St{\"o}hlker
  T, Vorobjev G and {the CRYRING@ESR working group} 2015 {\em Physica
  Scripta\/} {\bf T166} 014075
  \urlprefix\url{http://stacks.iop.org/1402-4896/2015/i=T166/a=014075}

\bibitem{Walker-IJMS2013}
Walker P, Litvinov Y~A and Geissel H 2013 {\em International Journal of Mass
  Spectrometry\/} {\bf 349-350} 247 -- 254
  \urlprefix\url{http://www.sciencedirect.com/science/article/pii/S1387380613001334}

\bibitem{Xia-NIM2002}
Xia J, Zhan W, Wei B, Yuan Y, Song M, Zhang W, Yang X, Yuan P, Gao D, Zhao H,
  Yang X, Xiao G, Man K, Dang J, Cai X, Wang Y, Tang J, Qiao W, Rao Y, He Y,
  Mao L and Zhou Z 2002 {\em Nuclear Instruments and Methods in Physics
  Research Section A: Accelerators, Spectrometers, Detectors and Associated
  Equipment\/} {\bf 488} 11--25
  \urlprefix\url{http://www.sciencedirect.com/science/article/pii/S0168900202004758}

\bibitem{Xiao-IJMPE2009}
Xiao G~Q, Xia J~W, Yuan Y~J, Mao R~S, Zheng J~H, Tu X~L, Wang M, Huang W~X, Xu
  H~S and Zhan W~L 2009 {\em International Journal of Modern Physics E\/} {\bf
  18} 405
  \urlprefix\url{http://www.worldscientific.com/doi/abs/10.1142/S0218301309012446}

\bibitem{Zhan-NPA2010}
Zhan W, Xu H, Xiao G, Xia J, Zhao H and Yuan Y 2010 {\em Nuclear Physics A\/}
  {\bf 834} 694c--700c
  \urlprefix\url{http://www.sciencedirect.com/science/article/pii/S0375947410001272}

\bibitem{Yuan-NIM2013}
Yuan Y, Yang J, Xia J, Yuan P, Qiao W, Gao D, Xiao G, Zhao H, Xu H, Song M,
  Yang X, Cai X, Ma L, Yang X, Man K, He Y, Zhou Z, Zhang J, Xu Z, Liu Y, Mao
  R, Zhang W, Xie D, Sun L, Yang Y, Yin D, Li P, Li J, Shi J, Chai W, Wei B and
  Zhan W 2013 {\em Nuclear Instruments and Methods in Physics Research Section
  B: Beam Interactions with Materials and Atoms\/} {\bf 317, Part B} 214--217
  \urlprefix\url{http://www.sciencedirect.com/science/article/pii/S0168583X13008537}

\bibitem{Xu-IJMPE2009}
Xu H 2009 {\em International Journal of Modern Physics E\/} {\bf 18} 335
  \urlprefix\url{http://www.worldscientific.com/doi/abs/10.1142/S0218301309012367}

\bibitem{Xu-IJMS2013}
Xu H, Zhang Y and Litvinov Y~A 2013 {\em International Journal of Mass
  Spectrometry\/} {\bf 349-350} 162--171
  \urlprefix\url{http://www.sciencedirect.com/science/article/pii/S138738061300170X}

\bibitem{Ozawa-PTEP2012}
Ozawa A, Uesaka T, Wakasugi M and {the Rare-RI Ring Collaboration} 2012 {\em
  Progress of Theoretical and Experimental Physics\/} {\bf 2012} 03C009
  \urlprefix\url{http://ptep.oxfordjournals.org/content/2012/1/03C009}

\bibitem{Geissel-ARNPS1995}
Geissel H, Munzenberg G and Riisager K 1995 {\em Annual Review of Nuclear and
  Particle Sciences\/} {\bf 45} 163
  \urlprefix\url{http://www.annualreviews.org/doi/abs/10.1146/annurev.ns.45.120195.001115}

\bibitem{Zhong-JPCS2010}
Zhong Q, Aumann T, Bishop S, Blaum K, Boretzky K, Bosch F, Br{\"a}uning H,
  Brandau C, Davinson T, Dillmann I, Ershova O, Geissel H, Gy{\"u}rky G, Heil
  M, K{\"a}ppeler F, Keli{\'c} A, Kozhuharov C, Langer C, Bleis T~L, Litvinov
  Y~A, Lotay G, Marganiec J, Petridis N, Plag R, Popp U, Reifarth R, Riese B,
  Rigollet C, Scheidenberger C, Simon H, St{\"o}hlker T, Sz{\"u}cs T, Weber G,
  Weick H, Winters D~F~A, Winters N and Woods P~J 2010 {\em Journal of Physics:
  Conference Series\/} {\bf 202} 2011
  \urlprefix\url{http://stacks.iop.org/1742-6596/202/i=1/a=012011}

\bibitem{Bo-PRC2015}
Mei B, Aumann T, Bishop S, Blaum K, Boretzky K, Bosch F, Brandau C, Br\"auning
  H, Davinson T, Dillmann I, Dimopoulou C, Ershova O, F\"ul\"op Z, Geissel H,
  Glorius J, Gy\"urky G, Heil M, K\"appeler F, Kelic-Heil A, Kozhuharov C,
  Langer C, Le~Bleis T, Litvinov Y, Lotay G, Marganiec J, M\"unzenberg G,
  Nolden F, Petridis N, Plag R, Popp U, Rastrepina G, Reifarth R, Riese B,
  Rigollet C, Scheidenberger C, Simon H, Sonnabend K, Steck M, St\"ohlker T,
  Sz\"ucs T, S\"ummerer K, Weber G, Weick H, Winters D, Winters N, Woods P and
  Zhong Q 2015 {\em Phys. Rev. C\/} {\bf 92}(3) 035803
  \urlprefix\url{http://link.aps.org/doi/10.1103/PhysRevC.92.035803}

\bibitem{Woods-PS2015}
Woods P, Blaum K, Bosch F, Heil M, Litvinov Y~A and Reifarth R 2015 {\em
  Physica Scripta\/} {\bf T166} 014002
  \urlprefix\url{http://stacks.iop.org/1402-4896/2015/i=T166/a=014002}

\bibitem{Doherty-PS2015}
Doherty D~T, Woods P~J, Litvinov Y~A, Najafi M~A, Bagchi S, Bishop S, Bo M,
  Brandau C, Davinson T, Dillmann I, Estrade A, Egelhof P, Evdokimov A,
  Gumberidze A, Heil M, Lederer C, Litvinov S~A, Lotay G, Kalantar-Nayestanaki
  N, Kiselev O, Kozhuharov C, Kr{\"o}ll T, Mahjour-Shafei M, Mutterer M, Nolden
  F, Petridis N, Popp U, Reifarth R, Rigollet C, Roy S, Steck M, St{\"o}hlker
  T, Streicher B, Trotsenko S, von Schmid M, Yan X~L and Zamora J~C 2015 {\em
  Physica Scripta\/} {\bf T166} 014007
  \urlprefix\url{http://stacks.iop.org/1402-4896/2015/i=T166/a=014007}

\bibitem{vonSchmid-PS2015}
von Schmid M, Bagchi S, B{\"o}nig S, Csatlós M, Dillmann I, Dimopoulou C,
  Egelhof P, Eremin V, Furuno T, Geissel H, Gernh{\"a}user R, Harakeh M~N,
  Hartig A~L, Ilieva S, Kalantar-Nayestanaki N, Kiselev O, Kollmus H,
  Kozhuharov C, Krasznahorkay A, Kr{\"o}ll T, Kuilman M, Litvinov S, Litvinov
  Y~A, Mahjour-Shafiei M, Mutterer M, Nagae D, Najafi M~A, Nociforo C, Nolden
  F, Popp U, Rigollet C, Roy S, Scheidenberger C, Steck M, Streicher B, Stuhl
  L, Th{\"u}rauf M, Uesaka T, Weick H, Winfield J~S, Winters D, Woods P~J,
  Yamaguchi T, Yue K, Zamora J~C, Zenihiro J and {the EXL collaboration} 2015
  {\em Physica Scripta\/} {\bf T166} 014005
  \urlprefix\url{http://stacks.iop.org/1402-4896/2015/i=T166/a=014005}

\bibitem{Zamora-PS2015}
Zamora J~C, Bagchi S, B{\"o}nig S, Csatlós M, Dillmann I, Dimopoulou C,
  Egelhof P, Eremin V, Furuno T, Geissel H, Gernh{\"a}user R, Harakeh M~N,
  Hartig A~L, Ilieva S, Kalantar-Nayestanaki N, Kiselev O, Kollmus K,
  Kozhuharov C, Krasznahorkay A, Kr{\"o}ll T, Kuilman M, Litvinov S, Litvinov
  Y~A, Mahjour-Shafiei M, Mutterer M, Nagae D, Najafi M~A, Nociforo C, Nolden
  F, Popp U, Rigollet C, Roy S, Scheidenberger C, von Schmid M, Steck M,
  Streicher B, Stuhl L, Th{\"u}rauf M, Uesaka T, Weick H, Winfield J~S, Winters
  D, Woods P~J, Yamaguchi T, Yue K, Zenihiro J and {the EXL collaboration} 2015
  {\em Physica Scripta\/} {\bf T166} 014006
  \urlprefix\url{http://stacks.iop.org/1402-4896/2015/i=T166/a=014006}

\bibitem{Botermann-PRL2014}
Botermann B, Bing D, Geppert C, Gwinner G, H\"ansch T~W, Huber G, Karpuk S,
  Krieger A, K\"uhl T, N\"ortersh\"auser W, Novotny C, Reinhardt S, S\'anchez
  R, Schwalm D, St\"ohlker T, Wolf A and Saathoff G 2014 {\em Phys. Rev.
  Lett.\/} {\bf 113}(12) 120405
  \urlprefix\url{http://link.aps.org/doi/10.1103/PhysRevLett.113.120405}

\bibitem{Tashenov-PRL2014}
Tashenov S, Bana\ifmmode~\acute{s}\else \'{s}\fi{} D, Beyer H, Brandau C,
  Fritzsche S, Gumberidze A, Hagmann S, Hillenbrand P~M, J\"org H, Kojouharov
  I, Kozhuharov C, Lestinsky M, Litvinov Y~A, Maiorova A~V, Schaffner H,
  Shabaev V~M, Spillmann U, St\"ohlker T, Surzhykov A and Trotsenko S 2014 {\em
  Phys. Rev. Lett.\/} {\bf 113}(11) 113001
  \urlprefix\url{http://link.aps.org/doi/10.1103/PhysRevLett.113.113001}

\bibitem{Lochmann-PRA2014}
Lochmann M, J\"ohren R, Geppert C, Andelkovic Z, Anielski D, Botermann B,
  Bussmann M, Dax A, Fr\"ommgen N, Hammen M, Hannen V, K\"uhl T, Litvinov Y~A,
  L\'opez-Coto R, St\"ohlker T, Thompson R~C, Vollbrecht J, Volotka A,
  Weinheimer C, Wen W, Will E, Winters D, S\'anchez R and N\"ortersh\"auser W
  2014 {\em Phys. Rev. A\/} {\bf 90}(3) 030501
  \urlprefix\url{http://link.aps.org/doi/10.1103/PhysRevA.90.030501}

\bibitem{Beyer-JPB2015}
Beyer H~F, Gassner T, Trassinelli M, He{\ss}Ÿ R, Spillmann U, Banaś D,
  Blumenhagen K~H, Bosch F, Brandau C, Chen W, Dimopoulou C, F{\"o}rster E,
  Grisenti R~E, Gumberidze A, Hagmann S, Hillenbrand P~M, Indelicato P,
  Jagodzinski P, K{\"a}mpfer T, Kozhuharov C, Lestinsky M, Liesen D, Litvinov
  Y~A, Loetzsch R, Manil B, M{\"a}rtin R, Nolden F, Petridis N, Sanjari M~S,
  Schulze K~S, Schwemlein M, Simionovici A, Steck M, St{\"o}hlker T, Szabo C~I,
  Trotsenko S, Uschmann I, Weber G, Wehrhan O, Winckler N, Winters D~F~A,
  Winters N and Ziegler E 2015 {\em Journal of Physics B: Atomic, Molecular and
  Optical Physics\/} {\bf 48} 144010
  \urlprefix\url{http://stacks.iop.org/0953-4075/48/i=14/a=144010}

\bibitem{Nortershauser-PS2015}
N{\"o}rtersh{\"a}user W and Sanchez R 2015 {\em Physica Scripta\/} {\bf T166}
  014020 \urlprefix\url{http://stacks.iop.org/1402-4896/2015/i=T166/a=014020}

\bibitem{Sanchez-PS2015}
Sanchez R, Ullmann J, Vollbrecht J, Andelkovic Z, Dax A, Geithner W, Geppert C,
  Gorges C, Hammen M, Hannen V, Kaufmann S, K{\"o}nig K, Litvinov Y~A, Lochmann
  M, Maa{\aa}Ÿ B, Meisner J, Murb{\"o}ck T, N{\"o}rtersh{\"a}user W, Schmidt S,
  Schmidt M, Steck M, St{\"o}hlker T, Thompson R~C and Weinheimer C 2015 {\em
  Physica Scripta\/} {\bf T166} 014021
  \urlprefix\url{http://stacks.iop.org/1402-4896/2015/i=T166/a=014021}

\bibitem{Jungmann-PS2015}
Jungmann K~P 2015 {\em Physica Scripta\/} {\bf T166} 014033
  \urlprefix\url{http://stacks.iop.org/1402-4896/2015/i=T166/a=014033}

\bibitem{Kubo-PTEP2012}
Kubo T, Kameda D, Suzuki H, Fukuda N, Takeda H, Yanagisawa Y, Ohtake M, Kusaka
  K, Yoshida K, Inabe N, Ohnishi T, Yoshida A, Tanaka K and Mizoi Y 2012 {\em
  Progress in Theoretical and Experimental Physics\/} {\bf 2012} 03C003
  \urlprefix\url{http://ptep.oxfordjournals.org/content/2012/1/03C003}

\bibitem{Scheidenberger-HI2006}
Scheidenberger C, Beckert K, Beller P, Bosch F, Brandau C, Boutin D, Chen L,
  Franzke B, Geissel H, Kn{\"o}bel R, Kozhuharov C, Kurcewicz J, Litvinov S,
  Litvinov Y, Mazzocco M, M{\"u}nzenberg G, Nolden F, Pla{\ss}Ÿ W, Steck M, Sun
  B, Weick H and Winkler M 2006 {\em Hyperfine Interactions\/} {\bf 173} 61--66
  \urlprefix\url{http://dx.doi.org/10.1007/s10751-007-9543-2}

\bibitem{Meshkov-PS2015}
Meshkov I 2015 {\em Physica Scripta\/} {\bf T166} 014037
  \urlprefix\url{http://stacks.iop.org/1402-4896/2015/i=T166/a=014037}

\bibitem{Brandau-HI2010}
Brandau C, Kozhuharov C, M{\"u}ller A, Bernhardt D, Bosch F, Boutin D, Currell
  F~J, Dimopoulou C, Franzke B, Fritzsche S, Gumberidze A, Harman Z, Jentschura
  U~D, Keitel C~H, Kozhedub Y~S, Kr{\"u}cken R, Litvinov Y~A, Nolden F,
  O'Rourke B, Reuschl R, Schippers S, Shabaev V~M, Spillmann U, Stachura Z,
  Steck M, St{\"o}hlker T, Tupitsyn I~I, Winters D~F~A and Wolf A 2010 {\em
  Hyperfine Interactions\/} {\bf 196} 115--127
  \urlprefix\url{http://dx.doi.org/10.1007/s10751-009-0142-2}

\bibitem{Doherty-PS2015a}
Doherty D~T, Kraft-Bermuth S and Litvinov S 2015 {\em Physica Scripta\/} {\bf
  T166} 014077
  \urlprefix\url{http://stacks.iop.org/1402-4896/2015/i=T166/a=014077}

\bibitem{Geissel-PRL1992}
Geissel H, Beckert K, Bosch F, Eickhoff H, Franczak B, Franzke B, Jung M,
  Klepper O, Moshammer R, M\"unzenberg G, Nickel F, Nolden F, Schaaf U,
  Scheidenberger C, Sp\"adtke P, Steck M, S\"ummerer K and Magel A 1992 {\em
  Physical Review Letters\/} {\bf 68}(23) 3412--3415
  \urlprefix\url{http://link.aps.org/doi/10.1103/PhysRevLett.68.3412}

\bibitem{Wakasugi-JPSJ2015}
Wakasugi M and {the RIKEN RI Ring Collaboration} 2015 {\em JPS Conference
  Proceedings\/} {\bf 6} 010020
  \urlprefix\url{http://dx.doi.org/10.7566/JPSCP.6.010020}

\bibitem{Yan-PS2015}
Yan X~L, Bosch F, Litvinov Y~A, Nolden F, Steck M, Tu X~L, Xu H~S, Zhou X~H and
  Zhang Y~H 2015 {\em Physica Scripta\/} {\bf T166} 014045
  \urlprefix\url{http://stacks.iop.org/1402-4896/2015/i=T166/a=014045}

\bibitem{Radon-PRL1997}
Radon T, Kerscher T, Schlitt B, Beckert K, Beha T, Bosch F, Eickhoff H, Franzke
  B, Fujita Y, Geissel H, Hausmann M, Irnich H, Jung H~C, Klepper O, Kluge H~J,
  Kozhuharov C, Kraus G, L\"obner K~E~G, M\"unzenberg G, Novikov Y, Nickel F,
  Nolden F, Patyk Z, Reich H, Scheidenberger C, Schwab W, Steck M, S\"ummerer K
  and Wollnik H 1997 {\em Phys. Rev. Lett.\/} {\bf 78}(25) 4701--4704
  \urlprefix\url{http://link.aps.org/doi/10.1103/PhysRevLett.78.4701}

\bibitem{Radon-NPA2000}
Radon T, Geissel H, M{\"u}nzenberg G, Franzke B, Kerscher T, Nolden F, Novikov
  Y, Patyk Z, Scheidenberger C, Attallah F, Beckert K, Beha T, Bosch F,
  Eickhoff H, Falch M, Fujita Y, Hausmann M, Herfurth F, Irnich H, Jung H,
  Klepper O, Kozhuharov C, Litvinov Y, L{\"o}bner K, Nickel F, Reich H, Schwab
  W, Schlitt B, Steck M, S{\"u}mmerer K, Winkler T and Wollnik H 2000 {\em
  Nuclear Physics A\/} {\bf 677} 75 -- 99
  \urlprefix\url{http://www.sciencedirect.com/science/article/pii/S0375947400003043}

\bibitem{Munzenberg-IJMS2013}
M{\"u}nzenberg G 2013 {\em International Journal of Mass Spectrometry\/} {\bf
  349-350} 9 -- 18
  \urlprefix\url{http://www.sciencedirect.com/science/article/pii/S1387380613000857}

\bibitem{Nolden-NIM2004}
Nolden F, Beckert K, Beller P, Franzke B, Peschke C and Steck M 2004 {\em
  Nuclear Instruments and Methods in Physics Research Section A\/} {\bf 532}
  329
  \urlprefix\url{http://www.sciencedirect.com/science/article/pii/S016890020401263X}

\bibitem{Steck-NIM2004}
Steck M, Beller P, Beckert K, Franzke B and Nolden F 2004 {\em Nuclear
  Instruments and Methods in Physics Research Section A: Accelerators,
  Spectrometers, Detectors and Associated Equipment\/} {\bf 532} 357--365
  \urlprefix\url{http://www.sciencedirect.com/science/article/pii/S0168900204012689}

\bibitem{Steck-PRL1996}
Steck M, Beckert K, Eickhoff H, Franzke B, Nolden F, Reich H, Schlitt B and
  Winkler T 1996 {\em Phys. Rev. Lett.\/} {\bf 77} 3803
  \urlprefix\url{http://link.aps.org/doi/10.1103/PhysRevLett.77.3803}

\bibitem{Hasse-JPB2003}
Hasse R~W 2003 {\em Journal of Physics B\/} {\bf 36} 1011
  \urlprefix\url{http://stacks.iop.org/0953-4075/36/i=5/a=320}

\bibitem{Borer-1974}
Borer J, Brabham P, Hereward H~G, H\"ubner K~H, Schnell W and Thorndah A 1974
  {\em Proc. 9th Int. Conf. on High Energy Accelerators (Stanford, CA)\/}  54

\bibitem{Schaaf-PHD}
Schaaf U 1991 Schottky-$\mathrm{D}$iagnose und
  $\mathrm{BTF}$-$\mathrm{M}$essungenan gek{\"u}hlten $\mathrm{S}$trahlen im
  $\mathrm{S}$chwerionen-$\mathrm{S}$peicherring $\mathrm{ESR}$ Phd thesis
  Johann-Wolfgang-Goethe-Universit{\"a}t Frankfurt

\bibitem{Nolden-AIP2006}
Nolden F, Beckert K, Beller P, Franzke B, Gostishchev V, Kozhuharov C, Litvinov
  Y~A, Schwinn A and Steck M 2006 {\em AIP Conference Proceedings\/} {\bf 821}
  211--220
  \urlprefix\url{http://scitation.aip.org/content/aip/proceeding/aipcp/10.1063/1.2190113}

\bibitem{Litvinov-NPA2005}
Litvinov Y, Geissel H, Radon T, Attallah F, Audi G, Beckert K, Bosch F, Falch
  M, Franzke B, Hausmann M, Hellstr{\"o}m M, Kerscher T, Klepper O, Kluge H~J,
  Kozhuharov C, L{\"o}bner K, M{\"u}nzenberg G, Nolden F, Novikov Y, Quint W,
  Patyk Z, Reich H, Scheidenberger C, Schlitt B, Steck M, S{\"u}mmerer K,
  Vermeeren L, Winkler M, Winkler T and Wollnik H 2005 {\em Nuclear Physics
  A\/} {\bf 756} 3--38
  \urlprefix\url{http://www.sciencedirect.com/science/article/pii/S037594740500357X}

\bibitem{Nolden-NIM2011}
Nolden F, H{\"u}lsmann P, Litvinov Y, Moritz P, Peschke C, Petri P, Sanjari M,
  Steck M, Weick H, Wu J, Zang Y, Zhang S and Zhao T 2011 {\em Nuclear
  Instruments and Methods in Physics Research Section A: Accelerators,
  Spectrometers, Detectors and Associated Equipment\/} {\bf 659} 69--77
  \urlprefix\url{http://www.sciencedirect.com/science/article/pii/S016890021101182X}

\bibitem{Sanjari-PS2013}
Sanjari M~S, H{\"u}llsmann P, Nolden F, Schempp A, Wu J~X, Atanasov D, Bosch F,
  Kozhuharov C, Litvinov Y~A, Moritz P, Peschke C, Petri P, Shubina D, Steck M,
  Weick H, Winckler N, Zang Y~D and Zhao T~C 2013 {\em Physica Scripta\/} {\bf
  T156} 014088
  \urlprefix\url{http://stacks.iop.org/1402-4896/2013/i=T156/a=014088}

\bibitem{Litvinov-HI2001}
Litvinov Y, Attallah F, Beckert K, Bosch F, Falch M, Franzke B, Geissel H,
  Hausmann M, Kerscher T, Klepper O, Kluge H~J, Kozhuharov C, L{\"o}bner K,
  M{\"u}nzenberg G, Nolden F, Novikov Y, Patyk Z, Quint W, Radon T,
  Scheidenberger C, Steck M, Vermeeren L and Wollnik H 2001 {\em Hyperfine
  Interactions\/} {\bf 132} 281--287
  \urlprefix\url{http://dx.doi.org/10.1023/A\%3A1011907602706}

\bibitem{Litvinov-RPP2011}
Litvinov Y~A and Bosch F 2011 {\em Reports on Progress in Physics\/} {\bf 74}
  016301 \urlprefix\url{http://stacks.iop.org/0034-4885/74/i=1/a=016301}

\bibitem{Litvinov-NPA2004}
Litvinov Y, Geissel H, Novikov Y, Patyk Z, Radon T, Scheidenberger C, Attallah
  F, Beckert K, Bosch F, Falch M, Franzke B, Hausmann M, Kerscher T, Klepper O,
  Kluge H~J, Kozhuharov C, L{\"o}bner K, M{\"u}nzenberg G, Nolden F, Steck M
  and Wollnik H 2004 {\em Nuclear Physics A\/} {\bf 734} 473--476
  \urlprefix\url{http://www.sciencedirect.com/science/article/pii/S0375947404001083}

\bibitem{Geissel-NPA2004}
Geissel H, Litvinov Y, Attallah F, Beckert K, Beller P, Bosch F, Boutin D,
  Faestermann T, Falch M, Franzke B, Hausmann M, Hellstr{\"o}m M, Kaza E,
  Kerscher T, Klepper O, Kluge H~J, Kozhuharov C, Kratz K~L, Litvinov S,
  L{\"o}bner K, Maier L, Mato{\v s} M, M{\"u}nzenberg G, Nolden F, Novikov Y,
  Ohtsubo T, Ostrowski A, Patyk Z, Pfeiffer B, Portillo M, Radon T,
  Scheidenberger C, Shishkin V, Stadlmann J, Steck M, Viera D, Weick H, Winkler
  M, Wollnik H and Yamaguchi T 2004 {\em Nuclear Physics A\/} {\bf 746}
  150c--155c
  \urlprefix\url{http://www.sciencedirect.com/science/article/pii/S0375947404009224}

\bibitem{Ohtsubo-PRL2005}
Ohtsubo T, Bosch F, Geissel H, Maier L, Scheidenberger C, Attallah F, Beckert
  K, Beller P, Boutin D, Faestermann T, Franczak B, Franzke B, Hausmann M,
  Hellstr\"om M, Kaza E, Kienle P, Klepper O, Kluge H~J, Kozhuharov C, Litvinov
  Y~A, Matos M, M\"unzenberg G, Nolden F, Novikov Y~N, Portillo M, Radon T,
  Stadlmann J, Steck M, St\"ohlker T, S\"ummerer K, Takahashi K, Weick H,
  Winkler M and Yamaguchi T 2005 {\em Phys. Rev. Lett.\/} {\bf 95}(5) 052501
  \urlprefix\url{http://link.aps.org/doi/10.1103/PhysRevLett.95.052501}

\bibitem{Atanasov-JPB2015}
Atanasov D, Blaum K, Bosch F, Brandau C, B{\"u}hler P, Chen X, Dillmann I,
  Faestermann T, Gao B, Geissel H, Gernh{\"a}user R, Hagmann S, Izumikawa T,
  Hillenbrand P~M, Kozhuharov C, Kurcewicz J, Litvinov S~A, Litvinov Y~A, Ma X,
  M{\"u}nzenberg G, Najafi M~A, Nolden F, Ohtsubo T, Ozawa A, Ozturk F~C, Patyk
  Z, Reed M, Reifarth R, Sanjari M~S, Schneider D, Steck M, St{\"o}hlker T, Sun
  B, Suzaki F, Suzuki T, Trageser C, Tu X, Uesaka T, Walker P, Wang M, Weick H,
  Winckler N, Woods P, Xu H, Yamaguchi T, Yan X, Zhang Y and {the FRS-ESR,
  ILIMA, SPARC and TBWD Collaborations} 2015 {\em Journal of Physics B: Atomic,
  Molecular and Optical Physics\/} {\bf 48} 144024
  \urlprefix\url{http://stacks.iop.org/0953-4075/48/i=14/a=144024}

\bibitem{Hausmann-HI2001}
Hausmann M, Stadlmann J, Attallah F, Beckert K, Beller P, Bosch F, Eickhoff H,
  Falch M, Franczak B, Franzke B, Geissel H, Kerscher T, Klepper O, Kluge H~J,
  Kozhuharov C, Litvinov Y~A, L{\"o}bner K~E~G, M{\"u}nzenberg G, Nankov N,
  Nolden F, Novikov Y~N, Ohtsubo T, Radon T, Schatz H, Scheidenberger C, Steck
  M, Sun Z, Weick H and Wollnik H 2001 {\em Hyperfine Interactions\/} {\bf 132}
  289--295 \urlprefix\url{http://dx.doi.org/10.1023/A\%3A1011911720453}

\bibitem{Hausmann-NIM2000}
Hausmann M, Attallah F, Beckert K, Bosch F, Dolinskiy A, Eickhoff H, Falch M,
  Franczak B, Franzke B, Geissel H, Kerscher T, Klepper O, Kluge H~J,
  Kozhuharov C, L{\"o}bner K, M{\"u}nzenberg G, Nolden F, Novikov Y, Radon T,
  Schatz H, Scheidenberger C, Stadlmann J, Steck M, Winkler T and Wollnik H
  2000 {\em Nuclear Instruments and Methods in Physics Research Section A:
  Accelerators, Spectrometers, Detectors and Associated Equipment\/} {\bf 446}
  569--580
  \urlprefix\url{http://www.sciencedirect.com/science/article/pii/S0168900299011924}

\bibitem{Bo-NIM2010}
Mei B, Tu X, Wang M, Xu H, Mao R, Hu Z, Ma X, Yuan Y, Zhang X, Geng P, Shuai P,
  Zang Y, Tang S, Ma P, Lu W, Yan X, Xia J, Xiao G, Guo Z, Zhang H and Yue K
  2010 {\em Nuclear Instruments and Methods in Physics Research Section A:
  Accelerators, Spectrometers, Detectors and Associated Equipment\/} {\bf 624}
  109--113
  \urlprefix\url{http://www.sciencedirect.com/science/article/pii/S0168900210019145}

\bibitem{Abe-PS2015}
Abe Y, Yamaguchi Y, Wakasugi M, Uesaka T, Ozawa A, Suzaki F, Nagae D, Miura H,
  Yamaguchi T and Yano Y 2015 {\em Physica Scripta\/} {\bf T166} 014047
  \urlprefix\url{http://stacks.iop.org/1402-4896/2015/i=T166/a=014047}

\bibitem{Trotscher-NIM1992}
Tr{\"o}tscher J, Balog K, Eickhoff H, Franczak B, Franzke B, Fujita Y, Geissel
  H, Klein C, Knollmann J, Kraft A, L{\"o}bner K, Magel A, M{\"u}nzenberg G,
  Przewloka A, Rosenauer D, Sch{\"a}fer H, Sendor M, Vieira D, Vogel B,
  Winkelmann T and Wollnik H 1992 {\em Nuclear Instruments and Methods in
  Physics Research Section B: Beam Interactions with Materials and Atoms\/}
  {\bf 70} 455--458
  \urlprefix\url{http://www.sciencedirect.com/science/article/pii/0168583X9295965T}

\bibitem{Nagae-NIM2013}
Nagae D, Abe Y, Okada S, Ozawa A, Yamaguchi T, Suzuki H, Moriguchi T, Ishibashi
  Y, Fukuoka S, Nishikiori R, Niwa T, Suzuki T, Suzaki F, Sato K, Furuki H,
  Ichihashi N, Miyazawa S, Yamaguchi Y, Uesaka T and Wakasugi M 2013 {\em
  Nuclear Instruments and Methods in Physics Research Section B-Beam
  Interactions With Materials and Atoms\/} {\bf 317} 640--643
  \urlprefix\url{http://www.sciencedirect.com/science/article/pii/S0168583X13009968}

\bibitem{Zhang-NIM2014b}
Zhang W, Tu X, Wang M, Zhang Y, Xu H, Litvinov Y~A, Blaum K, Chen X, Hu Z,
  Huang W, Ma X, Mao R, Mei B, Shuai P, Sun B, Yamaguchi T, Xia J, Xiao G, Xu
  X, Yan X, Yang J, Yuan Y, Zhou X, Zhao H and Zhao T 2014 {\em Nuclear
  Instruments and Methods in Physics Research Section A: Accelerators,
  Spectrometers, Detectors and Associated Equipment\/} {\bf 755} 38--43
  \urlprefix\url{http://www.sciencedirect.com/science/article/pii/S0168900214004276}

\bibitem{Shuai-PLB2014}
Shuai P, Xu H, Tu X, Zhang Y, Sun B, Wang M, Litvinov Y, Blaum K, Zhou X, He J,
  Sun Y, Kaneko K, Yuan Y, Xia J, Yang J, Audi G, Yan X, Chen X, Jia G, Hu Z,
  Ma X, Mao R, Mei B, Sun Z, Wang S, Xiao G, Xu X, Yamaguchi T, Yamaguchi Y,
  Zang Y, Zhao H, Zhao T, Zhang W and Zhan W 2014 {\em Physics Letters B\/}
  {\bf 735} 327--331
  \urlprefix\url{http://www.sciencedirect.com/science/article/pii/S0370269314004511}

\bibitem{Sun-NPA2008}
Sun B, Kn{\"o}bel R, Litvinov Y, Geissel H, Meng J, Beckert K, Bosch F, Boutin
  D, Brandau C, Chen L, Cullen I, Dimopoulou C, Fabian B, Hausmann M,
  Kozhuharov C, Litvinov S, Mazzocco M, Montes F, M{\"u}nzenberg G, Musumarra
  A, Nakajima S, Nociforo C, Nolden F, Ohtsubo T, Ozawa A, Patyk Z, Plass W,
  Scheidenberger C, Steck M, Suzuki T, Walker P, Weick H, Winckler N, Winkler M
  and Yamaguchi T 2008 {\em Nuclear Physics A\/} {\bf 812} 1--12
  \urlprefix\url{http://www.sciencedirect.com/science/article/pii/S0375947408006751}

\bibitem{Tu-NIM2011}
Tu X, Wang M, Litvinov Y, Zhang Y, Xu H, Sun Z, Audi G, Blaum K, Du C, Huang W,
  Hu Z, Geng P, Jin S, Liu L, Liu Y, Mei B, Mao R, Ma X, Suzuki H, Shuai P, Sun
  Y, Tang S, Wang J, Wang S, Xiao G, Xu X, Xia J, Yang J, Ye R, Yamaguchi T,
  Yan X, Yuan Y, Yamaguchi Y, Zang Y, Zhao H, Zhao T, Zhang X, Zhou X and Zhan
  W 2011 {\em Nuclear Instruments and Methods in Physics Research Section A:
  Accelerators, Spectrometers, Detectors and Associated Equipment\/} {\bf 654}
  213--218
  \urlprefix\url{http://www.sciencedirect.com/science/article/pii/S0168900211014471}

\bibitem{Yamaguchi-IJMS2013}
Yamaguchi T, Yamaguchi Y and Ozawa A 2013 {\em International Journal of Mass
  Spectrometry\/} {\bf 349-€"350} 240--246
  \urlprefix\url{http://www.sciencedirect.com/science/article/pii/S1387380613001541}

\bibitem{Yamaguchi-NIM2013}
Yamaguchi Y, Wakasugi M, Uesaka T, Ozawa A, Abe Y, Fujinawa T, Kase M, Komiyama
  M, Kubo T, Kumagai K, Maie T, Nagae D, Ohnishi J, Suzaki F, Tokuchi A,
  Watanabe Y, Yoshida K, Yamada K, Yamaguchi T, Yamasawa H, Yanagisawa Y,
  Zenihiro J and Yano Y 2013 {\em Nuclear Instruments and Methods in Physics
  Research Section B-Beam Interactions With Materials and Atoms\/} {\bf 317}
  629--635
  \urlprefix\url{http://www.sciencedirect.com/science/article/pii/S0168583X13006228}

\bibitem{Yamaguchi-PS2015}
Yamaguchi Y, Miura H, Wakasugi M, Abe Y, Ozawa A, Suzaki F, Tokuchi A, Uesaka
  T, Yamaguchi T and Yano Y 2015 {\em Physica Scripta\/} {\bf T166} 014056
  \urlprefix\url{http://stacks.iop.org/1402-4896/2015/i=T166/a=014056}

\bibitem{Zhang-PRL2012}
Zhang Y~H, Xu H~S, Litvinov Y~A, Tu X~L, Yan X~L, Typel S, Blaum K, Wang M,
  Zhou X~H, Sun Y, Brown B~A, Yuan Y~J, Xia J~W, Yang J~C, Audi G, Chen X~C,
  Jia G~B, Hu Z~G, Ma X~W, Mao R~S, Mei B, Shuai P, Sun Z~Y, Wang S~T, Xiao
  G~Q, Xu X, Yamaguchi T, Yamaguchi Y, Zang Y~D, Zhao H~W, Zhao T~C, Zhang W
  and Zhan W~L 2012 {\em Phys. Rev. Lett.\/} {\bf 109}(10) 102501
  \urlprefix\url{http://link.aps.org/doi/10.1103/PhysRevLett.109.102501}

\bibitem{Litvinov-APP2010}
Litvinov Y~A, Geissel H, Knoebel R, Sun B and Xu H 2010 {\em Acta Physica
  Polonica B\/} {\bf 41} 511--523
  \urlprefix\url{http://www.actaphys.uj.edu.pl/_cur/store/vol41/pdf/v41p0511.pdf}

\bibitem{Geissel-JPG2005}
Geissel H and Litvinov Y~A 2005 {\em Journal of Physics G: Nuclear and Particle
  Physics\/} {\bf 31} S1779
  \urlprefix\url{http://stacks.iop.org/0954-3899/31/i=10/a=072}

\bibitem{Geissel-HI2006}
Geissel H, Kn{\"o}bel R~K, Litvinov Y~A, Sun B, Beckert K, Beller P, Bosch F,
  Boutin D, Brandau C, Chen L, Fabian B, Hausmann M, Kozhuharov C, Kurcewicz J,
  Litvinov S~A, Mazzocco M, Montes F, M{\"u}nzenberg G, Musumarra A, Nociforo
  C, Nolden F, Plass W~R, Scheidenberger C, Steck M, Weick H and Winkler M 2006
  {\em Hyperfine Interactions\/} {\bf 173} 49--54
  \urlprefix\url{http://dx.doi.org/10.1007/s10751-007-9541-4}

\bibitem{Geissel-EPJST2007}
Geissel H, Litvinov Y~A, Beckert K, Beller P, Bosch F, Boutin D, Brandau C,
  Chen L, Hausmann M, Klepper O, Kn{\"o}bel R, Kozhuharov C, Kurcewicz J,
  Litvinov S~A, Mazzocco M, M{\"u}nzenberg G, Nociforo C, Nolden F, Patyk Z,
  Pf{\"u}tzner M, Pla{\ss}Ÿ W, Scheidenberger C, Steck M, Sun B, Takahashi K,
  Weick H, Winckler N and Winkler M 2007 {\em The European Physical Journal
  Special Topics\/} {\bf 150} 109--115
  \urlprefix\url{http://dx.doi.org/10.1140/epjst/e2007-00280-x}

\bibitem{Zhang-NIM2014a}
Zhang W, Tu X, Wang M, Zhang Y, Xu H, Litvinov Y~A, Blaum K, Chen R, Chen X, Fu
  C, Ge Z, Gao B, Hu Z, Huang W, Litvinov S, Liu D, Ma X, Mao R, Mei B, Shuai
  P, Sun B, Xia J, Xiao G, Xing Y, Xu X, Yamaguchi T, Yan X, Yang J, Yuan Y,
  Zeng Q, Zhang X, Zhao H, Zhao T and Zhou X 2014 {\em Nuclear Instruments and
  Methods in Physics Research Section A: Accelerators, Spectrometers, Detectors
  and Associated Equipment\/} {\bf 756} 1--5
  \urlprefix\url{http://www.sciencedirect.com/science/article/pii/S0168900214004562}

\bibitem{Xing-PS2015}
Xing Y~M, Wang M, Zhang Y~H, Shuai P, Xu X, Chen R~J, Yan X~L, Tu X~L, Zhang W,
  Fu C~Y, Xu H~S, Litvinov Y~A, Blaum K, Chen X~C, Ge Z, Gao B~S, Huang W~J,
  Litvinov S~A, Liu D~W, Ma X~W, Mao R~S, Xiao G~Q, Yang J~C, Yuan Y~J, Zeng Q
  and Zhou X~H 2015 {\em Physica Scripta\/} {\bf T166} 014010
  \urlprefix\url{http://stacks.iop.org/1402-4896/2015/i=T166/a=014010}

\bibitem{Geissel-AIP2006}
Geissel H, Litvinov Y~A, Pfeiffer B, Attallah F, Audi G, Beckert K, Beller P,
  Bosch F, Boutin D, B{\"u}rvenich T~J, Chen L, Faestermann T, Falch M, Franzke
  B, Hausmann M, Kaza E, Kerscher T, Kienle P, Klepper O, Kn{\"o}bel R,
  Kozhuharov C, Kratz K, Litvinov S~A, L{\"o}bner K~E~G, Maier L, Mato{\v s} M,
  Montes F, M{\"u}nzenberg G, Nociforo C, Nolden F, Novikov Y~N, Ohtsubo T,
  Ostrowski A, Patyk Z, Pla{\ss}Ÿ W, Portillo M, Radon T, Schatz H,
  Scheidenberger C, Stadlmann J, Steck M, Sun B, Takahashi K, Vorobjev G, Weick
  H, Winkler M, Wollnik H and Yamaguchi T 2006 {\em AIP Conference
  Proceedings\/} {\bf 831} 108--113
  \urlprefix\url{http://scitation.aip.org/content/aip/proceeding/aipcp/10.1063/1.2200908}

\bibitem{Sun-GSI2011}
Sun B, Kn{\"o}bel R, Kuzminchuk N, Weick H, Winckler N, Ayet S, Blaum K, Bosch
  F, Cakirli R, Dillmann I, Dimopoulou C, Estrad{\'e} A, Farinon F, Geissel H,
  lsmann P~H, Jesch C, Kozhuharov C, Kurcewicz J, Litvinov S, Litvinov Y, Mukha
  I, Nociforo C, Nolden F, Petri P, Pietri S, Pla{\ss} W, Prochazka A, Reed M,
  Riese B, Sanjari M, Shubina D, Scheidenberger C, Steck M, hlker T~S, Tu X,
  Walker P, Winkler M and Winfield J 2011 {\em GSI Scientific Report 2011\/}
  {\bf 2012-1} PHN--NUSTAR--FRS--21

\bibitem{Zang-CPC2011}
Zang Y~D, Wu J~X, Zhao T~C, Zhang S~H, Mao R~S, Xu H~S, Sun Z~Y, Ma X~W, Tu
  X~L, Xiao G~Q, Nolden F, H{\"u}lsmann P, Litvinov Y~A, Peschke C, Petri P,
  Sanjari M~S and Steck M 2011 {\em Chinese Physics C\/} {\bf 35} 1124--1129
  \urlprefix\url{http://stacks.iop.org/1674-1137/35/i=12/a=008}

\bibitem{Suzaki-NIM2013}
Suzaki F, Zenihiro J, Yamaguchi T, Ozawa A, Uesaka T, Wakasugi M, Yamada K,
  Yamaguchi Y and {the Rare-RI Ring Collaboration} 2013 {\em Nuclear
  Instruments and Methods in Physics Research Section B-Beam Interactions With
  Materials and Atoms\/} {\bf 317} 636--639
  \urlprefix\url{http://www.sciencedirect.com/science/article/pii/S0168583X13008811}

\bibitem{Suzaki-JPSJ2015}
Suzaki F, Zenihiro J, Abe Y, Ozawa A, Suzuki T, Uesaka T, Wakasugi M, Yamada K,
  Yamaguchi T and Yamaguchi Y 2015 {\em JPS Conference Proceedings\/} {\bf 6}
  030119 \urlprefix\url{http://dx.doi.org/10.7566/JPSCP.6.030119}

\bibitem{Suzaki-PS2015}
Suzaki F, Abe Y, Ozawa A, Suzuki T, Uesaka T, Wakasugi M, Yamada K, Yamaguchi
  T, Yamaguchi Y, Zenihiro J and {the Rare-RI Ring collaboration} 2015 {\em
  Physica Scripta\/} {\bf T166} 014059
  \urlprefix\url{http://stacks.iop.org/1402-4896/2015/i=T166/a=014059}

\bibitem{Chen-NPA2012}
Chen L, Pla{\ss} W, Geissel H, Kn{\"o}bel R, Kozhuharov C, Litvinov Y, Patyk Z,
  Scheidenberger C, Siegien-Iwaniuk K, Sun B, Weick H, Beckert K, Beller P,
  Bosch F, Boutin D, Caceres L, Carroll J, Cullen D, Cullen I, Franzke B, Gerl
  J, Gorska M, Jones G, Kishada A, Kurcewicz J, Litvinov S, Liu Z, Mandal S,
  Montes F, M{\"u}nzenberg G, Nolden F, Ohtsubo T, Podolyak Z, Propri R, Rigby
  S, Saito N, Saito T, Shindo M, Steck M, Walker P, Williams S, Winkler M,
  Wollersheim H~J and Yamaguchi T 2012 {\em Nuclear Physics A\/} {\bf 882}
  71--89
  \urlprefix\url{http://www.sciencedirect.com/science/article/pii/S0375947412001005}

\bibitem{Kurcewicz-APP2010}
Kurcewicz J, Bosch F, Geissel H, Litvinov Y, Winckler N, Beckert K, Beller P,
  Boutin D, Chen L, Dimopoulou C, Essel H, Fabian B, Faestermann T, Fragner A,
  Franzke B, Haettner E, Hausmann M, Hess S, Kienle P, Kn{\"o}bel R, Kozhuharov
  C, Litvinov S, Maier L, Mazzocco M, Montes F, Musumarra A, Nociforo C, Nolden
  F, Patyk Z, Pla{\ss} W, Prochazka A, Reda R, Reuschl R, Scheidenberger C,
  Steck M, St{\"o}hlker T, Sun B, KTakahashi, Torilov S, Trassinelli M, Weick H
  and Winkler M 2010 {\em Acta Physica Polonica B\/} {\bf 41} 525--536
  \urlprefix\url{http://www.actaphys.uj.edu.pl/_cur/store/vol41/pdf/v41p0525.pdf}

\bibitem{Reed-PRC2012}
Reed M~W, Walker P~M, Cullen I~J, Litvinov Y~A, Shubina D, Dracoulis G~D, Blaum
  K, Bosch F, Brandau C, Carroll J~J, Cullen D~M, Deo A~Y, Detwiler B,
  Dimopoulou C, Dong G~X, Farinon F, Geissel H, Haettner E, Heil M, Kempley
  R~S, Kn\"obel R, Kozhuharov C, Kurcewicz J, Kuzminchuk N, Litvinov S, Liu Z,
  Mao R, Nociforo C, Nolden F, Plass W~R, Podolyak Z, Prochazka A,
  Scheidenberger C, Steck M, St\"ohlker T, Sun B, Swan T~P~D, Trees G, Weick H,
  Winckler N, Winkler M, Woods P~J, Xu F~R and Yamaguchi T 2012 {\em Physical
  Review C\/} {\bf 86} 054321
  \urlprefix\url{http://link.aps.org/doi/10.1103/PhysRevC.86.054321}

\bibitem{Shubina-PRC2013}
Shubina D, Cakirli R~B, Litvinov Y~A, Blaum K, Brandau C, Bosch F, Carroll J~J,
  Casten R~F, Cullen D~M, Cullen I~J, Deo A~Y, Detwiler B, Dimopoulou C,
  Farinon F, Geissel H, Haettner E, Heil M, Kempley R~S, Kozhuharov C, Kn\"obel
  R, Kurcewicz J, Kuzminchuk N, Litvinov S~A, Liu Z, Mao R, Nociforo C, Nolden
  F, Patyk Z, Plass W~R, Prochazka A, Reed M~W, Sanjari M~S, Scheidenberger C,
  Steck M, St\"ohlker T, Sun B, Swan T~P~D, Trees G, Walker P~M, Weick H,
  Winckler N, Winkler M, Woods P~J, Yamaguchi T and Zhou C 2013 {\em Phys. Rev.
  C\/} {\bf 88}(2) 024310
  \urlprefix\url{http://link.aps.org/doi/10.1103/PhysRevC.88.024310}

\bibitem{Litvinov-HI2006}
Litvinov Y, Geissel H, Beckert K, Beller P, Bosch F, Boutin D, Brandau C, Chen
  L, Kn{\"o}bel R, Kozhuharov C, Kurcewicz J, Litvinov S, Mazzocco M,
  M{\"u}nzenberg G, Nociforo C, Nolden F, Pla{\ss}Ÿ W, Scheidenberger C, Steck
  M, Sun B, Weick H and Winkler M 2006 {\em Hyperfine Interactions\/} {\bf 173}
  55--60 \urlprefix\url{http://dx.doi.org/10.1007/s10751-007-9542-3}

\bibitem{Stadlmann-PLB2004}
Stadlmann J, Hausmann M, Attallah F, Beckert K, Beller P, Bosch F, Eickhoff H,
  Falch M, Franczak B, Franzke B, Geissel H, Kerscher T, Klepper O, Kluge H~J,
  Kozhuharov C, Litvinov Y, L{\"o}bner K, Mato{\v s} M, M{\"u}nzenberg G,
  Nankov N, Nolden F, Novikov Y, Ohtsubo T, Radon T, Schatz H, Scheidenberger
  C, Steck M, Weick H and Wollnik H 2004 {\em Physics Letters B\/} {\bf 586}
  27--33
  \urlprefix\url{http://www.sciencedirect.com/science/article/pii/S0370269304002989}

\bibitem{Tu-PRL2011}
Tu X~L, Xu H~S, Wang M, Zhang Y~H, Litvinov Y~A, Sun Y, Schatz H, Zhou X~H,
  Yuan Y~J, Xia J~W, Audi G, Blaum K, Du C~M, Geng P, Hu Z~G, Huang W~X, Jin
  S~L, Liu L~X, Liu Y, Ma X, Mao R~S, Mei B, Shuai P, Sun Z~Y, Suzuki H, Tang
  S~W, Wang J~S, Wang S~T, Xiao G~Q, Xu X, Yamaguchi T, Yamaguchi Y, Yan X~L,
  Yang J~C, Ye R~P, Zang Y~D, Zhao H~W, Zhao T~C, Zhang X~Y and Zhan W~L 2011
  {\em Phys. Rev. Lett.\/} {\bf 106}(11) 112501
  \urlprefix\url{http://link.aps.org/doi/10.1103/PhysRevLett.106.112501}

\bibitem{Zhang-JPSJ2015}
Zhang Y, Xu H and Litvinov Y~A 2015 {\em JPS Conference Proceedings\/} {\bf 6}
  010019 \urlprefix\url{http://dx.doi.org/10.7566/JPSCP.6.010019}

\bibitem{Yan-APJL2013}
Yan X~L, Xu H~S, Litvinov Y~A, Zhang Y~H, Schatz H, Tu X~L, Blaum K, Zhou X~H,
  Sun B~H, He J~J, Sun Y, Wang M, Yuan Y~J, Xia J~W, Yang J~C, Audi G, Jia G~B,
  Hu Z~G, Ma X~W, Mao R~S, Mei B, Shuai P, Sun Z~Y, Wang S~T, Xiao G~Q, Xu X,
  Yamaguchi T, Yamaguchi Y, Zang Y~D, Zhao H~W, Zhao T~C, Zhang W and Zhan W~L
  2013 {\em The Astrophysical Journal Letters\/} {\bf 766} L8
  \urlprefix\url{http://stacks.iop.org/2041-8205/766/i=1/a=L8}

\bibitem{Atanasov-PRL2015}
Atanasov D, Ascher P, Blaum K, Cakirli R~B, Cocolios T~E, George S, Goriely S,
  Herfurth F, Janka H~T, Just O, Kowalska M, Kreim S, Kisler D, Litvinov Y~A,
  Lunney D, Manea V, Neidherr D, Rosenbusch M, Schweikhard L, Welker A,
  Wienholtz F, Wolf R~N and Zuber K 2015 {\em Phys. Rev. Lett.\/} {\bf 115}(23)
  232501 \urlprefix\url{http://link.aps.org/doi/10.1103/PhysRevLett.115.232501}

\bibitem{Dillmann-PPNP2011}
Dillmann I and Litvinov Y~A 2011 {\em Progress in Particle and Nuclear
  Physics\/} {\bf 66} 358--362
  \urlprefix\url{http://www.sciencedirect.com/science/article/pii/S0146641011000354}

\bibitem{Caballero-Folch-arXiv2015}
Caballero-Folch R, Domingo-Pardo C, Agramunt J, Algora A, Ameil F, Arcones A,
  Ayyad Y, Benlliure J, Borzov I~N, Bowry M, Calvi{\~n}o F, Cano-Ott D,
  Cort{\`e}s G, Davinson T, Dillmann I, Estrade A, Evdokimov A, Faestermann T,
  Farinon F, Galaviz D, Garc{\'\i}a A~R, Geissel H, Gelletly W, Gernh{\"a}user
  R, G{\'o}mez-Hornillos M~B, Guerrero C, Heil M, Hinke C, Kn{\"o}bel R,
  Kojouharov I, Kurcewicz J, Kurz N, Litvinov Y, Maier L, Marganiec J, Marketin
  T, Marta M, Mart{\'\i}nez T, Mart{\'\i}nez-Pinedo G, Montes F, Mukha I,
  Napoli D~R, Nociforo C, Paradela C, Pietri S, Podoly{\'a}k Z, Prochazka A,
  Rice S, Riego A, Rubio B, Schaffner H, Scheidenberger C, Smith K, Sokol E,
  Steiger K, Sun B, Ta{\'\i}n J~L, Takechi M, Testov D, Weick H, Wilson E,
  Winfield J~S, Wood R, Woods P and Yeremin A 2015 {\em arXiv.org\/}
  arXiv:1511.01296 (\textit{Preprint} \eprint{1511.01296})
  \urlprefix\url{http://arxiv.org/abs/1511.01296}

\bibitem{Moller-ADNDT1995}
Moller P, Nix J, Myers W and Swiatecki W 1995 {\em Atomic Data and Nuclear Data
  Tables\/} {\bf 59} 185 -- 381
  \urlprefix\url{http://www.sciencedirect.com/science/article/pii/S0092640X85710029}

\bibitem{Goriely-PRC2010}
Goriely S, Chamel N and Pearson J~M 2010 {\em Phys. Rev. C\/} {\bf 82}(3)
  035804 \urlprefix\url{http://link.aps.org/doi/10.1103/PhysRevC.82.035804}

\bibitem{Goriely-PRL2009}
Goriely S, Hilaire S, Girod M and P\'eru S 2009 {\em Phys. Rev. Lett.\/} {\bf
  102}(24) 242501
  \urlprefix\url{http://link.aps.org/doi/10.1103/PhysRevLett.102.242501}

\bibitem{Wang-PLB2014}
Wang N, Liu M, Wu X and Meng J 2014 {\em Physics Letters B\/} {\bf 734} 215 --
  219
  \urlprefix\url{http://www.sciencedirect.com/science/article/pii/S037026931400358X}

\bibitem{Maripuu-ADNDT1976}
Maripuu S and Way K 1976 {\em Atomic Data and Nuclear Data Tables\/} {\bf 17} i
  -- ii
  \urlprefix\url{http://www.sciencedirect.com/science/article/pii/0092640X76900292}

\bibitem{Heyde-1999}
Heyde K 1999 {\em Basic Ideas and Concepts in Nuclear Physics\/} (Inst. Phys.
  Publ.) ISBN 9780750309806

\bibitem{Duflo-PRC1995}
Duflo J and Zuker A 1995 {\em Phys. Rev. C\/} {\bf 52}(1) R23--R27
  \urlprefix\url{http://link.aps.org/doi/10.1103/PhysRevC.52.R23}

\bibitem{Nayak-ADNDT2012}
Nayak R and Satpathy L 2012 {\em Atomic Data and Nuclear Data Tables\/} {\bf
  98} 616 -- 719
  \urlprefix\url{http://www.sciencedirect.com/science/article/pii/S0092640X12000137}

\bibitem{Litvinov-NPA2007}
Litvinov Y, Geissel H, Beckert K, Beller P, Bosch F, Boutin D, Brandau C, Chen
  L, Cullen I, Dimopoulou C, Fabian B, Hausmann M, Klepper O, Kn{\"o}bel R,
  Kozhuharov C, Kurcewicz J, Litvinov S, Liu Z, Mazzocco M, Montes F,
  M{\"u}nzenberg G, Musumarra A, Nakajima S, Nociforo C, Nolden F, Ohtsubo T,
  Ozawa A, Patyk Z, Plass W, Scheidenberger C, Steck M, Sun B, Suzuki T, Walker
  P, Weick H, Winckler N, Winkler M and Yamaguchi T 2007 {\em Nuclear Physics
  A\/} {\bf 787} 315--320
  \urlprefix\url{http://www.sciencedirect.com/science/article/pii/S0375947406010293}

\bibitem{Litvinov-PRL2005}
Litvinov Y~A, B\"urvenich T~J, Geissel H, Novikov Y~N, Patyk Z, Scheidenberger
  C, Attallah F, Audi G, Beckert K, Bosch F, Falch M, Franzke B, Hausmann M,
  Kerscher T, Klepper O, Kluge H~J, Kozhuharov C, L\"obner K~E~G, Madland D~G,
  Maruhn J~A, M\"unzenberg G, Nolden F, Radon T, Steck M, Typel S and Wollnik H
  2005 {\em Phys. Rev. Lett.\/} {\bf 95}(4) 042501
  \urlprefix\url{http://link.aps.org/doi/10.1103/PhysRevLett.95.042501}

\bibitem{Madland-NPA1988}
Madland D~G and Nix J 1988 {\em Nuclear Physics A\/} {\bf 476} 1--38
  \urlprefix\url{http://www.sciencedirect.com/science/article/pii/0375947488903703}

\bibitem{Litvinov-NPA2008}
Litvinov Y~A 2008 {\em Nuclear Physics A\/} {\bf 805} 260c -- 269c
  \urlprefix\url{http://www.sciencedirect.com/science/article/pii/S0375947408003527}

\bibitem{Alkhazov-ZPA1983}
Alkhazov G, Mezilev K, Novikov Y, Nurmukhamedov A, Ganbaatar N, Gromov K,
  Kalinnikov V, Potempa A and Tarkanyi F 1983 {\em Zeitschrift f{\"u}r Physik A
  Atoms and Nuclei\/} {\bf 311} 245--246
  \urlprefix\url{http://dx.doi.org/10.1007/BF01415110}

\bibitem{Bohm-PRC2014}
B\"ohm C, Borgmann C, Audi G, Beck D, Blaum K, Breitenfeldt M, Cakirli R~B,
  Cocolios T~E, Eliseev S, George S, Herfurth F, Herlert A, Kowalska M, Kreim
  S, Lunney D, Manea V, Minaya~Ramirez E, Naimi S, Neidherr D, Rosenbusch M,
  Schweikhard L, Stanja J, Wang M, Wolf R~N and Zuber K 2014 {\em Phys. Rev.
  C\/} {\bf 90}(4) 044307
  \urlprefix\url{http://link.aps.org/doi/10.1103/PhysRevC.90.044307}

\bibitem{Kreim-PRC2014}
Kreim S, Beck D, Blaum K, Borgmann C, Breitenfeldt M, Cocolios T~E, Gottberg A,
  Herfurth F, Kowalska M, Litvinov Y~A, Lunney D, Manea V, Mendonca T~M, Naimi
  S, Neidherr D, Rosenbusch M, Schweikhard L, Stora T, Wienholtz F, Wolf R~N
  and Zuber K 2014 {\em Phys. Rev. C\/} {\bf 90}(2) 024301
  \urlprefix\url{http://link.aps.org/doi/10.1103/PhysRevC.90.024301}

\bibitem{Chen-PRL2009}
Chen L, Litvinov Y~A, Pla\ss{} W~R, Beckert K, Beller P, Bosch F, Boutin D,
  Caceres L, Cakirli R~B, Carroll J~J, Casten R~F, Chakrawarthy R~S, Cullen
  D~M, Cullen I~J, Franzke B, Geissel H, Gerl J, G\'orska M, Jones G~A, Kishada
  A, Kn\"obel R, Kozhuharov C, Litvinov S~A, Liu Z, Mandal S, Montes F,
  M\"unzenberg G, Nolden F, Ohtsubo T, Patyk Z, Podoly\'ak Z, Propri R, Rigby
  S, Saito N, Saito T, Scheidenberger C, Shindo M, Steck M, Ugorowski P, Walker
  P~M, Williams S, Weick H, Winkler M, Wollersheim H~J and Yamaguchi T 2009
  {\em Phys. Rev. Lett.\/} {\bf 102}(12) 122503
  \urlprefix\url{http://link.aps.org/doi/10.1103/PhysRevLett.102.122503}

\bibitem{Zhang-PLB1989}
Zhang J~Y, Casten R~F and Brenner D~S 1989 {\em Physics Letters B\/} {\bf 227}
  1
  \urlprefix\url{http://www.sciencedirect.com/science/article/pii/0370269389912732}

\bibitem{Cakirli-PRL2006}
Cakirli R~B and Casten R~F 2006 {\em Phys. Rev. Lett.\/} {\bf 96}(13) 132501
  \urlprefix\url{http://link.aps.org/doi/10.1103/PhysRevLett.96.132501}

\bibitem{Bosch-IJMS2006}
Bosch F, Geissel H, Litvinov Y, Beckert K, Franzke B, Hausmann M, Kerscher T,
  Klepper O, Kozhuharov C, L{\"o}bner K, M{\"u}nzenberg G, Nolden F, Novikov Y,
  Patyk Z, Radon T, Scheidenberger C, Steck M and Wollnik H 2006 {\em
  International Journal of Mass Spectrometry\/} {\bf 251} 212--219
  \urlprefix\url{http://www.sciencedirect.com/science/article/pii/S1387380606000753}

\bibitem{Chen-PLB2010}
Chen L, Plass W~R, Geissel H, Kn{\"o}bel R, Kozhuharov C, Litvinov Y~A, Patyk
  Z, Scheidenberger C, Siegie{\'n}-Iwaniuk K, Sun B, Weick H, Beckert K, Beller
  P, Bosch F, Boutin D, Caceres L, Carroll J~J, Cullen D~M, Cullen I~J, Franzke
  B, Gerl J, G{\'o}rska M, Jones G~A, Kishada A, Kurcewicz J, Litvinov S~A, Liu
  Z, Mandal S, Montes F, M{\"u}nzenberg G, Nolden F, Ohtsubo T, Podoly{\'a}k Z,
  Propri R, Rigby S, Saito N, Saito T, Shindo M, Steck M, Ugorowski P, Walker
  P~M, Williams S, Winkler M, Wollersheim H~J and Yamaguchi T 2010 {\em Physics
  Letters B\/} {\bf 691} 234--237
  \urlprefix\url{http://www.sciencedirect.com/science/article/pii/S0370269310006945}

\bibitem{Sun-EPJ2007}
Sun B, Litvinov Y~A, Walker P~M, Beckert K, Beller P, Bosch F, Boutin D,
  Brandau C, Chen L, Dimopoulou C, Geissel H, Kn{\"o}bel R~K, Kozhuharov C,
  Kurcewicz J, Litvinov S~A, Mazzocco M, Meng J, Nociforo C, Nolden F, Plass
  W~R, Scheidenberger C, Steck M, Weick H and Winkler M 2007 {\em The European
  Physical Journal A\/} {\bf 31} 393
  \urlprefix\url{http://dx.doi.org/10.1140/epja/i2006-10252-0}

\bibitem{Walker-Nature1999}
Walker P and Dracoulis G 1999 {\em Nature\/} {\bf 399} 35--40
  \urlprefix\url{http://www.nature.com/doifinder/10.1038/19911}

\bibitem{Walker-HI2001}
Walker P~M and Dracoulis G~D 2001 {\em Hyperfine Interactions\/} {\bf 135}
  83--107 \urlprefix\url{http://dx.doi.org/10.1023/A\%3A1013915200556}

\bibitem{Dracoulis-PS2013}
Dracoulis G~D 2013 {\em Physica Scripta\/} {\bf T152} 014015
  \urlprefix\url{http://stacks.iop.org/1402-4896/2013/i=T152/a=014015}

\bibitem{Reed-PRL2010}
Reed M~W, Cullen I~J, Walker P~M, Litvinov Y~A, Blaum K, Bosch F, Brandau C,
  Carroll J~J, Cullen D~M, Deo A~Y, Detwiller B, Dimopoulou C, Dracoulis G~D,
  Farinon F, Geissel H, Haettner E, Heil M, Kempley R~S, Kn\"obel R, Kozhuharov
  C, Kurcewicz J, Kuzminchuk N, Litvinov S, Liu Z, Mao R, Nociforo C, Nolden F,
  Plass W~R, Prochazka A, Scheidenberger C, Steck M, St\"ohlker T, Sun B, Swan
  T~P~D, Trees G, Weick H, Winckler N, Winkler M, Woods P~J and Yamaguchi T
  2010 {\em Physical Review Letters\/} {\bf 105} 172501
  \urlprefix\url{http://link.aps.org/doi/10.1103/PhysRevLett.105.172501}

\bibitem{Reed-JPCS2012}
Reed M~W, Walker P~M, Cullen I~J, Litvinov Y~A, Blaum K, Bosch F, Brandau C,
  Carroll J~J, Cullen D~M, Deo A~Y, Detwiler B, Dimopoulou C, Dracoulis G~D,
  Farinon F, Geissel H, Haettner E, Heil M, Kempley R~S, Kn{\"o}bel R,
  Kozhuharov C, Kurcewicz J, Kuzminchuk N, Litvinov S, Liu Z, Mao R, Nociforo
  C, Nolden F, Plass W~R, Prochazka A, Scheidenberger C, Shubina D, Steck M,
  St{\"o}hlker T, Sun B, Swan T~P~D, Trees G, Weick H, Winckler N, Winkler M,
  Woods P~J and Yamaguchi T 2012 {\em Journal of Physics: Conference Series\/}
  {\bf 381} 2058
  \urlprefix\url{http://stacks.iop.org/1742-6596/381/i=1/a=012058}

\bibitem{Akber-PRC2015}
Akber A, Reed M~W, Walker P~M, Litvinov Y~A, Lane G~J, Kib\'edi T, Blaum K,
  Bosch F, Brandau C, Carroll J~J, Cullen D~M, Cullen I~J, Deo A~Y, Detwiler B,
  Dimopoulou C, Dracoulis G~D, Farinon F, Geissel H, Haettner E, Heil M,
  Kempley R~S, Kn\"obel R, Kozhuharov C, Kurcewicz J, Kuzminchuk N, Litvinov S,
  Liu Z, Mao R, Nociforo C, Nolden F, Pla\ss{} W~R, Podoly\'ak Z, Prochazka A,
  Scheidenberger C, Shubina D, Steck M, St\"ohlker T, Sun B, Swan T~P~D, Trees
  G, Weick H, Winckler N, Winkler M, Woods P~J and Yamaguchi T 2015 {\em Phys.
  Rev. C\/} {\bf 91}(3) 031301
  \urlprefix\url{http://link.aps.org/doi/10.1103/PhysRevC.91.031301}

\bibitem{Chen-PRL2013}
Chen L, Walker P~M, Geissel H, Litvinov Y~A, Beckert K, Beller P, Bosch F,
  Boutin D, Caceres L, Carroll J~J, Cullen D~M, Cullen I~J, Franzke B, Gerl J,
  G\'orska M, Jones G~A, Kishada A, Kn\"obel R, Kozhuharov C, Kurcewicz J,
  Litvinov S~A, Liu Z, Mandal S, Montes F, M\"unzenberg G, Nolden F, Ohtsubo T,
  Patyk Z, Pla\ss{} W~R, Podoly\'ak Z, Rigby S, Saito N, Saito T,
  Scheidenberger C, Simpson E~C, Shindo M, Steck M, Sun B, Williams S~J, Weick
  H, Winkler M, Wollersheim H~J and Yamaguchi T 2013 {\em Phys. Rev. Lett.\/}
  {\bf 110}(12) 122502
  \urlprefix\url{http://link.aps.org/doi/10.1103/PhysRevLett.110.122502}

\bibitem{Sun-NPA2010}
Sun B, Bosch F, Boutin D, Brandau C, Chen L, Dimopoulou C, Fabian B, Geissel H,
  Kn{\"o}bel R, Kozhuharov C, Kurcewicz J, Litvinov S~A, Litvinov Y~A, Meng J,
  M{\"u}nzenberg G, Nociforo C, Nolden F, Pla{\ss} W~R, Scheidenberger C, Steck
  M, Walker P~M, Weick H, Winckler N and Winkler M 2010 {\em Nuclear Physics
  A\/} {\bf 834} 476c--478c
  \urlprefix\url{http://www.sciencedirect.com/science/article/pii/S0375947410000709}

\bibitem{Sun-PLB2010}
Sun B, Kn{\"o}bel R, Geissel H, Litvinov Y, Walker P, Blaum K, Bosch F, Boutin
  D, Brandau C, Chen L, Cullen I, Dolinskii A, Fabian B, Hausmann M, Kozhuharov
  C, Kurcewicz J, Litvinov S, Liu Z, Mazzocco M, Meng J, Montes F,
  M{\"u}nzenberg G, Musumarra A, Nakajima S, Nociforo C, Nolden F, Ohtsubo T,
  Ozawa A, Patyk Z, Plass W, Scheidenberger C, Steck M, Suzuki T, Weick H,
  Winckler N, Winkler M and Yamaguchi T 2010 {\em Physics Letters B\/} {\bf
  688} 294--297
  \urlprefix\url{http://www.sciencedirect.com/science/article/pii/S0370269310004703}

\bibitem{Litvinov-PLB2003}
Litvinov Y~A, Attallah F, Beckert K, Bosch F, Boutin D, Falch M, Franzke B,
  Geissel H, Hausmann M, Kerscher T, Klepper O, Kluge H~J, Kozhuharov C,
  L{\"o}bner K~E~G, M{\"u}nzenberg G, Nolden F, Novikov Y~N, Patyk Z, Radon T,
  Scheidenberger C, Stadlmann J, Steck M, Trzhaskovskaya M~B and Wollnik H 2003
  {\em Physics Letters B\/} {\bf 573} 80--85
  \urlprefix\url{http://www.sciencedirect.com/science/article/pii/S0370269303013479}

\bibitem{Scheidenberger-APH2004}
Scheidenberger C, Attallah F, Beckert K, Beller P, Bosch F, Boutin D, Eickhoff
  H, Faestermann T, Falch M, Franczak B, Franzke B, Geissel H, Hausmann M,
  Hellstr{\"o}m M, Kaza E, Kerscher T, Klepper O, Kluge H~J, Koyama R,
  Kozhuharov C, Kratz K~L, Litvinov Y, L{\"o}bner K, Maier L, Matos M,
  M{\"u}nzenberg G, Nolden F, Novikov Y, Ohtsubo T, Ostrowski A, Ozawa A, Patyk
  Z, Pfeiffer B, Portillo M, Quint W, Radon T, Shishkin V, Stadlmann J, Steck
  M, S{\"u}mmerer M, Suzuki T, Trzhaskovskaja M, Vieira D, Watanabe S, Weick H,
  Winkler M, Wollnik H and Yamaguchi T 2004 {\em Acta Physica Hungarica Series
  A, Heavy Ion Physics\/} {\bf 19} 165--170
  \urlprefix\url{http://dx.doi.org/10.1556/APH.19.2004.1-2.27}

\bibitem{Wigner-1957}
Wigner E~P 1957 {\em R. A. Welch Foundation Conference on Chemical Research
  (Houston)\/} ed Millikan W~O p~67

\bibitem{Weinberg-PR1959}
Weinberg S and Treiman S~B 1959 {\em Phys. Rev.\/} {\bf 116}(2) 465--468
  \urlprefix\url{http://link.aps.org/doi/10.1103/PhysRev.116.465}

\bibitem{Lam-ADNDT2013}
Lam Y~H, Blank B, Smirnova N~A, Bueb J~B and Antony M~S 2013 {\em Atomic Data
  and Nuclear Data Tables\/} {\bf 99} 680 -- 703
  \urlprefix\url{http://www.sciencedirect.com/science/article/pii/S0092640X13000569}

\bibitem{Zhang-JPCS2013}
Zhang Y~H, Xu H~S, Litvinov Y~A, Tu X~L, Yan X~L, Typel S, Blaum K, Wang M,
  Zhou X~H, Sun Y, Brown B~A, Yuan Y~J, Xia J~W, Yang J~C, Audi G, Chen X~C,
  Jia G~B, Hu Z~G, Ma X~W, Mao R~S, Mei B, Shuai P, Sun Z~Y, Wang S~T, Xiao
  G~Q, Xu X, Yamaguchi T, Yamaguchi Y, Zang Y~D, Zhao H~W, Zhao T~C, Zhang W
  and Zhan W~L 2013 {\em Journal of Physics: Conference Series\/} {\bf 420}
  012054 \urlprefix\url{http://stacks.iop.org/1742-6596/420/i=1/a=012054}

\bibitem{Tu-JPG2014}
Tu X~L, Sun Y, Zhang Y~H, Xu H~S, Kaneko K, Litvinov Y~A and Wang M 2014 {\em
  Journal of Physics G: Nuclear and Particle Physics\/} {\bf 41} 025104
  \urlprefix\url{http://stacks.iop.org/0954-3899/41/i=2/a=025104}

\bibitem{Pyle-PRL2002}
Pyle M~C, Garc\'{\i}a A, Tatar E, Cox J, Nayak B~K, Triambak S, Laughman B,
  Komives A, Lamm L~O, Rolon J~E, Finnessy T, Knutson L~D and Voytas P~A 2002
  {\em Phys. Rev. Lett.\/} {\bf 88}(12) 122501
  \urlprefix\url{http://link.aps.org/doi/10.1103/PhysRevLett.88.122501}

\bibitem{Ringle-PRC2007}
Ringle R, Sun T, Bollen G, Davies D, Facina M, Huikari J, Kwan E, Morrissey
  D~J, Prinke A, Savory J, Schury P, Schwarz S and Sumithrarachchi C~S 2007
  {\em Phys. Rev. C\/} {\bf 75}(5) 055503
  \urlprefix\url{http://link.aps.org/doi/10.1103/PhysRevC.75.055503}

\bibitem{Yazidjian-PRC2007}
Yazidjian C, Audi G, Beck D, Blaum K, George S, Gu\'enaut C, Herfurth F,
  Herlert A, Kellerbauer A, Kluge H~J, Lunney D and Schweikhard L 2007 {\em
  Phys. Rev. C\/} {\bf 76}(2) 024308
  \urlprefix\url{http://link.aps.org/doi/10.1103/PhysRevC.76.024308}

\bibitem{Saastamoinen-PRC2009}
Saastamoinen A, Eronen T, Jokinen A, Elomaa V~V, Hakala J, Kankainen A, Moore
  I~D, Rahaman S, Rissanen J, Weber C, \"Ayst\"o J and Trache L 2009 {\em Phys.
  Rev. C\/} {\bf 80}(4) 044330
  \urlprefix\url{http://link.aps.org/doi/10.1103/PhysRevC.80.044330}

\bibitem{Kankainen-PRC2010}
Kankainen A, Eronen T, Gorelov D, Hakala J, Jokinen A, Kolhinen V~S, Reponen M,
  Rissanen J, Saastamoinen A, Sonnenschein V and \"Ayst\"o J 2010 {\em Phys.
  Rev. C\/} {\bf 82}(5) 052501
  \urlprefix\url{http://link.aps.org/doi/10.1103/PhysRevC.82.052501}

\bibitem{Schatz-PRL2001}
Schatz H, Aprahamian A, Barnard V, Bildsten L, Cumming A, Ouellette M, Rauscher
  T, Thielemann F~K and Wiescher M 2001 {\em Phys. Rev. Lett.\/} {\bf 86}(16)
  3471--3474
  \urlprefix\url{http://link.aps.org/doi/10.1103/PhysRevLett.86.3471}

\bibitem{Wallace-AJSS1981}
Wallace R~K and Woosley S~E 1981 {\em The Astrophysical Journal Supplement
  Series\/} {\bf 45} 389--420

\bibitem{Schatz-PR1998}
Schatz H, Aprahamian A, G{\"o}rres J, Wiescher M, Rauscher T, Rembges J,
  Thielemann F~K, Pfeiffer B, M{\"o}ller P, Kratz K~L, Herndl H, Brown B and
  Rebel H 1998 {\em Physics Reports\/} {\bf 294} 167--263
  \urlprefix\url{http://www.sciencedirect.com/science/article/pii/S0370157397000483}

\bibitem{Parikh-PRC2009}
Parikh A, Jos\'e J, Iliadis C, Moreno F and Rauscher T 2009 {\em Phys. Rev.
  C\/} {\bf 79}(4) 045802
  \urlprefix\url{http://link.aps.org/doi/10.1103/PhysRevC.79.045802}

\bibitem{Blank-PRL1995}
Blank B, Andriamonje S, Czajkowski S, Davi F, Del~Moral R, Dufour J~P, Fleury
  A, Musqu\`ere A, Pravikoff M~S, Grzywacz R, Janas Z, Pf\"utzner M, Grewe A,
  Heinz A, Junghans A, Lewitowicz M, Sauvestre J~E and Donzaud C 1995 {\em
  Phys. Rev. Lett.\/} {\bf 74}(23) 4611--4614
  \urlprefix\url{http://link.aps.org/doi/10.1103/PhysRevLett.74.4611}

\bibitem{Jokinen-ZP1996}
Jokinen A, Oinonen M, {\"A}yst{\"o} J, Baumann P, Didierjean F, Hoff P, Huck A,
  Knipper A, Marguier G, Novikov Y~N, Popov A~V, Ramdhane M, Seliverstov D~M,
  Van~Duppen P and Walter G 1996 {\em Zeitschrift f{\"u}r Physik\/} {\bf 355}
  227--230 \urlprefix\url{http://dx.doi.org/10.1007/BF02769690}

\bibitem{Schury-PRC2007}
Schury P, Bachelet C, Block M, Bollen G, Davies D~A, Facina M, Folden~III C~M,
  Gu\'enaut C, Huikari J, Kwan E, Kwiatkowski A, Morrissey D~J, Ringle R, Pang
  G~K, Prinke A, Savory J, Schatz H, Schwarz S, Sumithrarachchi C~S and Sun T
  2007 {\em Phys. Rev. C\/} {\bf 75}(5) 055801
  \urlprefix\url{http://link.aps.org/doi/10.1103/PhysRevC.75.055801}

\bibitem{Winger-PRC1993}
Winger J~A, Bazin D~P, Benenson W, Crawley G~M, Morrissey D~J, Orr N~A, Pfaff
  R, Sherrill B~M, Thoennessen M, Yennello S~J and Young B~M 1993 {\em Phys.
  Rev. C\/} {\bf 48}(6) 3097--3105
  \urlprefix\url{http://link.aps.org/doi/10.1103/PhysRevC.48.3097}

\bibitem{Audi-NPA2003}
Audi G, Wapstra A and Thibault C 2003 {\em Nuclear Physics A\/} {\bf 729}
  337--676
  \urlprefix\url{http://www.sciencedirect.com/science/article/pii/S0375947403018098}

\bibitem{VanWormer-APJ1994}
Van~Wormer L, Goerres J, Iliadis C, Wiescher M and Thielemann F~K 1994 {\em The
  Astrophysical Journal\/} {\bf 432} 326--350

\bibitem{Tu-NPA2016}
Tu X, Litvinov Y, Blaum K, Mei B, Sun B, Sun Y, Wang M, Xu H and Zhang Y 2016
  {\em Nuclear Physics A\/} {\bf 945} 89 -- 94
  \urlprefix\url{http://www.sciencedirect.com/science/article/pii/S0375947415002237}

\bibitem{Sanjari-PS2015}
Sanjari M~S, Chen X, H{\"u}lsmann P, Litvinov Y~A, Nolden F, Piotrowski J,
  Steck M and St{\"o}hlker T 2015 {\em Physica Scripta\/} {\bf T166} 014060
  \urlprefix\url{http://stacks.iop.org/1402-4896/2015/i=T166/a=014060}

\bibitem{Yamaguchi-PS2015a}
Yamaguchi T and {the Rare-RI Ring collaboration} 2015 {\em Physica Scripta\/}
  {\bf T166} 014039
  \urlprefix\url{http://stacks.iop.org/1402-4896/2015/i=T166/a=014039}

\bibitem{Litvinov-NIM2013}
Litvinov Y, Bishop S, Blaum K, Bosch F, Brandau C, Chen L, Dillmann I, Egelhof
  P, Geissel H, Grisenti R, Hagmann S, Heil M, Heinz A, Kalantar-Nayestanaki N,
  Kn{\"o}bel R, Kozhuharov C, Lestinsky M, Ma X, Nilsson T, Nolden F, Ozawa A,
  Raabe R, Reed M, Reifarth R, Sanjari M, Schneider D, Simon H, Steck M,
  St{\"o}hlker T, Sun B, Tu X, Uesaka T, Walker P, Wakasugi M, Weick H,
  Winckler N, Woods P, Xu H, Yamaguchi T, Yamaguchi Y and Zhang Y 2013 {\em
  Nuclear Instruments and Methods in Physics Research Section B: Beam
  Interactions with Materials and Atoms\/} {\bf 317, Part B} 603--616
  \urlprefix\url{http://www.sciencedirect.com/science/article/pii/S0168583X13008197}

\bibitem{Yan-JPCS2016}
Yan X~L, Blaum K, Litvinov Y~A, Tu X~L, Xu H~S, Zhang Y~H, Zhou X~H and
  on~In-Ring Mass~Measurements E~~C~C 2016 {\em Journal of Physics: Conference
  Series\/} {\bf 665} 012053
  \urlprefix\url{http://stacks.iop.org/1742-6596/665/i=1/a=012053}

\bibitem{Nilsson-PS2015}
Nilsson T and {the NUSTAR collaboration} 2015 {\em Physica Scripta\/} {\bf
  T166} 014070
  \urlprefix\url{http://stacks.iop.org/1402-4896/2015/i=T166/a=014070}

\bibitem{Krucken-AIP2006}
Kr{\"u}cken R, Bosch F, Cargnelli M, Fabbietti L, Faestemann T, Franzke B,
  Fuhrmann H, Hayano R~S, Hirtl A, Homolka J, Kienle P, Kozhuharov C, Lenske H,
  Litvinov Y, Marton J, Nolden F, Ring P, Shatunov Y, Skrinsky A~N, Suzuki K,
  Vostrikov V~A, Yamaguchi T, Widmann E, Wycech S and Zmeskal J 2006 {\em AIP
  Conference Proceedings\/} {\bf 831} 3--7
  \urlprefix\url{http://scitation.aip.org/content/aip/proceeding/aipcp/10.1063/1.2200890}

\bibitem{Moeini-NIM2011}
Moeini H, Ilieva S, Aksouh F, Boretzky K, Chatillon A, Corsi A, Egelhof P,
  Emling H, Ickert G, Jourdan J, Nayestanaki N~K, Kiselev D, Kiselev O,
  Kozhuharov C, Bleis T~L, Le X, Litvinov Y, Mahata K, Meier J, Nolden F,
  Paschalis S, Popp U, Simon H, Steck M, St{\"o}hlker T, Weick H,
  Werthm{\"u}ller D and Zalite A 2011 {\em Nuclear Instruments and Methods in
  Physics Research Section A: Accelerators, Spectrometers, Detectors and
  Associated Equipment\/} {\bf 634} 77 -- 84
  \urlprefix\url{http://www.sciencedirect.com/science/article/pii/S0168900211000878}

\bibitem{Antonov-NIM2011}
Antonov A, Gaidarov M, Ivanov M, Kadrev D, Aiche M, Barreau G, Czajkowski S,
  Jurado B, Belier G, Chatillon A, Granier T, Taieb J, Dore D, Letourneau A,
  Ridikas D, Dupont E, Berthoumieux E, Panebianco S, Farget F, Schmitt C,
  Audouin L, Khan E, Tassan-Got L, Aumann T, Beller P, Boretzky K, Dolinskii A,
  Egelhof P, Emling H, Franzke B, Geissel H, Kelic-Heil A, Kester O, Kurz N,
  Litvinov Y, M{\"u}nzenberg G, Nolden F, Schmidt K~H, Scheidenberger C, Simon
  H, Steck M, Weick H, Enders J, Pietralla N, Richter A, Schrieder G, Zilges A,
  Distler M, Merkel H, M{\"u}ller U, Junghans A, Lenske H, Fujiwara M, Suda T,
  Kato S, Adachi T, Hamieh S, Harakeh M, Kalantar-Nayestanaki N, W{\"o}rtche H,
  Berg G, Koop I, Logatchov P, Otboev A, Parkhomchuk V, Shatilov D, Shatunov P,
  Shatunov Y, Shiyankov S, Shvartz D, Skrinsky A, Chulkov L, Danilin B,
  Korsheninnikov A, Kuzmin E, Ogloblin A, Volkov V, Grishkin Y, Lisin V,
  Mushkarenkov A, Nedorezov V, Polonski A, Rudnev N, Turinge A, Artukh A,
  Avdeichikov V, Ershov S, Fomichev A, Golovkov M, Gorshkov A, Grigorenko L,
  Klygin S, Krupko S, Meshkov I, Rodin A, Sereda Y, Seleznev I, Sidorchuk S,
  Syresin E, Stepantsov S, Ter-Akopian G, Teterev Y, Vorontsov A, Kamerdzhiev
  S, Litvinova E, Karataglidis S, Rodriguez R~A, Borge M, Ramirez C~F, Garrido
  E, Sarriguren P, Vignote J, Prieto L~F, Herraiz J~L, de~Guerra E~M,
  Udias-Moinelo J, Soriano J~A, Rojo A~L, Caballero J, Johansson H, Jonson B,
  Nilsson T, Nyman G, Zhukov M, Golubev P, Rudolph D, Hencken K, Jourdan J,
  Krusche B, Rauscher T, Kiselev D, Trautmann D, Al-Khalili J, Catford W,
  Johnson R, Stevenson P, Barton C, Jenkins D, Lemmon R, Chartier M, Cullen D,
  Bertulani C and Heinz A 2011 {\em Nuclear Instruments and Methods in Physics
  Research Section A: Accelerators, Spectrometers, Detectors and Associated
  Equipment\/} {\bf 637} 60 -- 76
  \urlprefix\url{http://www.sciencedirect.com/science/article/pii/S0168900211002300}

\bibitem{Stohlker-NIM2015}
St{\"o}hlker T, Bagnoud V, Blaum K, Blazevic A, Br{\"a}uning-Demian A, Durante
  M, Herfurth F, Lestinsky M, Litvinov Y, Neff S, Pleskac R, Schuch R,
  Schippers S, Severin D, Tauschwitz A, Trautmann C, Varentsov D and Widmann E
  2015 {\em Nuclear Instruments and Methods in Physics Research Section B: Beam
  Interactions with Materials and Atoms\/} {\bf 365, Part B} 680 -- 685
  \urlprefix\url{http://www.sciencedirect.com/science/article/pii/S0168583X15006552}

\bibitem{Geissel-NIM2003}
Geissel H, Weick H, Winkler M, M{\"u}nzenberg G, Chichkine V, Yavor M, Aumann
  T, Behr K, B{\"o}hmer M, Br{\"u}nle A, Burkard K, Benlliure J, Cortina-Gil D,
  Chulkov L, Dael A, Ducret J~E, Emling H, Franczak B, Friese J, Gastineau B,
  Gerl J, Gernh{\"a}user R, Hellstr{\"o}m M, Jonson B, Kojouharova J, Kulessa
  R, Kindler B, Kurz N, Lommel B, Mittig W, Moritz G, M{\"u}hle C, Nolen J,
  Nyman G, Roussell-Chomaz P, Scheidenberger C, Schmidt K~H, Schrieder G,
  Sherrill B, Simon H, S{\"u}mmerer K, Tahir N, Vysotsky V, Wollnik H and
  Zeller A 2003 {\em Nuclear Instruments and Methods in Physics Research
  Section B: Beam Interactions with Materials and Atoms\/} {\bf 204} 71 -- 85
  \urlprefix\url{http://www.sciencedirect.com/science/article/pii/S0168583X02018931}

\bibitem{Dolinskii-PS2015}
Dolinskii A, Berkaev D, Blell U, Dimopoulou C, Gorda O, Leibrock H, Litvinov S,
  Laier U, Koop I, Schurig I, Starostenko A, Shatunov P and Weinrich U 2015
  {\em Physica Scripta\/} {\bf T166} 014040
  \urlprefix\url{http://stacks.iop.org/1402-4896/2015/i=T166/a=014040}

\bibitem{Dolinskii-NIM2007}
Dolinskii A, Litvinov S, Steck M and Weick H 2007 {\em Nuclear Instruments and
  Methods in Physics Research Section A: Accelerators, Spectrometers, Detectors
  and Associated Equipment\/} {\bf 574} 207 -- 212
  \urlprefix\url{http://www.sciencedirect.com/science/article/pii/S0168900207002525}

\bibitem{SLitvinov-NIM2013}
Litvinov S, Toprek D, Weick H and Dolinskii A 2013 {\em Nuclear Instruments and
  Methods in Physics Research Section A: Accelerators, Spectrometers, Detectors
  and Associated Equipment\/} {\bf 724} 20 -- 26
  \urlprefix\url{http://www.sciencedirect.com/science/article/pii/S0168900213006633}

\bibitem{Kovalenko-PS2015}
Kovalenko O, Dolinskii O, Litvinov Y~A, Maier R, Prasuhn D and St{\"o}hlker T
  2015 {\em Physica Scripta\/} {\bf T166} 014042
  \urlprefix\url{http://stacks.iop.org/1402-4896/2015/i=T166/a=014042}

\bibitem{Yang-NIM2013}
Yang J, Xia J, Xiao G, Xu H, Zhao H, Zhou X, Ma X, He Y, Ma L, Gao D, Meng J,
  Xu Z, Mao R, Zhang W, Wang Y, Sun L, Yuan Y, Yuan P, Zhan W, Shi J, Chai W,
  Yin D, Li P, Li J, Mao L, Zhang J and Sheng L 2013 {\em Nuclear Instruments
  and Methods in Physics Research Section B: Beam Interactions with Materials
  and Atoms\/} {\bf 317, Part B} 263--265
  \urlprefix\url{http://www.sciencedirect.com/science/article/pii/S0168583X13009877}

\bibitem{Gao-CPC2014}
Gao X, Yang J~C, Xia J~W, Chai W~P, Shi J and Shen G~D 2014 {\em Chinese
  Physics C\/} {\bf 38} 047002
  \urlprefix\url{http://stacks.iop.org/1674-1137/38/i=4/a=047002}

\bibitem{Grieser-EPJST2012}
Grieser M, Litvinov Y~A, Raabe R, Blaum K, Blumenfeld Y, Butler P~A, Wenander
  F, Woods P~J, Aliotta M, Andreyev A, Artemyev A, Atanasov D, Aumann T,
  Balabanski D, Barzakh A, Batist L, Bernardes A~P, Bernhardt D, Billowes J,
  Bishop S, Borge M, Borzov I, Bosch F, Boston A~J, Brandau C, Catford W,
  Catherall R, Cederk{\"a}ll J, Cullen D, Davinson T, Dillmann I, Dimopoulou C,
  Dracoulis G, D{\"u}llmann C~E, Egelhof P, Estrade A, Fischer D, Flanagan K,
  Fraile L, Fraser M~A, Freeman S~J, Geissel H, Gerl J, Greenlees P, Grisenti
  R~E, Habs D, von Hahn R, Hagmann S, Hausmann M, He J~J, Heil M, Huyse M,
  Jenkins D, Jokinen A, Jonson B, Joss D~T, Kadi Y, Kalantar-Nayestanaki N, Kay
  B~P, Kiselev O, Kluge H~J, Kowalska M, Kozhuharov C, Kreim S, Kr{\"o}ll T,
  Kurcewicz J, Labiche M, Lemmon R~C, Lestinsky M, Lotay G, Ma X~W, Marta M,
  Meng J, M{\"u}cher D, Mukha I, M{\"u}ller A, Murphy A~S~J, Neyens G, Nilsson
  T, Nociforo C, N{\"o}rtersh{\"a}user W, Page R~D, Pasini M, Petridis N,
  Pietralla N, Pf{\"u}tzner M, Podoly{\'a}k Z, Regan P, Reed M~W, Reifarth R,
  Reiter P, Repnow R, Riisager K, Rubio B, Sanjari M~S, Savin D~W,
  Scheidenberger C, Schippers S, Schneider D, Schuch R, Schwalm D, Schweikhard
  L, Shubina D, Siesling E, Simon H, Simpson J, Smith J, Sonnabend K, Steck M,
  Stora T, St{\"o}hlker T, Sun B, Surzhykov A, Suzaki F, Tarasov O, Trotsenko
  S, Tu X~L, van Duppen P, Volpe C, Voulot D, Walker P~M, Wildner E, Winckler
  N, Winters D~F~A, Wolf A, Xu H~S, Yakushev A, Yamaguchi T, Yuan Y~J, Zhang
  Y~H and Zuber K 2012 {\em The European Physical Journal Special Topics\/}
  {\bf 207} 1--117
  \urlprefix\url{http://dx.doi.org/10.1140/epjst/e2012-01599-9}

\bibitem{Reifarth-PRST2014}
Reifarth R and Litvinov Y~A 2014 {\em Phys. Rev. ST Accel. Beams\/} {\bf 17}(1)
  014701 \urlprefix\url{http://link.aps.org/doi/10.1103/PhysRevSTAB.17.014701}

\bibitem{Glorius-PS2015}
Glorius J, Litvinov Y~A and Reifarth R 2015 {\em Physica Scripta\/} {\bf T166}
  014008 \urlprefix\url{http://stacks.iop.org/1402-4896/2015/i=T166/a=014008}

\end{thebibliography}
\end{document}